\documentclass[fleqn,usenatbib]{mnras}

\usepackage{newtxtext,newtxmath}
\usepackage[T1]{fontenc}

\DeclareRobustCommand{\VAN}[3]{#2}
\let\VANthebibliography\thebibliography
\def\thebibliography{\DeclareRobustCommand{\VAN}[3]{##3}\VANthebibliography}

\usepackage{amsfonts,amsmath,amssymb,mathrsfs}
\usepackage{graphicx}
\usepackage{subfigure}
\usepackage{epstopdf}
\usepackage{float}
\usepackage{color}
\usepackage{cancel}
\usepackage{ulem}
\usepackage{threeparttable}
\usepackage{smartref}
\usepackage{makecell}
\usepackage{multirow}
\usepackage{verbatim}
\usepackage{textcomp}
\usepackage{hyperref}
\usepackage[misc]{ifsym} %% Letter symbol
\usepackage{changepage}
\usepackage{tabularx}
\usepackage{booktabs}
\usepackage{CJK}

\title[Reconstructing the Speed of Light with GPR]{A Stochastic Approach to Reconstructing the Speed of Light in Cosmology}

\author[Cheng-Yu Zhang et al.]{
Cheng-Yu Zhang,$^{1,2}$
Wei Hong,$^{1,2}$
Yu-Chen Wang$^{3,4}$
and Tong-Jie Zhang$^{1,2}\thanks{tjzhang@bnu.edu.cn}$
\\
$^{1}$Institute for Frontiers in Astronomy and Astrophysics, Beijing Normal University, Beijing 102206, China\\
$^{2}$School of Physics and Astronomy, Beijing Normal University, Beijing 100875, China\\
$^{3}$Kavli Institute for Astronomy and Astrophysics, Peking University, Beijing 100871, China\\
$^{4}$Department of Astronomy, School of Physics, Peking University, Beijing 100871, China
}

\date{Accepted XXX. Received YYY; in original form ZZZ}

\pubyear{\the\year{}}

\begin{document}
\begin{CJK*}{UTF8}{gbsn}
\label{firstpage}
\pagerange{\pageref{firstpage}--\pageref{lastpage}}
\maketitle

\begin{abstract}
The Varying Speed of Light (VSL) model describes how the speed of light in a vacuum changes with cosmological redshift. Despite numerous models, there is little observational evidence for this variation. While the speed of light can be accurately measured by physical means, cosmological methods are rarely used. Previous studies quantified the speed of light at specific redshifts using Gaussian processes and reconstructed the redshift-dependent function $c(z)$. It is crucial to quantify the speed of light across varying redshifts. We use the latest data on angular diameter distances $D_A(z)$ and Hubble parameters $H(z)$ from baryon acoustic oscillation (BAO) and cosmic chronometer measurements in the redshift interval $z\in[0.07,1.965]$. The speed of light $c(z)$ is determined using Gaussian and deep Gaussian processes to reconstruct $H(z)$, $D_A(z)$, and $D^{\prime}_A(z)$. Furthermore, we conduct comparisons across three distinct models, encompassing two renowned VSL models. We get the result of the parameters constraints in the models (1) for the ``$c$-c" model, $c_0=29492.6 \pm^{6.2}_{5.3} \mathrm{~km} \mathrm{~s}^{-1}$. (2) For the ``$c$-cl" model, $c_0=29665.5 \pm^{11.2}_{11.4}\mathrm{~km} \mathrm{~s}^{-1}$ and $n=0.05535 \pm^{0.00008}_{0.00007}$. (3) For the ``$c$-CPL" model, $c_0=29555.7 \pm^{13.3}_{13.2} \mathrm{~km} \mathrm{~s}^{-1}$ and $n=-0.0607 \pm 0.0001$. Based on our findings, it may be inferred that Barrow's classical VSL model is not a suitable fit for our data. In contrast, the widely recognized Chevallier-Polarski-Linder (CPL) VSL model, under some circumstances, as well as the universal ``c is constant" model, demonstrate a satisfactory ability to account for our findings.
\end{abstract}

\begin{keywords}
Cosmology--methods: data analysis
\end{keywords}

%%%%%%%%%%%%%%%%%%%%%%%%%%%%%%%%%%%%%%%%%%%%%%%%%%

%%%%%%%%%%%%%%%%% BODY OF PAPER %%%%%%%%%%%%%%%%%%

\section{Introduction} \label{sec:intro}
The foundation and subsequent development of the current standard cosmological model (SCM) stands as a paramount accomplishment in the field of astronomy throughout the 20th century. Such a universe model might be regarded as the ``ground state" within the framework of general relativity. Within this cosmological framework, it is essential to acknowledge that the speed of light is a constant. This is an inevitable consequence due to the fundamental concept of Lorentz invariance in general relativity. Lorentz invariance, in turn, arises from two distinct postulates: the principle of relativity and the principle of constancy of the speed of light. Although the model demonstrates applicability to numerous phenomena inside our universe, there is some aspect that defies explanation \citep{2022NewAR..9501659P}. The issues pertaining to the horizon and flatness are currently under active discussion. The inflation hypothesis is a widely accepted paradigm that aims to address these issues. Conversely, some theories suggest that the speed of light increases as the universe evolves, leading to the proposal of the varying speed of light (VSL) model as a way to address these challenges. The foundational framework of this approach was initially suggested by \cite{1911AnP...340..898E}. Subsequently, the contemporary form of VSL was introduced by \cite{1993IJMPD...2..351M} in 1992. Albrecht, Barrow, and Magueijo have together developed a model that demonstrates a process for transforming the Einstein de Sitter model into a cosmological attractor \citep{1999PhRvD..59d3516A,Barrow:1998df,PhysRevD.59.043515,Barrow:1998he,Barrow:1999jq,Barrow:1999st}. This model has been established for a certain length of time. In the subsequent study, \cite{Magueijo:2000zt} presents a theoretical framework that introduces concepts for covariance and local Lorentz invariance in the context of the varying speed of light. The aforementioned approach possesses the advantage of selectively preserving the elements of conventional definitions that remain unchanged under unit transformations, thereby enabling a valid representation of experimental results. In 2003, \cite{2003RPPh...66.2025M} presented a comprehensive evaluation of their research endeavors pertaining to the plausibility of VSL. The model is coming into prominence, but without enough observational evidence. Any alteration in the speed of light ultimately culminates in a dissonance between two velocities, potentially giving rise to anomalous Cherenkov radiation, a phenomenon meticulously delimited by empirical observations \citep{2009ARNPS..59..245L}.

The vastness of the universe provides a plethora of observational data. Baryonic acoustic oscillations (BAO), in conjunction with additional observational datasets such as Type Ia supernovae (SNe Ia), observational Hubble data (OHD), large-scale structures, the cosmic microwave background, among others, can serve as valuable tools for constraining cosmological parameters. An alternative approach involves the computation of the differential ages of galaxies undergoing passive evolution at various redshifts. This method yields measurements of the Hubble parameter $H(z)$ that are not reliant on any specific model \citep{2002ApJ...573...37J}. This approach allows for the determination of the change rate $\Delta z/ \Delta t$, which can then be used to express the Hubble parameter $H(z)$ as $H(z) \simeq-\frac{1}{1+z} \frac{\Delta z}{\Delta t}$. The technique commonly referred to as cosmic chronometers (CCs) is typically employed in this context, with the corresponding $H(z)$ data derived from this method being denoted as CC $H(z)$. Several galaxy redshift surveys, including the Sloan Digital Sky Survey (SDSS) \citep{2023ApJS..267...44A,2022ApJS..259...35A,2020ApJS..249....3A,2019ApJS..240...23A,2018ApJS..235...42A,2017ApJS..233...25A,2015ApJS..219...12A,2014ApJS..211...17A,2012ApJS..203...21A,2011ApJS..193...29A,2009ApJS..182..543A,2008ApJS..175..297A,2007ApJS..172..634A,2006ApJS..162...38A,2005AJ....129.1755A,2004AJ....128..502A,2003AJ....126.2081A,2002AJ....123..485S}, the 6dF Galaxy Survey \citep{2005PASA...22..277J,2004MNRAS.355..747J,2009MNRAS.399..683J,2011MNRAS.416.3017B}, the Baryon Oscillation Sky Survey (BOSS) \citep{2013JCAP...04..026S,2013AJ....145...10D,2017MNRAS.464.3409B,2017MNRAS.469.1369S,2017MNRAS.464.1640S,2017MNRAS.467.2085G,2017MNRAS.466.2242B} provide the opportunity to measure the angular diameter distance $D_A(z)$, and the Hubble parameter $H(z)$ can be derived from the data of the WiggleZ Dark Energy Survey \citep{2010MNRAS.401.1429D,2011MNRAS.418.1707B,2014MNRAS.441.3524K,2012PhRvD..86j3518P,blake-2012-33470,10.1093/mnras/stx2963}, the third generation Slogan Digital Sky Survey (SDSS-\uppercase\expandafter{\romannumeral3}), strong gravitational lenses \citep{2015JCAP...11..033J,2019ApJ...883....3L}, gravitational waves \citep{2017ApJ...849L..16I}, galaxy clusters \citep{2006ApJ...647...25B}, etc., which makes it possible for us to use a larger $D_A(z)$ and $H(z)$ data set to measure the speed of light $c(z)$.

The advancement of machine learning and its widespread application in cosmology have led to the development of various methods aimed at improving the precision of data constraints. The Gaussian Process (GP) is widely recognized as a prominent technique in the field of astronomy. It serves as a non-parametric machine learning model that effectively captures the characteristics of functions within a stochastic statistical process \cite{2006gpml.book.....R}. Through the utilization of this method, it becomes possible to effectively accommodate the data set and obtain the projected value at any given point. A method utilizing GP was presented in \cite{PhysRevLett.114.101304} to determine the speed of light at a specific redshift. In this study, the authors \cite{2022JCAP...07..029R} employ a particular methodology that involves the utilization of two distinct covariance functions in order to obtain the value of $c(z)$ at a specific redshift. Subsequently, in accordance with this viewpoint, the authors proceed to reconstruct the function $c(z)$ inside the redshift interval $z\in[0,2]$. \cite{2016JCAP...08..016C} proposes a novel approach that is independent of any specific model to address the issue of degeneracy between cosmic curvature and the speed of light. The aim is to investigate the constancy of the speed of light, denoted as c. In this study, we adopt the approach outlined in the work of \citep{PhysRevLett.114.101304} to reconstruct the function $c(z)$ within the redshift interval $z\in[0.07,1.965]$. Our objective is to examine the relationship between the redshift $z$ and the corresponding changes in the quantity $c$. We present a visual representation of this relationship in the form of a figure. It is important to note that our ability to enhance the amount of information utilized in this analysis is limited by the constraints imposed by the selection and combination of observational data. The inaccuracy of predictions beyond the existing observational data stems from the inherent uncertainty associated with unknown future observational outcomes. We utilized a total of 35 data points for the evaluation of $H(z)$ using the CC approach, in addition to the 64 data points for $D_A(z)$ obtained from BAO and other observations \citep{2019ApJ...883....3L,2017ApJ...849L..16I,2015JCAP...11..033J}. The inclusion of these data points significantly enhances the accuracy and reliability of the Gaussian Process. The GP is extensively employed in several domains. The accurate determination of its computation, encompassing hyperparameters, the number of hyperparameters, and the selection of kernels, can significantly influence the reconstruction of cosmological data and the accuracy of our predictions. Hence, it is imperative to engage in a comprehensive discussion of GP \citep{2021ApJ...915..123S,2012PhRvD..85l3530S,2023JCAP...02..014H,2023ApJS..266...27Z}.

The rest of the paper is organized as follows: In Section \ref{sec:THEORETICAL BASIS}, we provide the theoretical basis for the cosmological measurement of $c$, along with various models of the VSL and GP. In Section \ref{sec:data analysis}, we describe how we use the GP to fit the data points. In Section \ref{sec:results}, we provide the variation tendency of $c$ and compare three models to discuss whether the trend conforms to the VSL model or not. Finally, in Section \ref{sec:conclusion}, we conclude our work and discuss some possible future work.

\section{Theoretical Basis} \label{sec:THEORETICAL BASIS}

\subsection{The Measurement of $c$ from Angular Diameter Distance}
The methodology employed in this paper is predicated on the literature referenced as \cite{PhysRevLett.114.101304}. Our endeavor is to constrain the speed of light by utilizing the latest dataset encompassing the angular diameter distance $D_A(z)$, in conjunction with observational Hubble data $H(z)$. The ensuing section will expound upon the meticulous theoretical underpinnings.

Firstly, in the VSL, the expression for the angular diameter distance can be derived as follows with assuming no spatial curvature and speed of light is no longer constant
\begin{equation}\label{eq1}
D_{\mathrm{A}}(z)=\frac{1}{(1+z)} \int_{0}^{z} \frac{c(z) d z}{H(z)}.
\end{equation}
A clear distinction can be observed between the functions $H(z)$ and $D_A(z)$ in that the former serves as a direct limitation on the Hubble parameter, while the latter imposes a constraint on the integral of the reciprocal of the Hubble parameter. Given that $H(z)$ exhibits a strictly rising behavior with respect to redshift, it follows that the integral in question displays a higher sensitivity to fluctuations in $H(z)$ in the vicinity of $z = 0$, whereas its sensitivity diminishes as the value of $z$ increases. We can then proceed to differentiate the function of speed of light with respect to $z$
\begin{equation}\label{eq2}
c(H(z),D_A(z),D_{\mathrm{A}}^{\prime}(z);z)=H(z)\left[(1+z) D_{\mathrm{A}}^{\prime}(z)+D_{\mathrm{A}}(z)\right].
\end{equation}
$c(H(z),D_A(z),D_{\mathrm{A}}^{\prime}(z);z)$'s uncertainty can be obtained through the standard error propagation as we assume that the $H(z)$ and $D_A(z)$ datasets are independent of each other, and due to the lack of error in redshift data, the redshift error term is not considered
\begin{equation}\label{eq3}
	\begin{aligned}
		\sigma_{c(H(z),D_A(z),D_{\mathrm{A}}^{\prime}(z);z)}^{2}=\left[(1+z) D_{\mathrm{A}}^{\prime}(z)+D_{\mathrm{A}}(z)\right]^{2} \sigma_{H(z)}^{2}
		\\+[H(z)(1+z)]^{2} \sigma_{D_{\mathrm{A}}^{\prime}(z)}^{2} 
		+H(z)^{2} \sigma_{D_{\mathrm{A}}(z)}^{2}.
	\end{aligned}
\end{equation}
It should be noted that our formulas here are different from similar formulas in \citep{2022JCAP...07..029R}, and their understanding of error propagation is unusual. 

Finally, it is worth noting that $D_A(z)$ has a maximum where $D^{\prime}_A(z_m)$ = 0, so we assume that at the maximum point $z_m$, we can get
\begin{equation}\label{eq4}
c\left(H(z),D_A(z),D_{\mathrm{A}}^{\prime}(z);z_{\mathrm{m}}\right)=D_{\mathrm{A}}\left(z_{\mathrm{m}}\right) H\left(z_{\mathrm{m}}\right).
\end{equation}
According to Equation (\ref{eq4}), \cite{PhysRevLett.114.101304} reconstructs the $H(z)$ and $D_A(z)$ and find the $z_m$ to get the $c\left(H(z),D_A(z),D_{\mathrm{A}}^{\prime}(z);z_{\mathrm{m}}\right)$. From a mathematical and empirical point of view, the maximum point $z_m$ is critical to the fitting of the final curve, as it is more sensitive to the data and contains more cosmological information than other points on the $D_A(z)$ curve \citep{2023ApJS..268...67H}. Nevertheless, this approach alone provides the opportunity to quantify the parameter $c$ at a single redshift value denoted as $z_{M}$. It is important to exercise caution when utilizing the variable $z_{M}$ in order to facilitate the simplification of the equation, hence enabling a more precise measurement of the variable $c$. It is noted that equations (\ref{eq2}) and (\ref{eq3}) also apply to other $c(H(z),D_A(z),D_{\mathrm{A}}^{\prime}(z);z)$, so in our research, we try to get more $c(H(z),D_A(z),D_{\mathrm{A}}^{\prime}(z);z)$ at different redshifts according to Equation (\ref{eq2}). we reconstruct the $H(z)$, $D_A(z)$, and $D^{\prime}_A(z)$, and by using Equation (\ref{eq2}), we obtain $c(H(z),D_A(z),D_{\mathrm{A}}^{\prime}(z);z)$ with errors in the redshift range $[0.07,1.965]$.

\subsection{The Model of VSL}
The proposal of the VSL model emerged as an attempt to address the horizon and flatness issues within the field of cosmology. In this section, we provide a concise overview of two VSL models. The first model, referred to as the ``$c$-cl model", is documented in the \citep{PhysRevD.59.043515}. The second model discussed in this study is derived from the widely recognized Chevallier-Polarski-Linder (CPL) model \citep{2001IJMPD..10..213C,PhysRevLett.90.091301}. The CPL model is commonly employed as the benchmark model for dynamical dark energy theories, and hence, it is referred to as the ``$c$-CPL" model in this context.

In the minimally coupled theory, the substitution of the constant $c$ with a field is performed inside the framework of the preferred frame for the $c$-cl model. Hence, the action remains as \citep{Barrow:1998he}
\begin{equation}
	S=\int d x^4\left(\sqrt{-g}\left(\frac{\psi(R+2 \Lambda)}{16 \pi G}+\mathcal{L}_M\right)+\mathcal{L}_\psi\right)
\end{equation}
with $\psi\left(x^\mu\right)=c^4$. The dynamical variables consist of the metric tensor $g_{\mu \nu}$, any matter field variables present in the matter Lagrangian $\mathcal{L}_M$, and the scalar field $\psi$ itself. From this, the Friedmann, the acceleration, and the fluid equation can be expressed as
\begin{equation}
	\begin{aligned}
	&\frac{\dot{a}^2}{a^2}=\frac{8 \pi G(t) \rho}{3}-\frac{K c^2(t)}{a^2},\\
	&\ddot{a}=-\frac{4 \pi G(t)}{3}\left(\rho+\frac{3 p}{c^2(t)}\right) a,\\
	&\dot{\rho}+3 \frac{\dot{a}}{a}\left(\rho+\frac{p}{c^2}\right)=-\rho \frac{\dot{G}}{G}+\frac{3 K c \dot{c}}{4 \pi G a^2},
	\end{aligned}
\end{equation}
with the remaining matter obeys an equation of state of the form
\begin{equation}
	p=(\gamma-1) \rho c^2(t),
\end{equation}
where $\rho$ and $p$ represent the density and pressure of the matter, respectively. The metric curvature parameter is denoted as $K$, whereas $\gamma$ is a constant. Consequently, the speed of light, denoted as $c$, undergoes variations within the local Lorentzian frames that are associated with the cosmological expansion. Additionally, a minimal coupling arises in Einstein's equations due to the omission of surface factors, which can be attributed to a special-relativistic effect.

In order to solve the generalized conservation equation, \cite{PhysRevD.59.043515} assumes that the rate of variation of $c(t)$ is proportional to the expansion rate of the universe
\begin{equation}
	c(t)=c_0 a(t)^n=c_0\left(\frac{a_0}{1+z}\right)^n,
\end{equation}
where $c_0$ and $n$ are constant, $a_0=1$, and $z$ denotes the redshift. The flatness problem and the horizon problem can be resolved irrespective of the behavior of $G(t)$ when $n \leqslant \frac{1}{2}(2-3 \gamma)$. The Lambda problem can be resolved when $n<-\frac{3 \gamma}{2}$ and the rate of variation $G(t)$ is proportional to the expansion rate of the universe, expressed as $G(t)=G_0 a^q$, where $G_0$ and $q$ are constants. However, it should be noted that the model has its limitations. If $c$ varies, there may be potential issues with the perturbations to the isotropic expansion of the universe, which manifest as powers of $v/c$. If no other modifications to physics exist, this phenomenon results in alterations to the fine structure constant and other gauge couplings during the initial stages of the universe. One may need a special tuning of the initial sizes of these terms in the Friedmann equation with respect to the density term in order for their effects to just start to become significant close to the present epoch.

The second comes from the well-known CPL model \citep{2001IJMPD..10..213C,PhysRevLett.90.091301}, which was introduced to solve the problem of the evolution of dark energy during the evolution of the VSL model. Based on the CPL model, the fluid equation of dark energy can be expressed as
\begin{equation}
	\dot{\rho}_{\mathrm{DE}}(a)+\frac{3}{a}\left[1+w_{\mathrm{DE}}(a)\right] \rho_{\mathrm{DE}}(a)=\frac{3 K c(a)\dot{c}(a)}{4 \pi G a^2}.
\end{equation}
Inspired by the equation of state $w(a)=w_0+w_a(1-a)$, a new hypothesis of variable velocity of light is introduced to solve the generalized conservation equation
\begin{equation}
	c(t)=c_0\left[1+n(1-a(t))\right] =c_0\left[1+n\left(1-\frac{a_0}{1+z}\right)\right],
\end{equation}
where $c_0$ and $n$ are constants.

\subsection{Gaussian Process}
The Gaussian Process (GP) is a machine learning technique employed for regression, specifically for estimating the value at a new location based on a given set of prior values. The underlying principle of this approach is based on the assumption that all values are drawn from a joint Gaussian distribution within the context of function space \citep{2006gpml.book.....R}. By employing the aforementioned assumption, along with a specification of the anticipated mean and an assumption on the covariance between data points, it becomes possible to derive estimations for a given set of observational data points. More precisely, the Gaussian random variable associated with a reconstructed point $z$ denotes the anticipated value for the GP.

In the scope of our research, it is necessary to undertake the task of reconstructing three functions, namely $H(z)$, $D_{A}(z)$, and $D_{A}^{\prime}(z)$. Hence, it is advisable to organize the two sets of observational data on redshift into two vectors, denoted as $\boldsymbol{X_1}=\left\lbrace z \mid H(z)\right\rbrace$ and $\boldsymbol{X_2}=\left\lbrace z \mid D_A(z)\right\rbrace$. In order to streamline the writing process, we have merged $\boldsymbol{X_1}$ and $\boldsymbol{X_2}$ into a single variable denoted as $\boldsymbol{X_n}$, ensuring consistency throughout. The reconstructed function and predicted data points are hypothesized to originate from a multivariate Gaussian distribution, characterized by a mean vector denoted as $\boldsymbol{\bar{f}^{*}_{n}}$ and a covariance matrix denoted as $\operatorname{cov}\left(\boldsymbol{f^{*}_{n}}\right)$.The value was determined using the methodology described in \citep{2006gpml.book.....R}
\begin{equation}
	\begin{aligned}
		&\boldsymbol{f^{*}_{n}} \mid \boldsymbol{X_n}, \boldsymbol{y_{n}}, \boldsymbol{X_n^*}  \sim \mathcal{N}\left(\boldsymbol{\bar{f}^{*}_{n}}, \operatorname{cov}\left(\boldsymbol{f^{*}_{n}}\right)\right), \\
		&\boldsymbol{\bar{f}^{*}_{n}} =K\left(\boldsymbol{X^{*}_n}, \boldsymbol{X_n}\right)\left[K\left(\boldsymbol{X_n}, \boldsymbol{X_n}\right)+\sigma_{n}^2 \mathcal{I}\right]^{-1} \boldsymbol{y_{n}},\\
		 &\operatorname{cov}\left(\boldsymbol{f}_*\right)=K\left(\boldsymbol{X^{*}_n}, \boldsymbol{X^{*}_n}\right)\\&-K\left(\boldsymbol{X^{*}_n}, \boldsymbol{X_n}\right)\left[K\left(\boldsymbol{X_n}, \boldsymbol{X_n}\right)+\sigma_{n}^2 \mathcal{I}\right]^{-1} K\left(\boldsymbol{X_n}, \boldsymbol{X^{*}_n}\right),
	\end{aligned}
\end{equation}
where $\boldsymbol{X_n^*}$ represents the predicted vector of redshifts, $\boldsymbol{y_{n}}$ denotes the observational data vector, namely the $\left\lbrace H(z)\right\rbrace$, and $\sigma_{n}^2=\left(\boldsymbol{\sigma}_n^i\right)^T \cdot \boldsymbol{\sigma}_n^i$ is the standard error of the observational data, and $\mathcal{I}$ is the identity matrix. $K\left(\boldsymbol{X_n}, \boldsymbol{X_n}\right)$ represents the covariance of the observational data, $K\left(\boldsymbol{X^{*}_n}, \boldsymbol{X^{*}_n}\right)$ is the covariance of the new predicted points, and $K\left(\boldsymbol{X_n}, \boldsymbol{X^{*}_n}\right)$ and $K\left(\boldsymbol{X^{*}_n}, \boldsymbol{X_n}\right)$ are the covariances between these groups of points. The computation of these covariance matrices can be performed by utilizing a selected covariance function, denoted as $k(\cdot)$, which is commonly referred to as the kernel function. The kernel function is characterized by the hyperparameters $\left(\sigma_f^2, l\right)$ \citep{2012JCAP...06..036S}. The length scale $l$ determines the length in the $z$-direction, which corresponds to a meaningful change of $f(z)$; ${\sigma}_f$ determines the typical change of $f(z)$, which can be considered as the amplitude of the function. In order to reconstruct $D_{A}^{\prime}(z)$ from observational data, it is necessary to modify the covariance metrics. The variables under consideration are transformed to represent the covariance between two specific points of the derivative function, as well as the covariance between a point of the observational data and the derivative function
\begin{equation}
	\begin{aligned}
		&K\left(\boldsymbol{X^{*}_n}, \boldsymbol{X^{*}_n}\right)=\frac{\partial^2 k\left(\boldsymbol{X^{*}_n}_i, \boldsymbol{X^{*}_n}_j\right)}{\partial \tilde{d}\boldsymbol{X^{*}_n}_{i} \partial \tilde{e}\boldsymbol{X^{*}_n}_{j}},\\
		&K\left(\boldsymbol{X_n}, \boldsymbol{X^{*}_n}\right)=\frac{\partial k\left(\boldsymbol{X_n}_i, \boldsymbol{X^{*}_n}_j\right)}{\partial \tilde{e}\boldsymbol{X^{*}_n}_{j}},
	\end{aligned}\label{GP-ker}
\end{equation}
where $\boldsymbol{X^{*}_n}_i$ and $\boldsymbol{X^{*}_n}_j$ are the corresponding redshift vectors, while $\tilde{d}\boldsymbol{X^{*}_n}_{i}$ and $\tilde{e}\boldsymbol{X^{*}_n}_{j}$ denote the value of the $d$-th and $e$-th dimensions of the redshift vectors, respectively.

It is crucial to consider the influence of hyperparameters on the construction of the covariance matrix. The best values of these hyperparameters need to be determined through training in order to achieve a comprehensive GP. The log marginal likelihood (LML) is a commonly employed technique in cosmological research for the purpose of hyperparameter training. The objective of hyperparameter optimization is to identify the optimal combination of hyperparameters that maximizes the LML. This optimal set of hyperparameters is subsequently employed in the GP to obtain the outcome. The LML can be expressed as
\begin{equation}
	\begin{aligned}
		\ln \mathcal{L}= & -\frac{1}{2} \boldsymbol{y_{n}}^{\top}[K\left(\boldsymbol{X_n}, \boldsymbol{X_n}\right)+\sigma_{n}^2 \mathcal{I}]^{-1} \boldsymbol{y_{n}} \\
		& -\frac{1}{2} \ln |K\left(\boldsymbol{X_n}, \boldsymbol{X_n}\right)+\sigma_{n}^2 \mathcal{I}|-\frac{m}{2} \ln 2 \pi,
	\end{aligned}
\end{equation}
where $m=dim\left( \boldsymbol{X_n}\right) $ is the dimension of $\boldsymbol{X_n}$. It is imperative to acknowledge that alternative approaches can also be employed for acquiring hyperparameters. When the LML reaches its maximum value, the corresponding hyperparameters produce the most probable representation of the function. In practical applications, the majority of GPs are implemented by optimizing the LML function.

In our study, we employ the approximate Bayesian computation (ABC) rejection method, which offers the advantage of not necessitating the definition of a likelihood function \citep{TURNER201269}, for the purpose of selecting several commonly used kernel functions: (1) Radial basis function (RBF) kernel. It is parameterized by a length-scale parameter $l>0$, which can take the form of either a scalar (representing the isotropic variation of the kernel) or a vector with the same number of dimensions. (2) Mat\'{e}rn kernel. It is a generalization of the RBF kernel and incorporates an extra parameter denoted as $\nu$ ($\nu=3/2,5/2,7/2,9/2$, we label they as M32, M52, M72, and M92) which controls the smoothness of the resulting function. (3) Rational quadratic (RQ) kernel, also known as Cauchy kernel (CHY). It can be seen as a scale mixing, namely an infinite sum, of RBF kernels with different characteristic length-scales. (4) Exp-Sine-Squared (ESS) kernel. It allows for modeling periodic functions. It is parameterized by a length-scale parameter and a periodicity parameter. The approximation of the likelihood function in ABC rejection is achieved through the utilization of frequencies for the estimation of probabilities, hence enabling the derivation of the posterior distribution. In this study, the model's parameters are repeatedly sampled, with each sample being denoted as a particle. Next, appropriate screening criteria are established, and the proportion of particles that successfully pass the screening is computed in relation to the total number of samples. This allows us to determine the frequency and subsequently the likelihood. In order to implement the ABC rejection algorithm, the kernel function is seen as a model denoted as $\mathcal{T}$. The hyperparameters $\sigma_f^2$ and $l$ are then treated as parameters within the model $\mathcal{T}$, as described by \citep{2009arXiv0910.4472T}.

The appropriate selection of a distance function is fundamental in ABC analysis, as the choice can impact the levels of statistical significance observed in comparisons between mock and observational data sets. One often employed distance functions are: (1) The likelihood function (LML). The utilization of this method is common for assessing the influence of hyperparameter values on the model's fit, hence establishing its suitability as one of the distance functions \citep{10.3389/fbuil.2017.00052,2021JCAP...08..027B}. (2) The $\chi^2$ estimation. The approach takes into consideration the objective of minimizing the sum of squared residuals while also accounting for the weighting of the inverse error. Hence, it offers a standard by which the model's quality may be evaluated, with a lower value of $\chi^2$ indicating a stronger alignment between the mock and observational data \citep{2021JCAP...08..027B}. (3) The Bias estimation. It provides the average of Euclidean distances between the mock and observational data sets and serves as an estimate for the anticipated disparity between the predicted and true values of the model, sometimes referred to as bias. The bias of a model performs the role of an indicator of its goodness of fit to the data, with a lower bias value suggesting a tighter alignment between the mean value of the mock data and the observational data \citep{2017A&C....19...16J,2023ApJS..266...27Z}. By integrating these three distance functions, we present three distinct approaches for particle filtration. The ABC rejection outcomes derived from these approaches provide a more thorough response to the inquiry regarding the optimal kernel performance in ABC analysis.

By comparing the likelihoods of two statistical models, we may calculate the Bayes factor, denoted as $\mathcal{B}_f$, which involves the comparison of the likelihoods of two statistical models. This factor quantifies the extent to which we prefer one model over the other based on the ratio of their likelihoods \citep{MOREY20166}. In this study, the Bayes factor is employed to evaluate the degree of reliance between various data sets and the kernel. In contrast to conventional hypothesis testing, which solely permits the acceptance or rejection of a hypothesis, the Bayes factor assesses the strength of evidence in favor of a hypothesis. Therefore, the Bayes factor serves the purpose of not only determining the optimal model among a set of competing kernels but also quantifying the extent to which it outperforms the alternative models. The plausibility of two alternative models, denoted as $\mathcal{T}_1$ and $\mathcal{T}_2$, is assessed using the Bayes factor, given observational data $\boldsymbol{y_{n}}$. The prior probability for both kernels is computed identically during the calculation of the Bayes factor. The approach solely considers the ratio of the posterior distributions of the two kernels as empirical evidence. And the scale of $\mathcal{B}_f$ has a quantitative interpretation based on probability theory \citep{10.1093/oso/9780198503682.001.0001}.

\section{Data Analysis} \label{sec:data analysis}
The data set includes 64 $D_A(z)$ data points and 35 groups of $H(z)$ data obtained from the cosmic chronometer, which are enumerated in Tables \ref{table:I} and \ref{table:II}, respectively.
\begin{table*}
	\centering
	\caption{The compiled independent $D_{A}(z)$ dataset.}
	%\begin{threeparttable}
	\begin{tabular}{ccccccc}
		\hline \hline Specification & Redshift $z$ & $D_{A}(z) \pm 1 \sigma$ error\tnote{a} & References &Redshift $z$ & $D_{A}(z)\pm 1 \sigma$ error\tnote{a}& References \\
		\hline Strong &$0.0098$ & $37.7 \pm 8.7$ & \cite{2017ApJ...849L..16I}&$0.6304$ & $1423.3 \pm 199.26$ & \cite{2015JCAP...11..033J}\\
		Lenses&$0.295$ & $876.5 \pm 113.95$ & \cite{2015JCAP...11..033J}&$1.789$ & $1805 \pm 2388$ & \cite{2019ApJ...883....3L}\\
		\hline  &$0.142$ & $780 \pm 150$ &\cite{2006ApJ...647...25B}&$0.282$ & $880 \pm 265$ & \cite{2006ApJ...647...25B}\\
		&$0.152$ & $610 \pm 65$ &\cite{2006ApJ...647...25B}&$0.288$ & $780 \pm 180$ & \cite{2006ApJ...647...25B}\\
		&$0.164$ & $580 \pm 270$ & \cite{2006ApJ...647...25B}&$0.291$ & $830 \pm 20$ & \cite{2006ApJ...647...25B}\\
		&$0.171$ & $520 \pm 135$ &\cite{2006ApJ...647...25B}&$0.322$ & $1190 \pm 145$ & \cite{2006ApJ...647...25B}\\
		&$0.171$ & $440 \pm 45$ & \cite{2006ApJ...647...25B}&$0.327$ & $1130 \pm 95$ & \cite{2006ApJ...647...25B}\\
		&$0.176$ & $660 \pm 125$ &\cite{2006ApJ...647...25B}&$0.375$ & $1080 \pm 195$ & \cite{2006ApJ...647...25B}\\
		&$0.182$ & $660 \pm 95$ &\cite{2006ApJ...647...25B}&$0.412$ & $1220 \pm 235$ & \cite{2006ApJ...647...25B}\\
		Galaxy&$0.183$ & $650 \pm 90$ &\cite{2006ApJ...647...25B}&$0.451$ & $960 \pm 70$ &\cite{2006ApJ...647...25B} \\
		Clusters&$0.202$ & $520 \pm 45$ & \cite{2006ApJ...647...25B}&$0.483$ & $1440 \pm 250$ & \cite{2006ApJ...647...25B}\\
		&$0.217$ & $980 \pm 155$ &\cite{2006ApJ...647...25B}&$0.545$ & $1490 \pm 45$ & \cite{2006ApJ...647...25B}\\
		&$0.224$ & $730 \pm 165$ &\cite{2006ApJ...647...25B}&$0.55$ & $1420 \pm 245$ & \cite{2006ApJ...647...25B}\\
		&$0.229$ & $640 \pm 185$ & \cite{2006ApJ...647...25B}&$0.686$ & $1680 \pm 430$ &\cite{2006ApJ...647...25B} \\
		&$0.23$ & $600 \pm 100$ & \cite{2006ApJ...647...25B}&$0.813$ & $1040 \pm 470$ & \cite{2006ApJ...647...25B}\\
		&$0.235$ & $460 \pm 95$ & \cite{2006ApJ...647...25B}&$0.826$ & $1330 \pm 270$ &\cite{2006ApJ...647...25B}\\
		&$0.252$ & $1070 \pm 50$ & \cite{2006ApJ...647...25B}&$0.89$ & $1080 \pm 350$ &\cite{2006ApJ...647...25B} \\
		&$0.255$ & $630 \pm 175$ & \cite{2006ApJ...647...25B}& & & \\
		\hline &$0.35$ & $1037 \pm 44$ &\cite{2014MNRAS.445.3737H}&$0.73$ & $1534 \pm 107$ &\cite{2012MNRAS.425..405B}\\
		&$0.35$ & $1050 \pm 38$ & \cite{2013MNRAS.431.2834X}&$0.77$ & $1573.39 \pm 31.72$ &\cite{2020MNRAS.498.3470W}\\
		&$0.38$ & $1090.90 \pm 18.13$ & \cite{2021PhRvD.103h3533A}&$0.835$ & $1521.85 \pm 41.02$ &\cite{2022PhRvD.105d3512A}\\
		&$0.44$ & $1205 \pm 114$ &\cite{2012MNRAS.425..405B}&$1.48$ & $1826.50 \pm 52.42$ & \cite{2021MNRAS.500.1201H}\\
		Baryonic&$0.51$ & $1302.02 \pm 20.47$ &\cite{2021PhRvD.103h3533A}&$1.48$ & $1821.11 \pm 47.47$ &\cite{2021PhRvD.103h3533A}\\
		Acoustic&$0.57$ & $1408 \pm 45$ & \cite{2014MNRAS.439.3504S}&$1.52$ & $1850.0 \pm 102.5$ &\cite{2018MNRAS.477.1639Z}\\
		Oscillations&$0.57$ & $2190 \pm 61$ &\cite{2012MNRAS.426.2719R}&$1.52$ & $1850 \pm 110$ &\cite{2018MNRAS.477.1604G}\\
		&$0.57$ & $1380 \pm 23$ & \cite{2014MNRAS.439.3504S}&$2.33$ & $1648.37 \pm 75.13$ &\cite{2021PhRvD.103h3533A}\\
		&$0.6$ & $1380 \pm 95$ &\cite{2012MNRAS.425..405B}&$2.34$ & $1650.18 \pm 82.05$ &\cite{deSainteAgathe:2019voe}\\
		&$0.698$ & $1473.23 \pm 25.04$ &\cite{2021MNRAS.500..736B}&$2.35$ & $1596.44 \pm 79.16$ &\cite{2019AA...629A..86B} \\
		&$0.7$ & $1546.05 \pm 28.57$ & \cite{2021PhRvD.103h3533A}&$2.36$ & $1590 \pm 60$ &\cite{2014JCAP...05..027F}\\
		&$0.72$ & $1466.5 \pm 136.6$ &\cite{2020MNRAS.492.4189I}& & &\\
		\hline
	\end{tabular}
	\begin{tablenotes}
		\footnotesize
		\item[a] $D_{A}(z)$ are in the unit of $\mathrm{Mpc}$.
	\end{tablenotes}
	\label{table:I}
\end{table*}

\begin{table}
	\centering
	\caption{The compiled independent $H(z)$ dataset.}
	\begin{tabular}{ccc}
		\hline\hline  $z$ & $H(z)$\tnote{a}  & References \\
		\hline $0.07$ & $69 \pm 19.6$ & \cite{Zhang:2012mp} \\
		$0.09$ & $69 \pm 12$ &\cite{2005PhRvD..71l3001S} \\
		$0.12$ & $68.6 \pm 26.2$ &\cite{Zhang:2012mp} \\
		$0.17$ & $83 \pm 8$ & \cite{2005PhRvD..71l3001S} \\
		$0.179$ & $75 \pm 4$ & \cite{2012JCAP...08..006M} \\
		$0.199$ & $75 \pm 5$ &  \cite{2012JCAP...08..006M} \\
		$0.20$ & $72.9 \pm 29.6$ &\cite{Zhang:2012mp} \\
		$0.27$ & $77 \pm 14$ &  \cite{2005PhRvD..71l3001S} \\
		$0.28$ & $88.8 \pm 36.6$ & \cite{Zhang:2012mp} \\
		$0.352$ & $83 \pm 14$ & \cite{2012JCAP...08..006M} \\
		$0.4$ & $95 \pm 17$ & \cite{2005PhRvD..71l3001S} \\
		$0.4004$ & $77 \pm 10.2$ &  \cite{2016JCAP...05..014M} \\
		$0.425$ & $87.1 \pm 11.2$ & \cite{2016JCAP...05..014M} \\
		$0.445$ & $92.8 \pm 12.9$ & \cite{2016JCAP...05..014M} \\
		$0.47$ & $89 \pm 67$ & \cite{2017MNRAS.467.3239R} \\
		$0.4783$ & $80.9 \pm 9$ & \cite{2016JCAP...05..014M} \\
		$0.48$ & $97 \pm 62$ &  \cite{2010ApJS..188..280S} \\
		$0.593$ & $104 \pm 13$ & \cite{2012JCAP...08..006M} \\
		$0.68$ & $92 \pm 8$ & \cite{2012JCAP...08..006M} \\
		$0.75$ & $98.8 \pm 33.6$ & \cite{2022ApJ...928L...4B} \\
		$0.75$ & $105 \pm 10.76$ & \cite{2023JCAP...11..047J} \\
		$0.781$ & $105 \pm 12$ &\cite{2012JCAP...08..006M} \\
		$0.8$ & $113.1 \pm 15.1$ &\cite{2023ApJS..265...48J}\\
		$0.875$ & $125 \pm 17$ &\cite{2012JCAP...08..006M} \\
		$0.88$ & $90 \pm 40$ &  \cite{2010ApJS..188..280S} \\
		$0.9$ & $117 \pm 23$ &\cite{2005PhRvD..71l3001S} \\
		$1.037$ & $154 \pm 20$ &\cite{2012JCAP...08..006M} \\
		$1.26$ & $135 \pm 65$ & \cite{2023AA...679A..96T}\\
		$1.3$ & $168 \pm 17$ & \cite{2005PhRvD..71l3001S} \\
		$1.363$ & $160 \pm 33.6$ & \cite{2015MNRAS.450L..16M} \\
		$1.43$ & $177 \pm 18$ & \cite{2005PhRvD..71l3001S} \\
		$1.53$ & $140 \pm 14$ &\cite{2005PhRvD..71l3001S} \\
		$1.75$ & $202 \pm 40$ &\cite{2005PhRvD..71l3001S} \\
		$1.965$ & $186.5 \pm 50.4$ & \cite{2015MNRAS.450L..16M} \\
		\hline
	\end{tabular}
\begin{tablenotes}
	\footnotesize
	\item[a] $H(z)$ in the unit of $\mathrm{km~s}^{-1}\;\mathrm{Mpc}^{-1}$.
\end{tablenotes}
	\label{table:II}
	% }
\end{table}

\begin{figure*}
	\centering
	\subfigure[]{
		\includegraphics[width=0.45\linewidth]{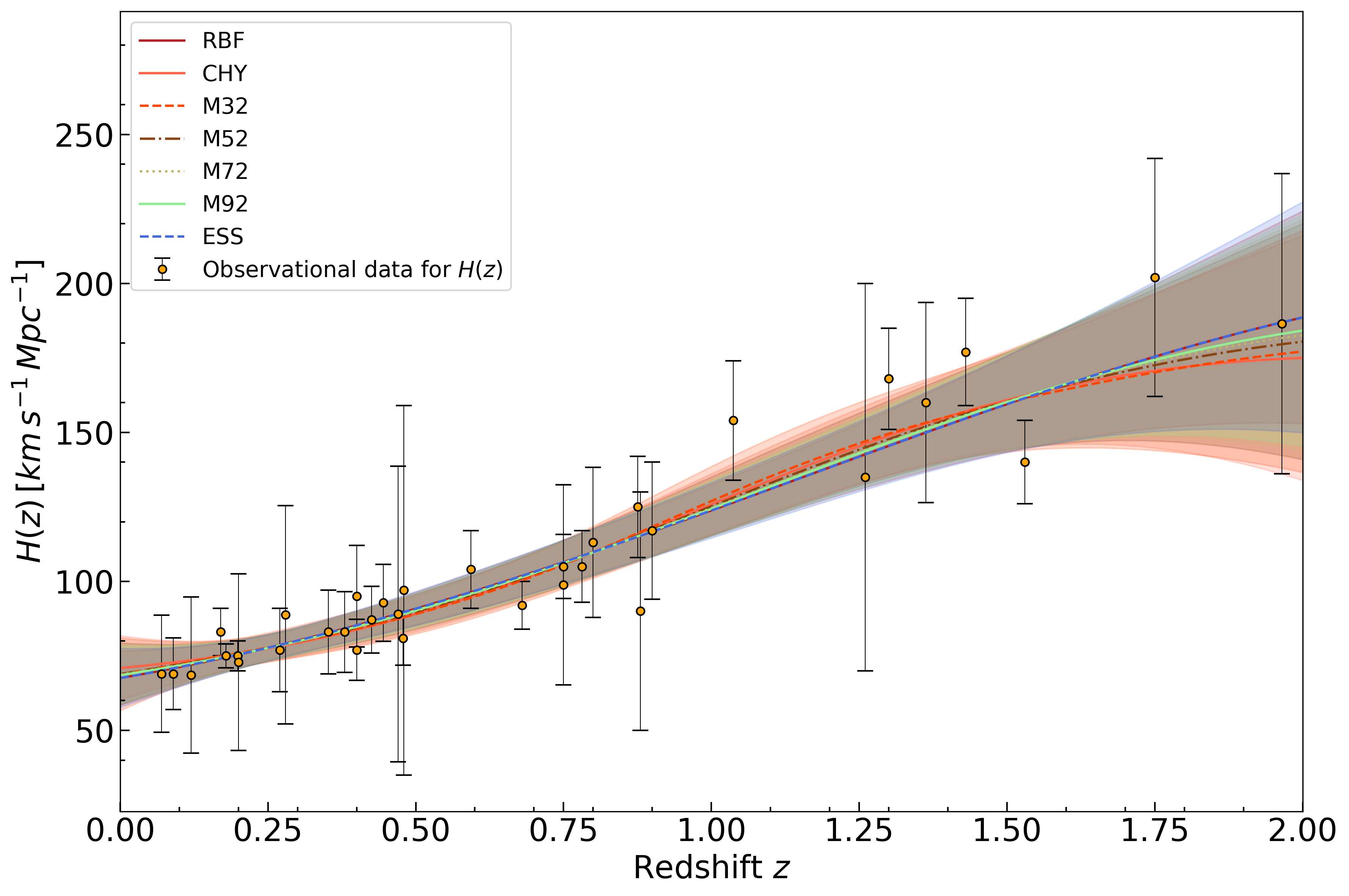}
	}
	\subfigure[]{
		\includegraphics[width=0.45\linewidth]{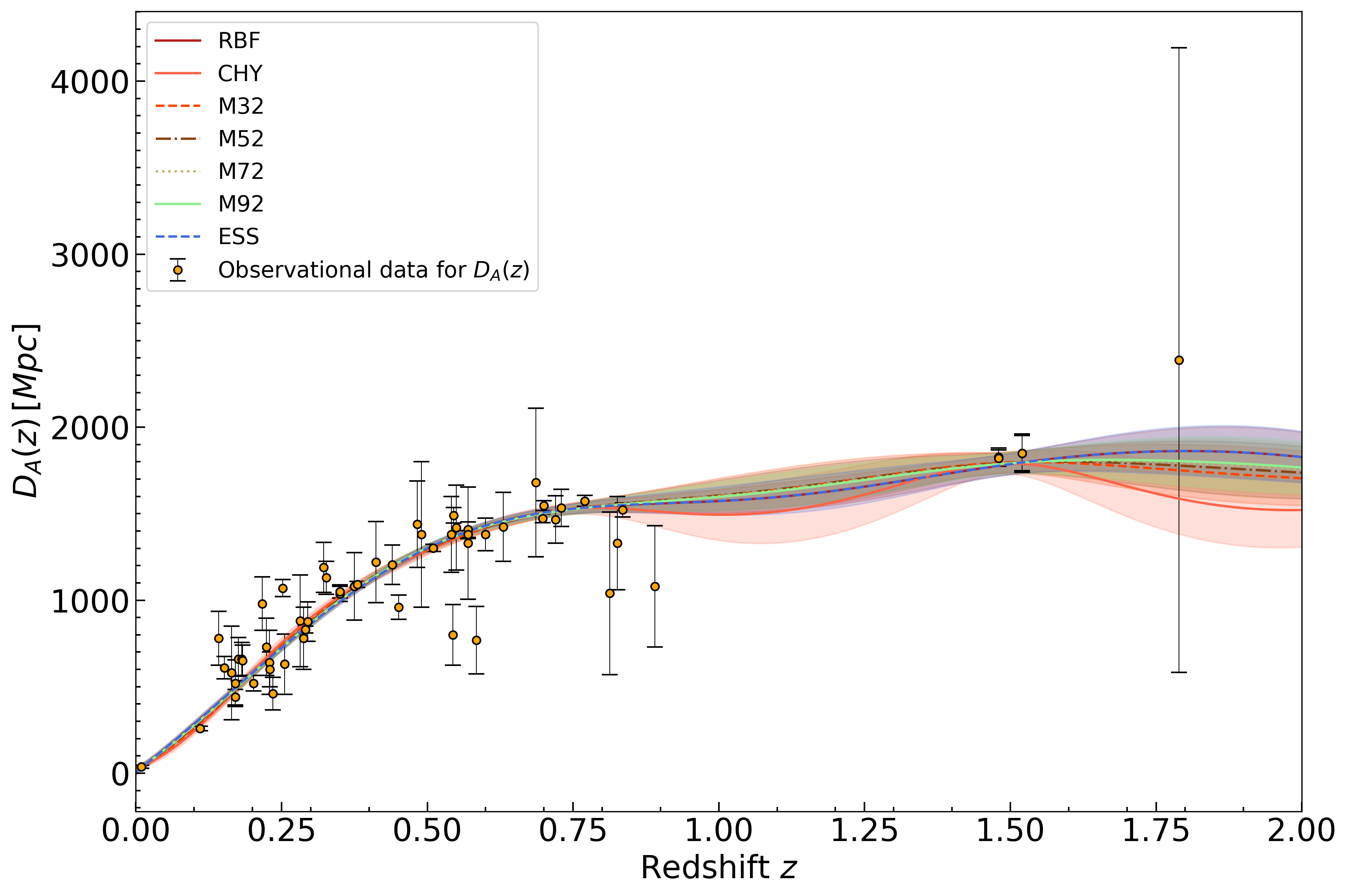}
	}
	\caption{The Gaussian Process results of (a) Hubble parameter $H(z)$, and (b) angular diameter distance $D_{A}(z)$ before select kernel functions with redshift from 0 to 2.0 labeled by the different color curves and the corresponding $1\sigma$ error with the different kernel functions that have one hidden layer shown by the same color shaded regions, respectively. And the yellow dots with black error bars represent the observational data.}
	\label{fig:select-rec-before}
\end{figure*}

We use the \href{https://scikit-learn.org/stable/index.html}{scikit-learn} module \citep{scikit-learn,sklearn-api} to demonstrate the general GP reconstruction generated using LML training hyperparameters. This package provides a convenient, powerful and extensible implementation of Gaussian Process Regression (GPR) which makes it possible for us to reconstruct the speed of light more accurately as it provide simple and efficient tools for predicted data analysis. The GP method has been discussed and applied in several cosmological papers \citep{PhysRevD.85.123530,2014PhRvD..89b3503Y,2016JCAP...04..016G,2022PDU....3600998M,2021ApJ...915..123S,2021ApJS..254...43W,2023ApJS..266...27Z,2023arXiv230605492K,2023arXiv230518873Y,2023ApJS..267...42D,2023MNRAS.524.3724C}.  Figure \ref{fig:select-rec-before} shows that different kernel selections result in distinct curves after reconstruction, but it is challenging to infer which performs better from the graphs. In addition, we can also obviously find that different observables have different degrees of agreement with the kernel function. For example, the kernel function CHY seems to agree fairly well with an observable $H(z)$ that shows obvious monotonicity with redshift, but not so well with an observable $D_{A}(z)$ that shows non-monotonicity with redshift.

As described earlier, in order to quantify the difference between different kernel functions for different data, we use ABC rejection method with a special threshold $\epsilon$ to select kernel functions for different observables. Threshold value is very important for ABC rejection method. When the results of a single calculation of the three distances mentioned above are less than the threshold value, we believe that this method will not reject the results of this calculation. So the value of the threshold is definitely not randomly selected. Setting the threshold too high would obscure the differences between specific kernel functions, while setting the threshold too low would result in not only a small number of particles in each kernel function but also particles that are very close to one another as we reduce as much randomness as possible for sampling these kernel functions. To address this issue, we continuously adjust the threshold until we reach the final result. When the posterior distributions of the individual kernels undergo significant changes when the threshold is set to $\varepsilon$, but do not differ significantly when the threshold is greater than $\varepsilon$, we consider $\varepsilon$ to be the appropriate threshold. When the previously observed differences are preserved when the threshold is set to a value less than $\varepsilon$, we consider $\varepsilon$ to be the correct threshold. It is worth mentioning that there are no circumstances where the differences in the posterior distributions of the kernels change when the threshold is decreased further, as we want to gradually lower the threshold to conserve computational resources for the ABC rejection procedure.

\begin{figure*}
	\centering
	\subfigure[]{
		\includegraphics[width=1\linewidth]{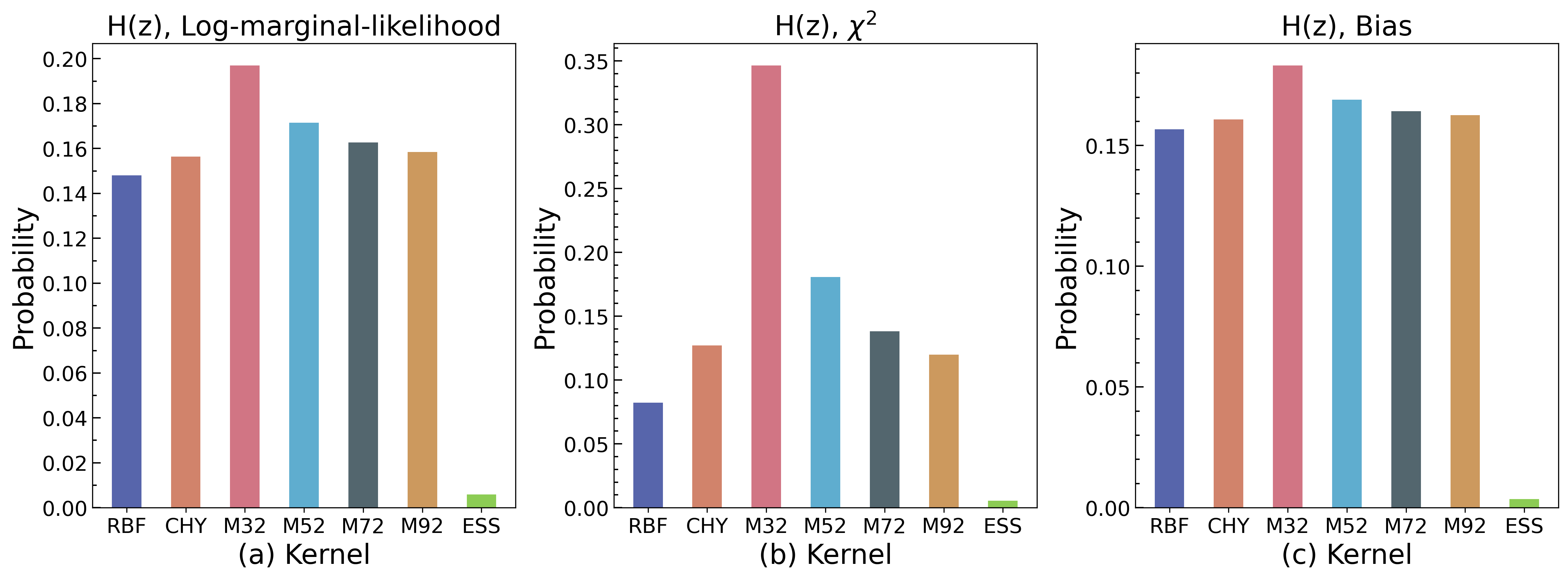}
	}
	\subfigure[]{
		\includegraphics[width=1\linewidth]{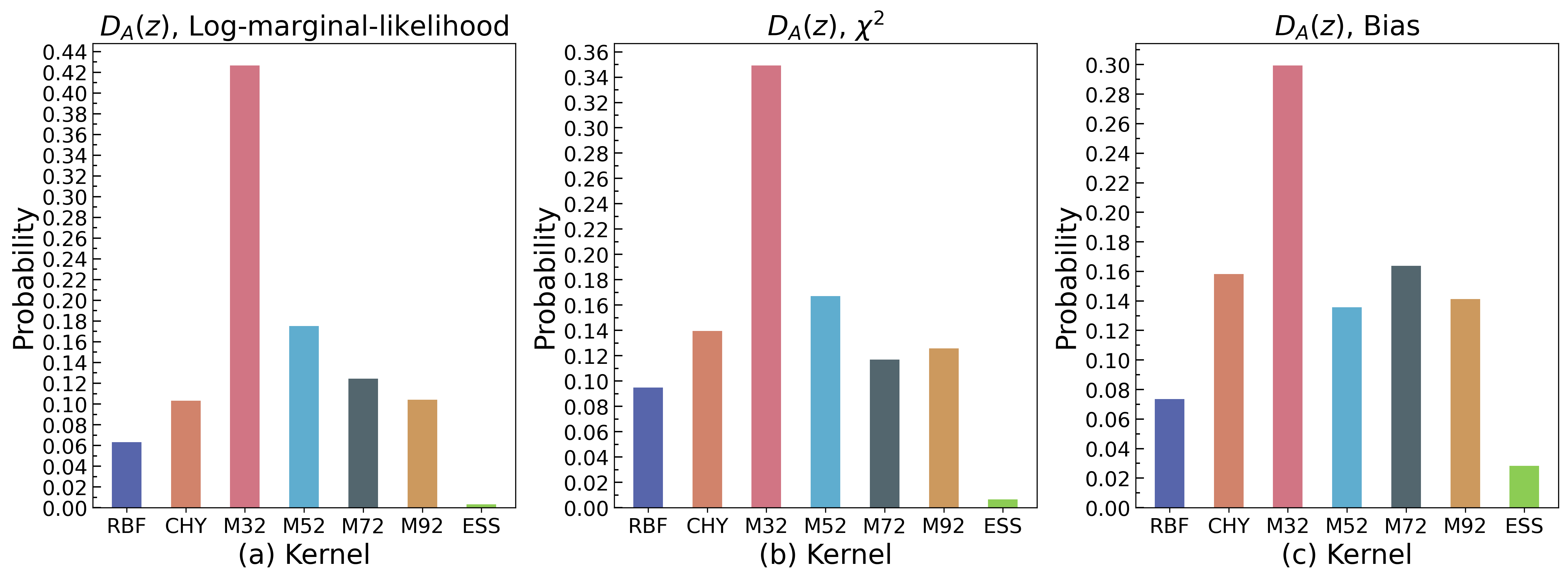}
	}
	\caption{The ABC rejection posterior probability histogram of (a) observational Hubble data $H(z)$, and (b) angular diameter distance data $D_{A}(z)$ with three different distance functions LML, $\chi^2$, and Bias under seven different kernel functions RBF, CHY, M32, M52, M72, M92, and ESS, respectively.}
	\label{fig:select-distance}
\end{figure*}

\begin{figure*}
	\centering
	\subfigure[]{
		\includegraphics[width=1\linewidth]{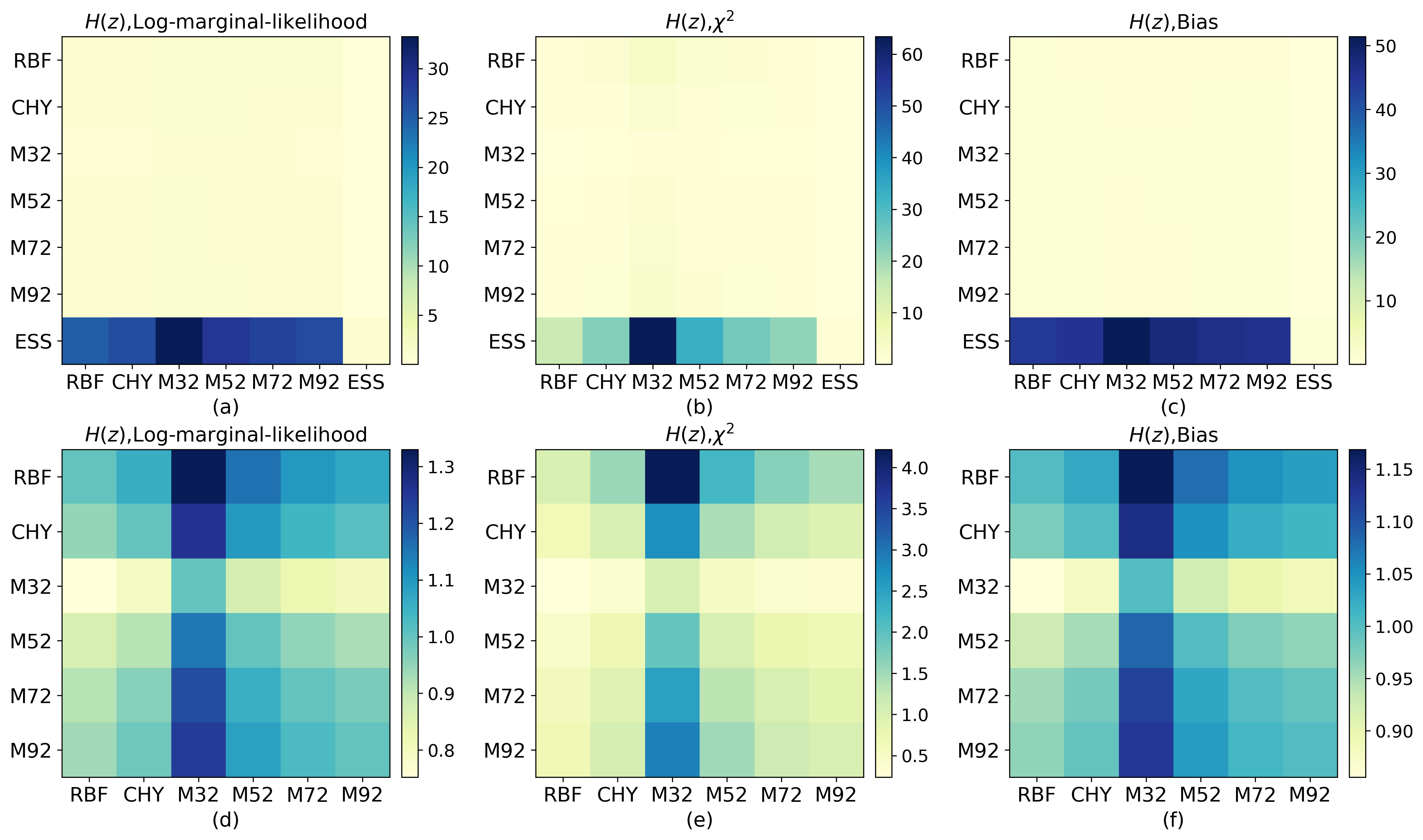}
	}
	\subfigure[]{
		\includegraphics[width=1\linewidth]{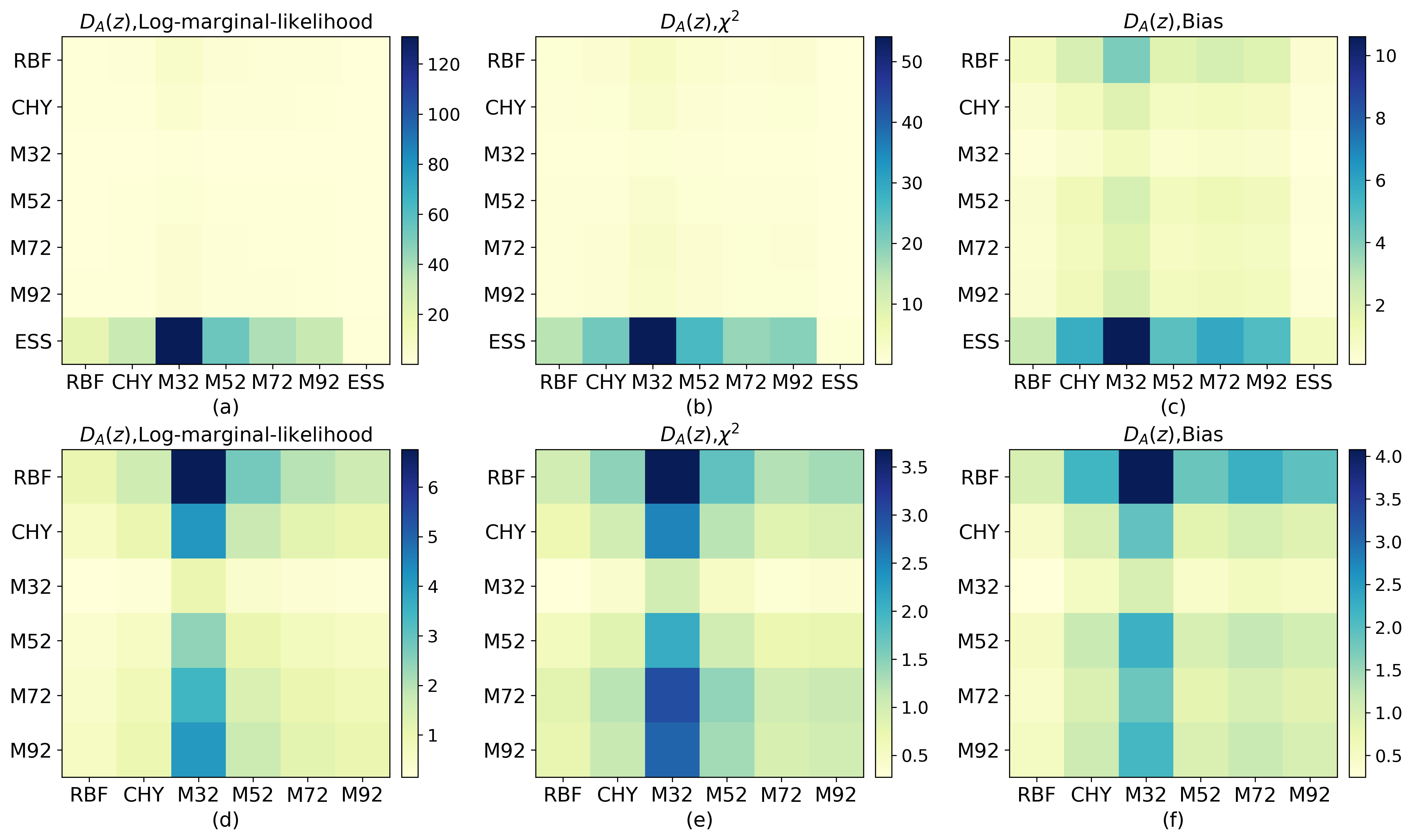}
	}
	\caption{The Bayes factor $\mathcal{B}_f$ between every two kernels is visualized using this heat map to represent the strength of evidence level of (a) observational Hubble data $H(z)$, and (b) angular diameter distance data $D_{A}(z)$ with three different distance functions LML, $\chi^2$, and Bias between any two kernels.}
	\label{fig:select-kernel}
\end{figure*}

Hereto, we employ three different types of data, apply the $\mathrm{ABC}$ rejection method to each type of data, and use three different distance functions in the computations, resulting as presented in Figure \ref{fig:select-distance}. The posterior distribution for each kernel function in Figure \ref{fig:select-distance} is derived by averaging 100 posterior probabilities. We observe that for both data sets, across different distance functions, M32 consistently shows the highest probability, while ESS consistently shows the lowest probability. In order to more clearly compare the advantages and disadvantages of the two kernel functions, we further transform the posterior distribution histogram into a Bayes factor $\mathcal{B}_f$ between the two kernel functions displayed in the form of heatmap in Figure \ref{fig:select-kernel}. And the darker the color, the larger the Bayes factor. In Figure \ref{fig:select-kernel}(a) and Figure \ref{fig:select-kernel}(b), the three subgraphs in the upper show all of our selected kernel functions, while the three subgraphs in the lower show the Bayes factors between the remaining six kernel functions after removing the very terrible ESS kernel function. This heatmap can be read like this, from the X-axis to the Y-axis. For example, the first row and third column in the concrete result of each graph should be interpreted as the Bayes factor of M32 (X-axis) with respect to RBF (Y-axis). And the scale of $\mathcal{B}_f$ has a quantitative interpretation based on probability theory \citep{10.1093/oso/9780198503682.001.0001}, as well as the strength of evidence. We can find that: (1) for $H(z)$ (a) with the LML distance function, M32 is at the ``Decisive'' level compared with other kernels. (b) With the $\chi^2$ distance function, M32 is at the ``Decisive'' level compared with other kernels. (c) With the Bias distance function, M32 is at the ``Decisive'' level compared with other kernels. (2) For $D_{A}(z)$ (a) with the LML distance function, M32 is at the ``Very strong'' level compared with RBF and at the ``Strong" level compared with other kernels. (b) With the $\chi^2$ distance function, M32 is at the ``Strong'' level compared with M52 and at the ``Very strong" level compared with other kernels. (c) With the Bias distance function, M32 is at the ``Very Strong'' level compared with RBF and at the ``Strong" level compared with other kernels. Therefore, we use M32 to reconstruct our two sets of data.

\section{Results and Discussions} \label{sec:results}

We allocate a total of 1000 reconstruction bins within the redshift range of $[0,2.0]$. This choice is made based on the belief that the entire observation atlas provides the most comprehensive and informative dataset. We do not opt for a specific selection and combination of observational data, as it does not have increased the amount of information available. Our objective is to obtain the functions $D_A(z)$, $H(z)$, and $D^{\prime}_A(z)$ using the M32 kernel function and the LML to train hyperparameters. To achieve this, we allow the GP to randomly initialize and optimize the hyperparameters 10,000 times. This approach aims to ensure that the resulting hyperparameter values fall within a reasonable range. Once the reconstructions of $H(z)$, $D_A(z)$, and $D^{\prime}_A(z)$ are obtained, we proceed to fit the function $c(z)$ using Equations (\ref{eq2}) and (\ref{eq3}).

\begin{figure*}
	\centering
	\subfigure[]{
		\includegraphics[width=0.47\linewidth]{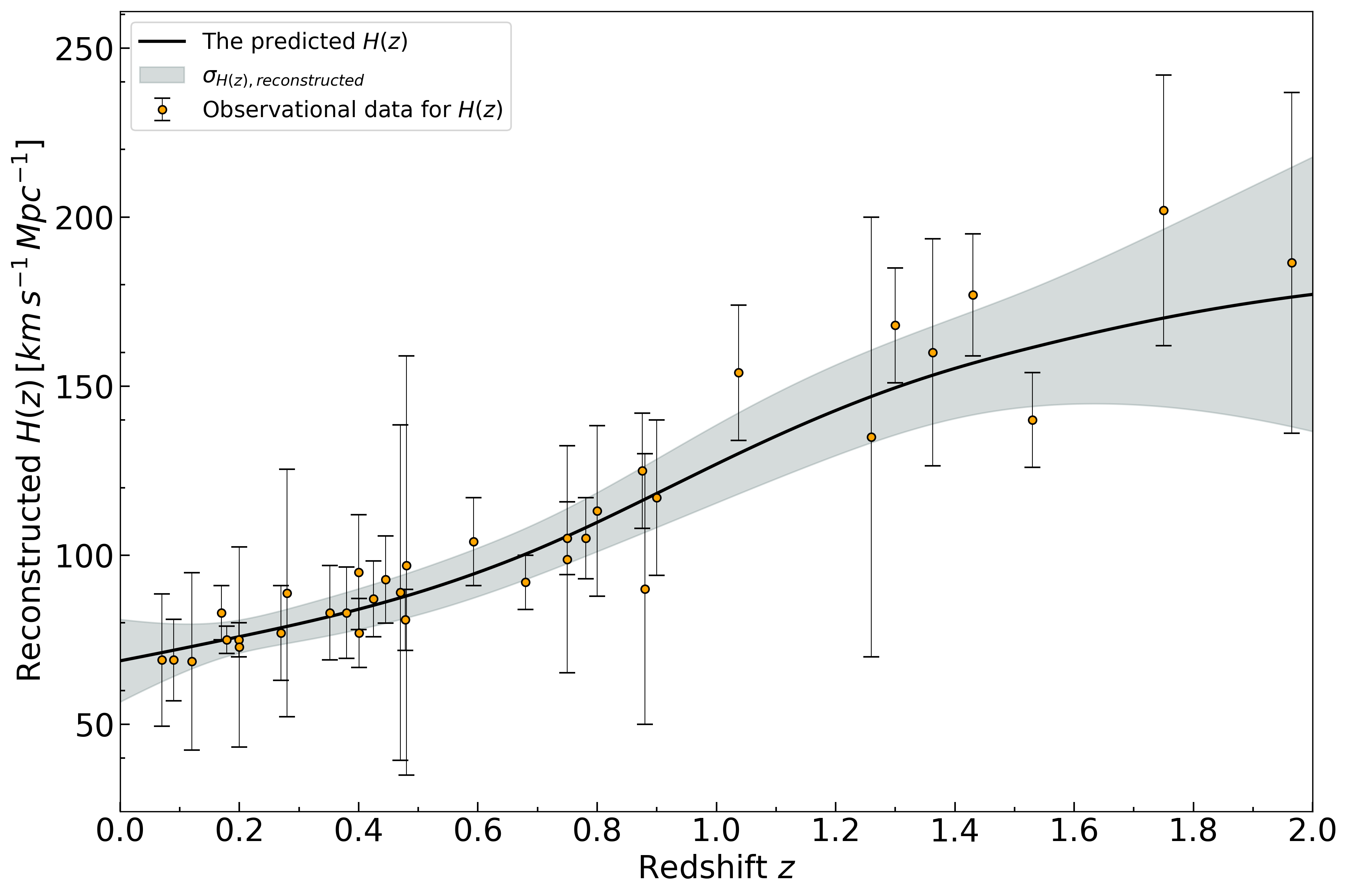}
	}
	\subfigure[]{
		\includegraphics[width=0.47\linewidth]{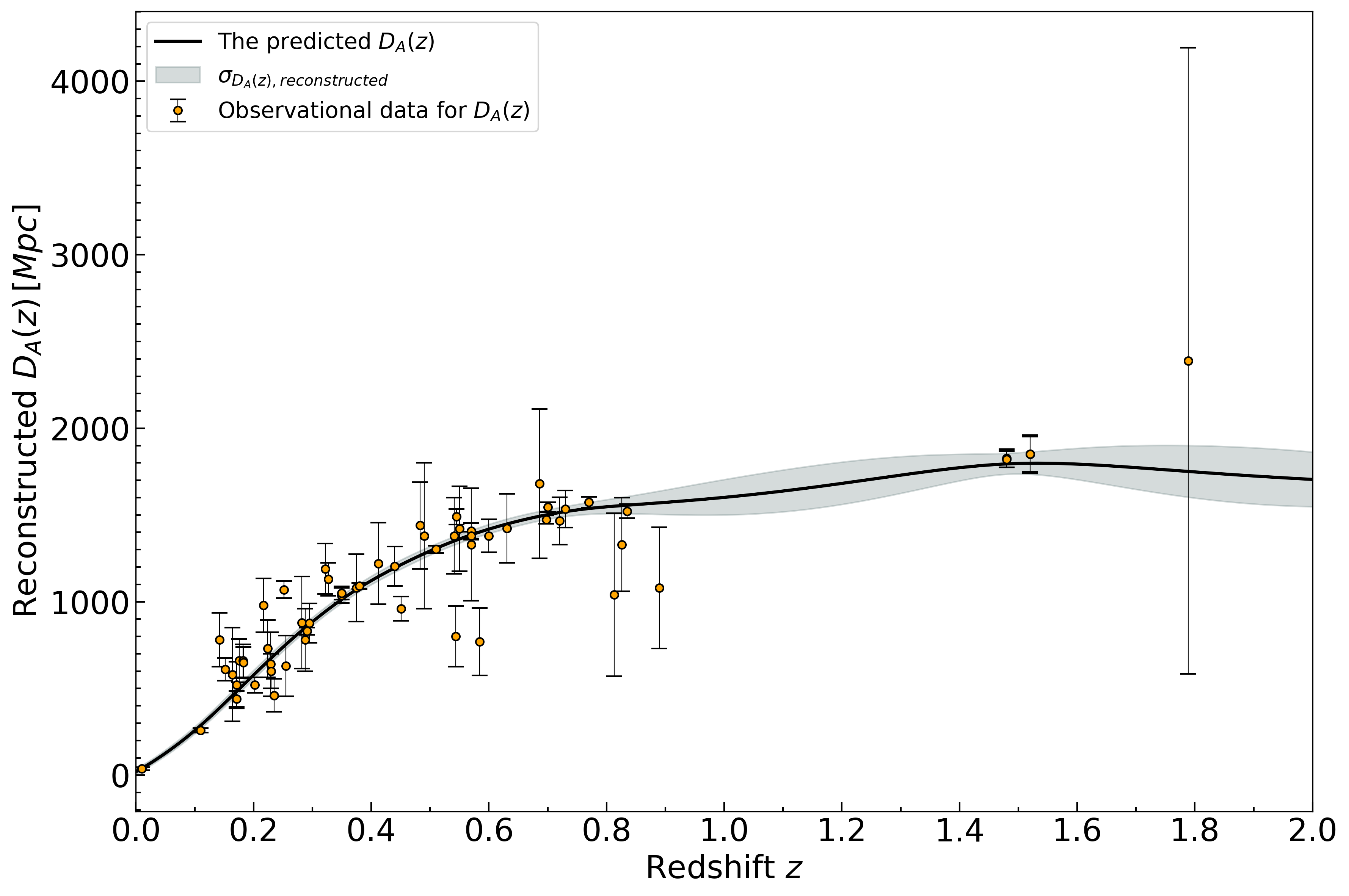}
	}
	\subfigure[]{
		\includegraphics[width=0.47\linewidth]{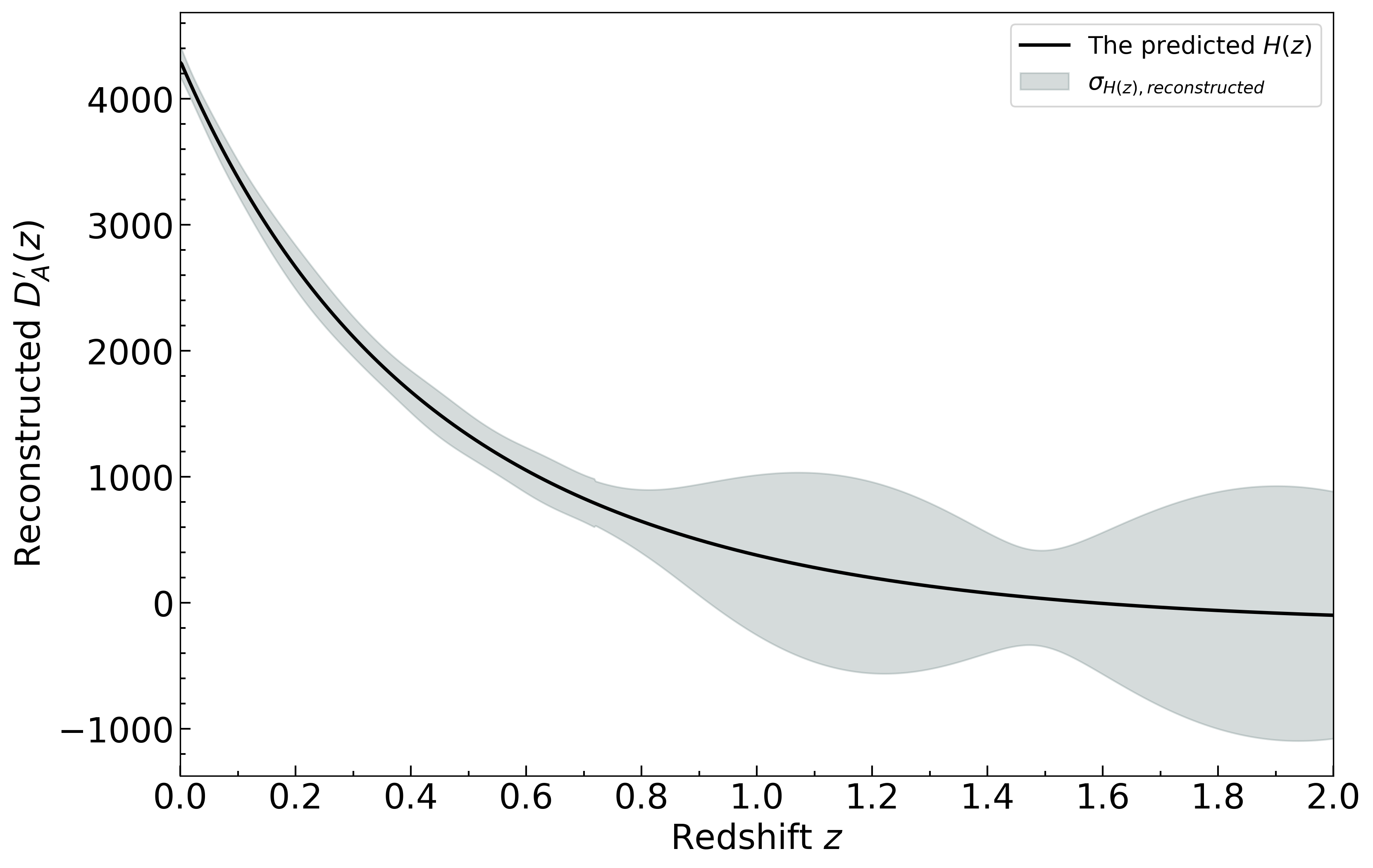}
	}
	\subfigure[]{
		\includegraphics[width=0.47\linewidth]{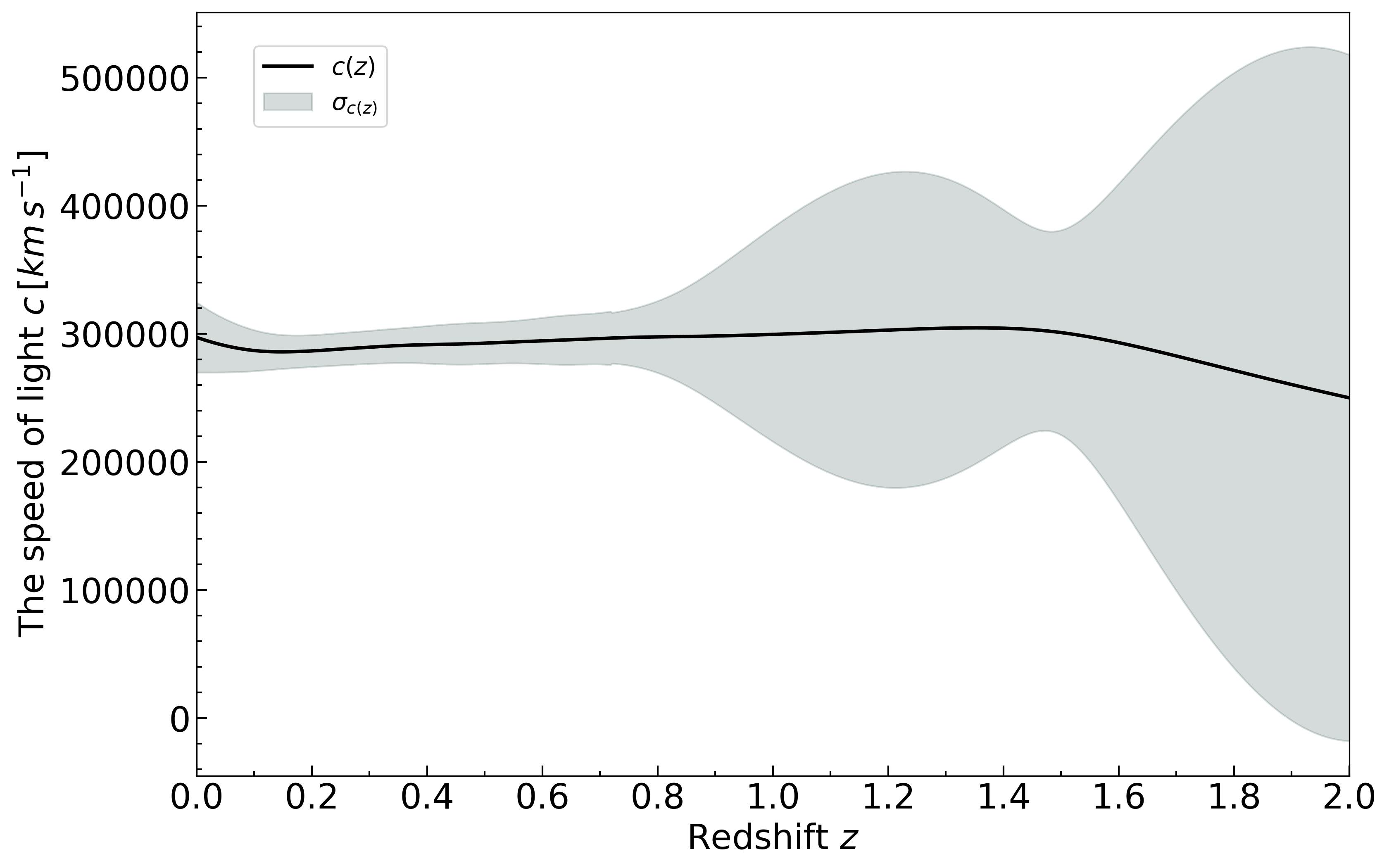}
	}
	\caption{The reconstructed results of (a) Hubble parameter $H(z)$, (b) angular diameter distance $D_{A}(z)$, (c) the derivative of angular diameter distance $D^{\prime}_A(z)$, and (d) the speed of light $c(z)$ with redshift $z\in[0.11,2.36]$ labeled by the black color curves and the corresponding $1\sigma$ error with the different kernel functions that have one hidden layer shown by the darkslategrey color shaded regions, respectively. And the yellow dots with black error bars represent the observational data.}
	\label{fig:after-re}
\end{figure*}

The reconstructed results of $H(z)$, $D_A(z)$, $D^{\prime}_A(z)$, and $c(z)$ together with their corresponding $1\sigma$ errors are shown in Figure \ref{fig:after-re}. A peculiar fluctuation is seen in the vicinity of the point $z \sim 1.5$, which cannot be accounted for by any theoretical models of VSL. Consequently, we hypothesize that this anomaly is due to the absence of data for the angular diameter distance $D_A(z)$ within the redshift range of $1.52$ to $2.33$. The value of the data in this redshift interval shows a clear downward trend. As evident from Equation (\ref{eq2}), this phenomenon occurs when the value of $D_A(z)$ surpasses the maximum value $D_A(z_m)$, resulting in a downward trajectory accompanied by a negative derivative $D^{\prime}_A(z)$. This leads to our calculations of the speed of light reveal that there is a discernible decrease in its value at high redshift. However, it is worth noting that our technique has a distinct benefit in that it avoids the introduction of novel cosmological models and information derived from the data into the final reconstructed structure. As a consequence, our findings strive to accurately represent the inherent facts without undue influence. In addition, due to the limited amount of data available at high redshift, the derivative value obtained in the reconstruction process is is quite tiny, resulting in the phenomenon of the reconstructed speed of light decreasing, which can promote the release of BAO and OHD data at high redshift. Therefore, it is imperative that we should not overlook any potential implicit possibilities, and we must continue to give them due thought.

\begin{figure*}
	\centering
	\subfigure[]{
		\includegraphics[width=0.47\linewidth]{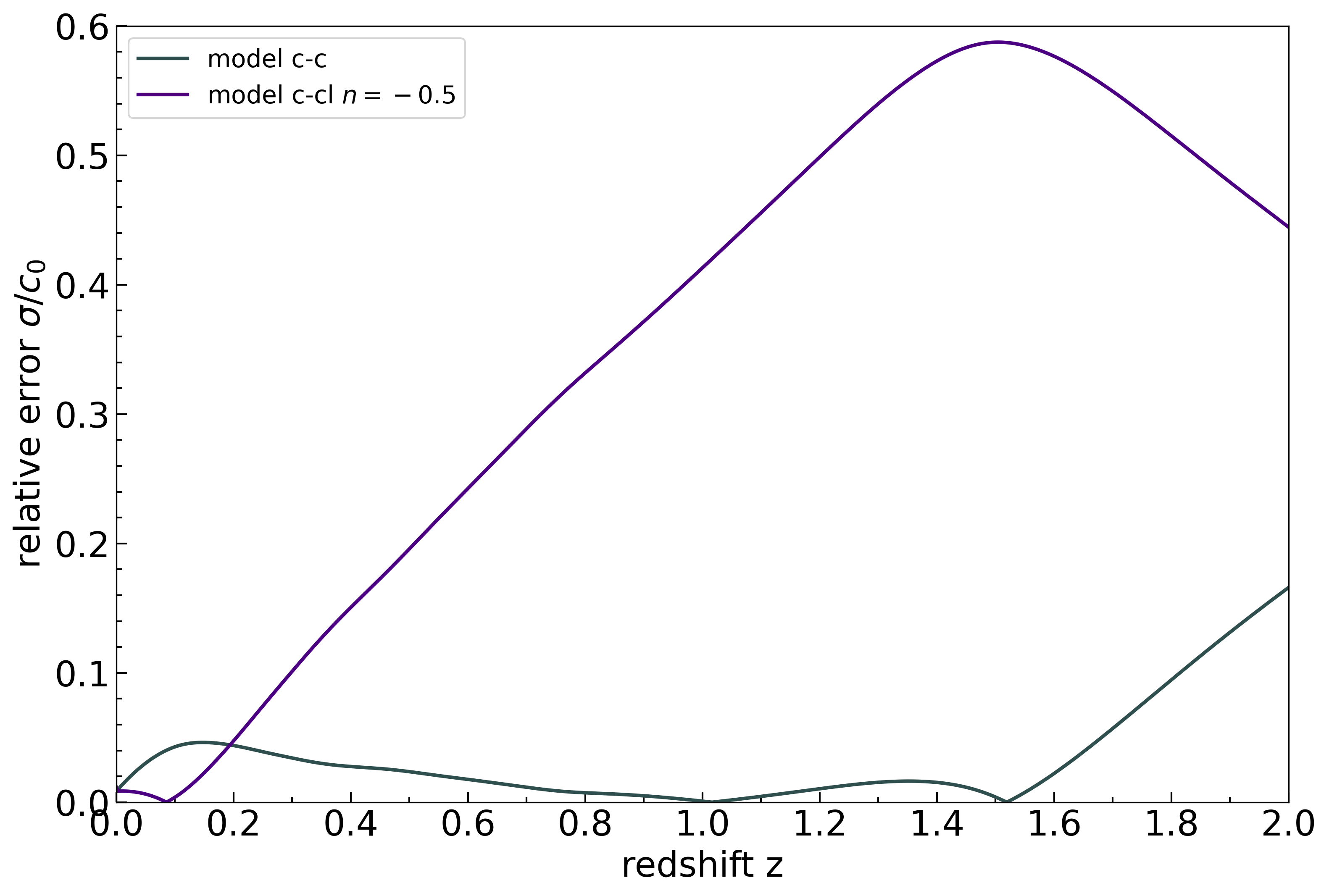}
	}
	\subfigure[]{
		\includegraphics[width=0.47\linewidth]{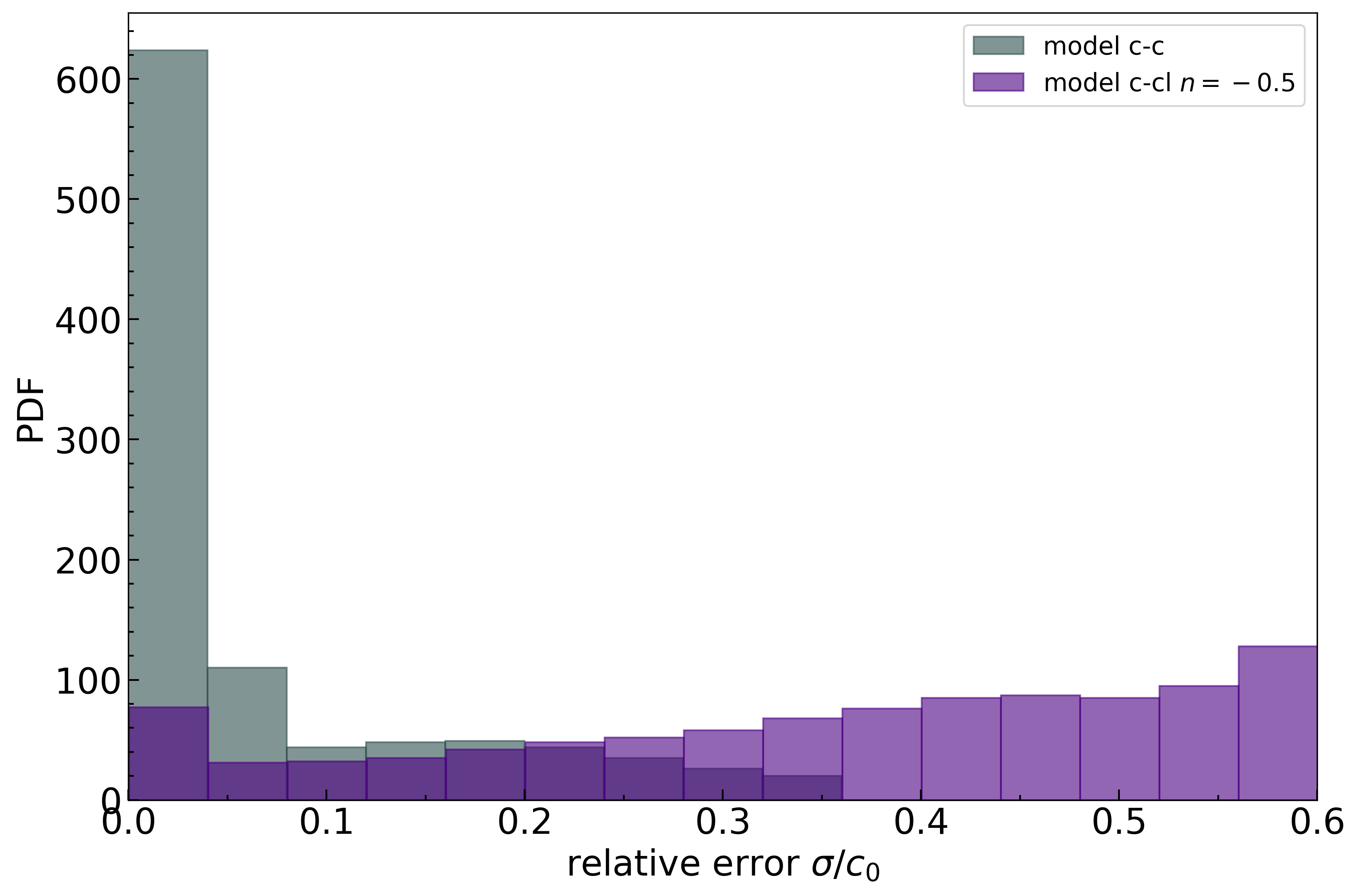}
	}
	\subfigure[]{
		\includegraphics[width=0.47\linewidth]{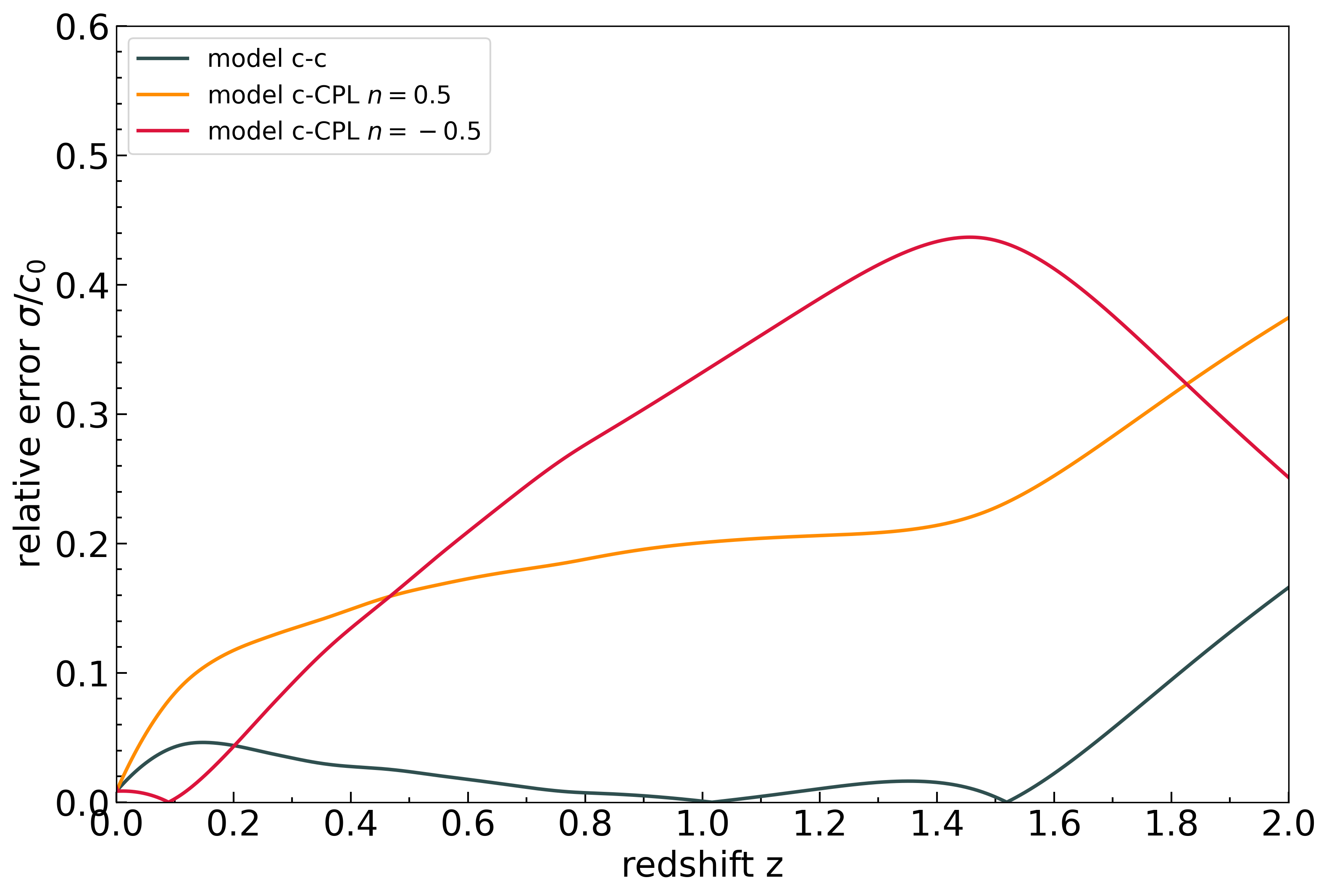}
	}
	\subfigure[]{
		\includegraphics[width=0.47\linewidth]{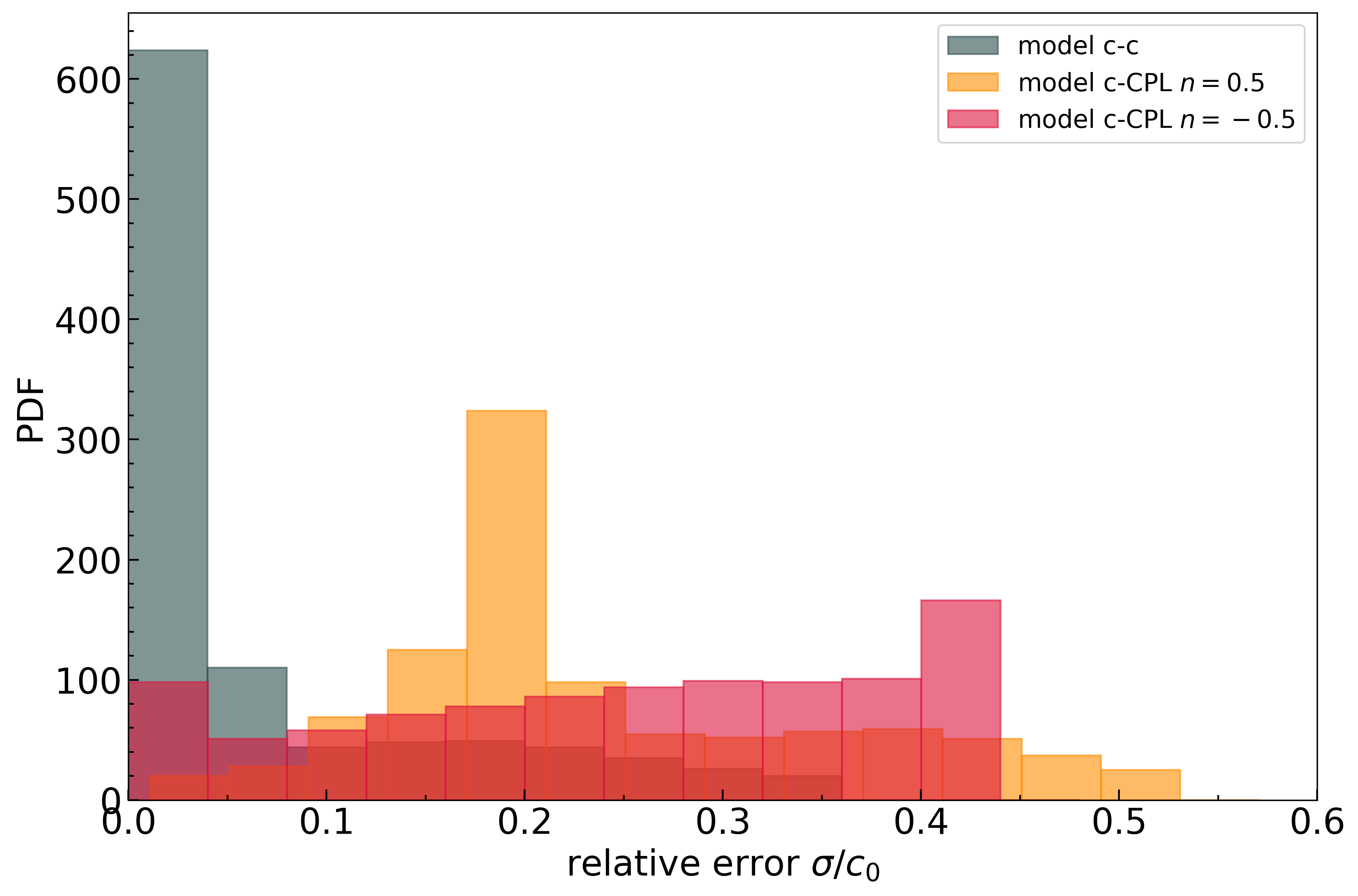}
	}
	\caption{The comparison of four ansatzes in redshift $z\in[0,2]$ with relative errors of (a) the ``$c$-c" model and the ``$c$-cl" model with $n = -0.5$ , (c) the ``$c$-c" model, the ``$c$-CPL" model with $n = 0.5$ and the ``$c$-CPL" model with $n = -0.5$, and probability density function of the relative errors (b) and (d), respectively. The ``$c$-c" model, the ``$c$-cl" model, the ``$c$-CPL" model with $n = 0.5$, and the ``$c$-CPL" model with $n = -0.5$ are labeled in color darkslategray, indigo, darkorange, and crimson in sequence.}
	\label{fig:model-check}
\end{figure*}

Then, we compare two VSL models with the universal ``$c$ is constant" model. For our analysis, we consider the following scenarios: the speed of light $c$ is constant(``$c$-c" model) , ${c} = {c_0}{a^n}$(``c-cl" model) with $n=0.5$, $c$ = ${c_0}[1+n(1-a)]$(``$c$-CPL" model) with $n = 0.5$, and $c$ = ${c_0}[1+n(1-a)]$(“$c$-CPL" model) with $n = -0.5$. For the ``$c$-cl" model, \cite{PhysRevD.59.043515} has given an upper bound on $n$, which is $-0.5$; for the ``$c$-CPL" model, we just assume two possibilities of $n$. Moreover, to compare the fit of four models,  we provide the relative errors ${\sigma}/{c_{model}}$ \citep{2013PhRvD..88j3528Y} the redshift range $z\in[0.07,1.965]$ and the probability density function (PDF) of the relative errors in Figure \ref{fig:model-check}, where ${c_{model}}$ are the theoretical values of the models $(2.9979 \pm 0.19) \times 10^5 \mathrm{~km} / \mathrm{s}$. 

The upper panels provide a comparison between Barrow's traditional VSL model and the universal constant speed of light model in the Figure. \ref{fig:model-check}. It is easy to draw the conclusion that the ``$c$-c" model fits our results much better since the relative errors of it center on a smaller value of number. On the other hand, the classical VSL model does not fit well with our results. Furthermore, it is noteworthy to remark that the value of $n = -0.5$ serves as an upper limit for $n$ in order to provide an explanation for the flatness issue, as discussed in \cite{PhysRevD.59.043515}. If we assume a smaller value of $n$, the fitted result will be worse. The lower panels make a comparison between the well-known CPL model and the ``$c$-c" model in the Figure. \ref{fig:model-check}. If we assume $n = 0.5$ in the ``$c$-CPL" model, the fitted result seems even better than the ``$c$-c" model when using this judging method, since the relative errors of it center closer to the small value of the number; but if we assume $n = -0.5$ in the ``$c$-CPL" model, the result will be no longer credible. By virtue of the result, we cannot robustly exclude the CPL model with strong confidence.

\begin{figure*}
	\centering
	\subfigure[]{
		\includegraphics[width=0.47\linewidth]{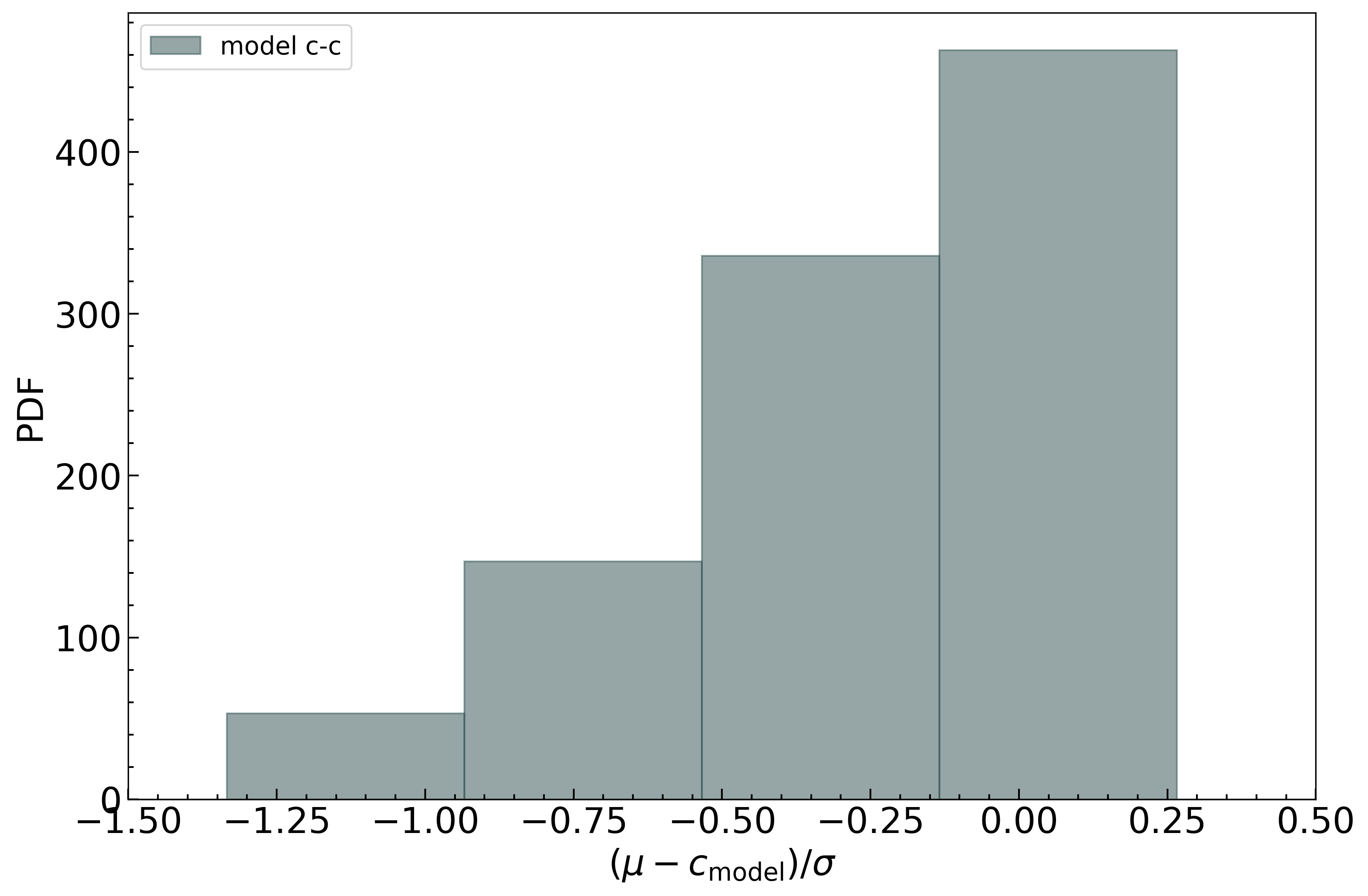}
	}
	\subfigure[]{
		\includegraphics[width=0.47\linewidth]{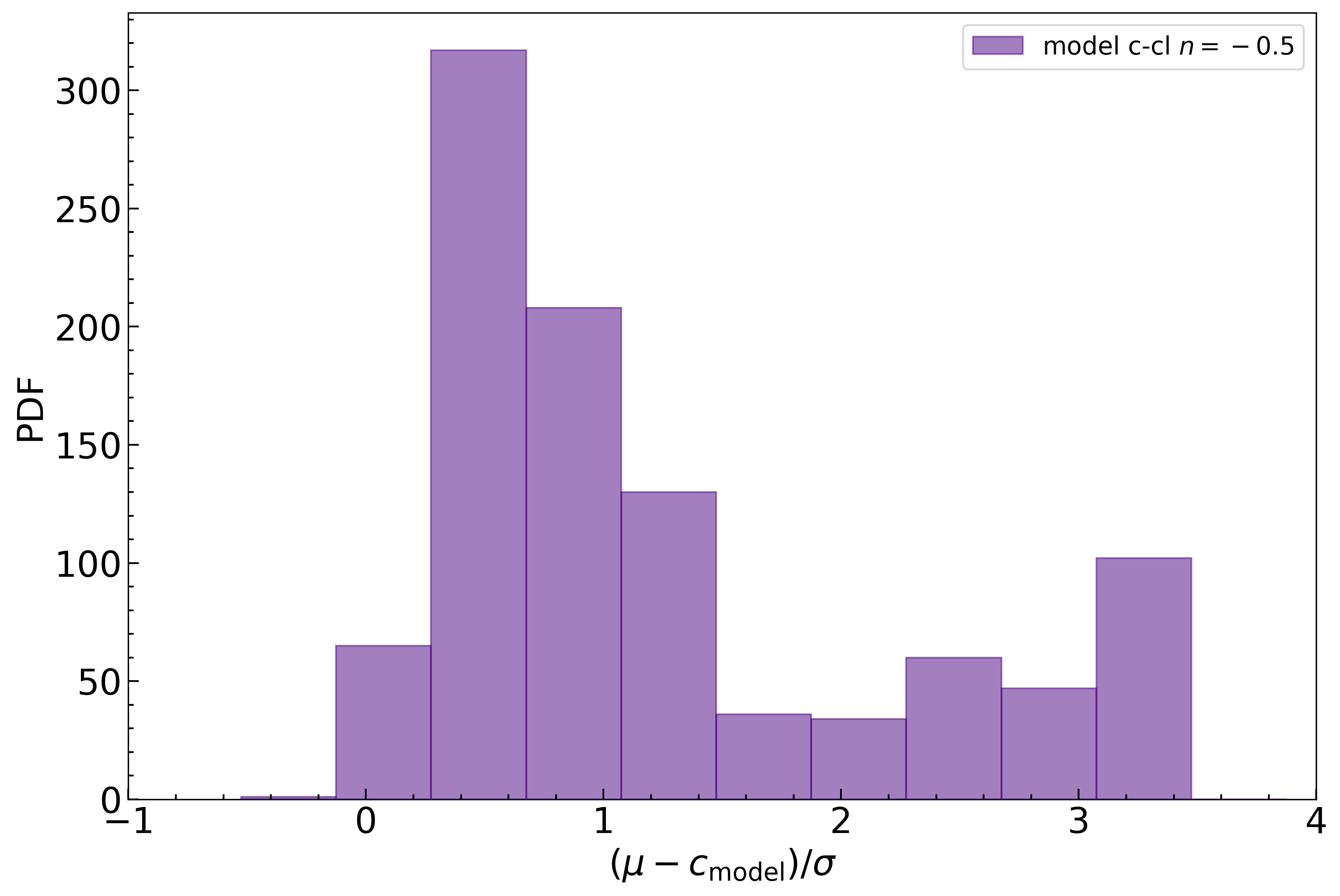}
	}
	\subfigure[]{
		\includegraphics[width=0.47\linewidth]{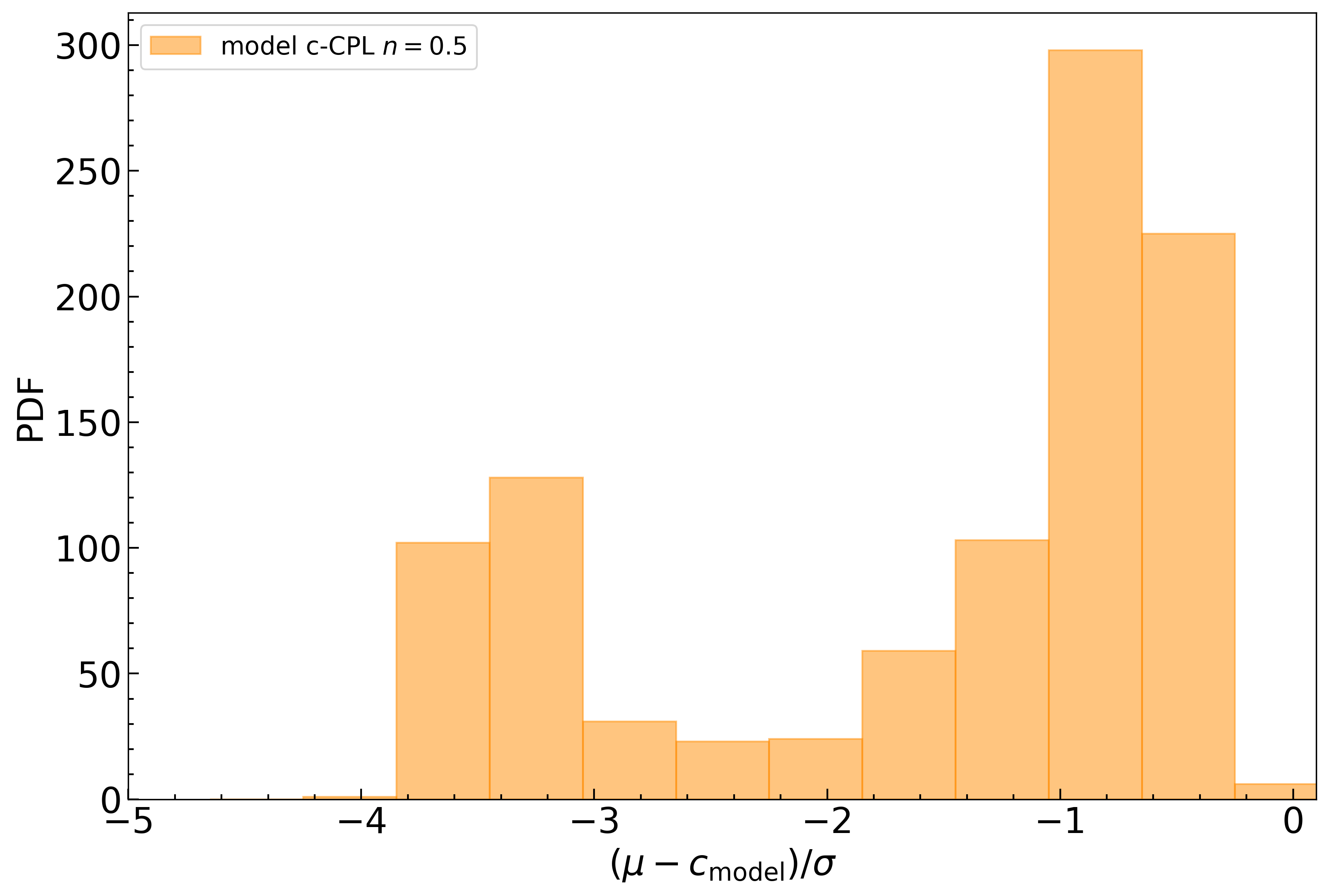}
	}
	\subfigure[]{
		\includegraphics[width=0.47\linewidth]{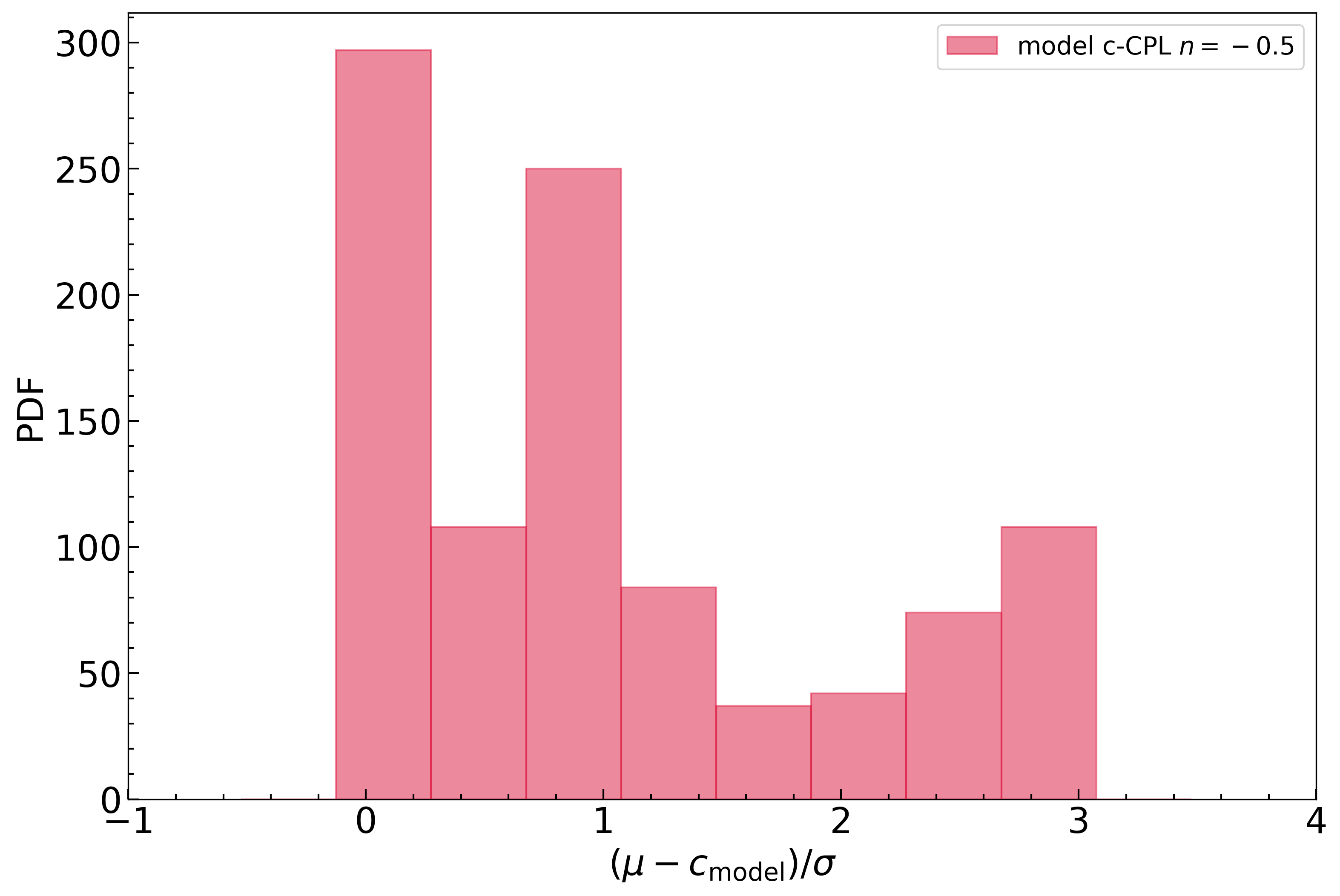}
	}
	\caption{The consistency of four four ansatzes in redshift $z\in[0.11,2.36]$ (a) the ``$c$-c" model, (b) the ``$c$-cl" model with $n = -0.5$, (c) the ``$c$-CPL" model with $n = 0.5$, and the ``$c$-CPL" model with $n = -0.5$. And the ``$c$-c" model, the ``$c$-cl" model, the ``$c$-CPL" model with $n = 0.5$, and the ``$c$-CPL" model with $n = -0.5$ are labeled in color darkslategray, indigo, darkorange, and crimson, respectively.
}
	\label{fig:model-fit}
\end{figure*}

In order to provide more evidence supporting the consistency of our findings with the ``$c$-c" model, we provide Figure \ref{fig:model-fit}. The calculation involves determining the quotient of the difference between the reconstructed speed of light, denoted as $\mu$, and the theoretical model's speed of light, denoted as $c_\mathrm{model}$, by the standard deviation $\sigma$. This is expressed as $(\mu-c_\mathrm{model})/\sigma$. If, at a certain redshift, the measured value of $\mu$ significantly deviates from the theoretical value but the Gaussian process at that redshift yields a bigger error, it does not imply that the theoretical model significantly differs from the observed result. In this analysis, we will compute the disparity in relative error. In the "$c$-c" model, the proportions of frequencies for which the standardized residuals $(\mu-c_\mathrm{model})/\sigma$ lie within the intervals $[-0.75, -0.3]$, $[-0.95, -0.1]$, and $[-1.25, 0.25]$ are around 68\%, 95\%, and 99\%, respectively. These proportions closely align with the predicted values of a Gaussian distribution. This observation suggests that the findings are broadly consistent with the ``$c$-c" model.

To check the consistency of our results with the models, we further calculate the reduced chi-square
\begin{equation}\label{eq6}
	\chi_N=\frac{1}{N} \sum_{i=1}^N \frac{\left(c_{\text {model }, i}-\mu_i\right)^2}{\sigma_i^2},
\end{equation}
which degrees of proximity between the obtained results and the theoretical models are assessed. As the value decreases, the observed outcome approaches the theoretical model more closely. We collect a sample of $N=1000$ observations of the variable $z$ in a uniform manner in order to get the statistic $\chi_N$, which primarily serves as a tool for making comparisons. The results indicate that the ``$c$-c" model is associated with a decreased 0.17. In contrast, the other three models, namely the ``$c$-cl" model with $n = -0.5$, the ``$c$-CPL" model with $n = 0.5$, and the ``$c$-CPL" model with $n = -0.5$, are associated with reduced chi-square values of 2.50, 3.74, and 1.92, respectively. Based on numerical analysis, it may be inferred that the ``$c$-c" model exhibits more consistency with our findings.

\begin{figure*}
	\centering
	\subfigure[]{
		\includegraphics[width=0.16\linewidth]{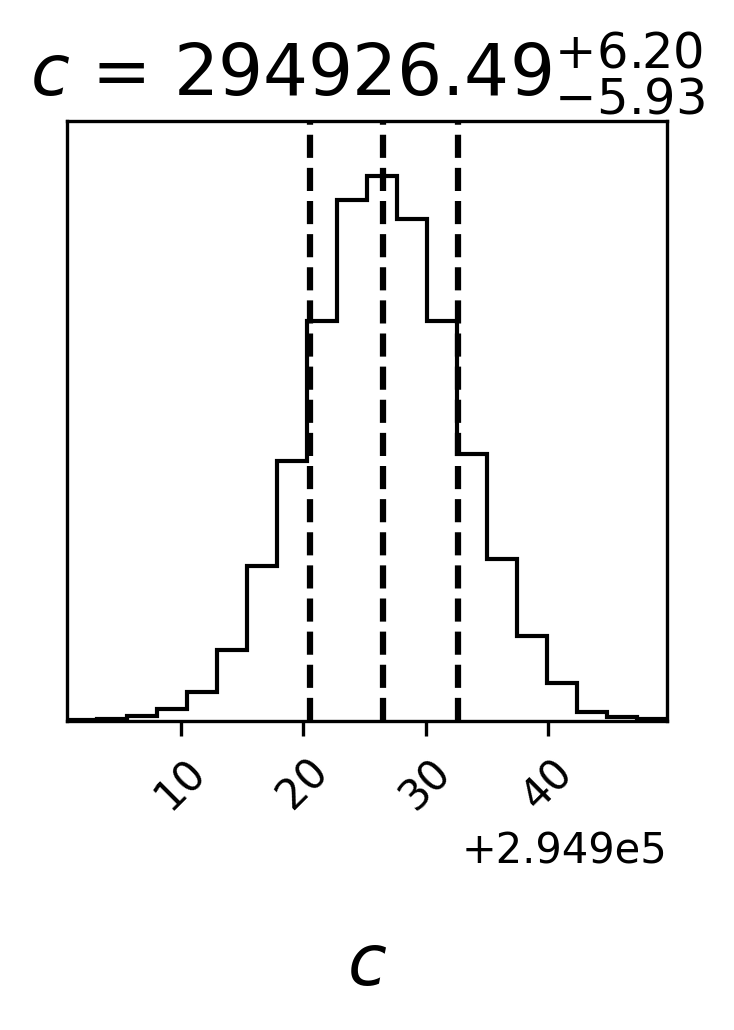}
	}
	\subfigure[]{
		\includegraphics[width=0.31\linewidth]{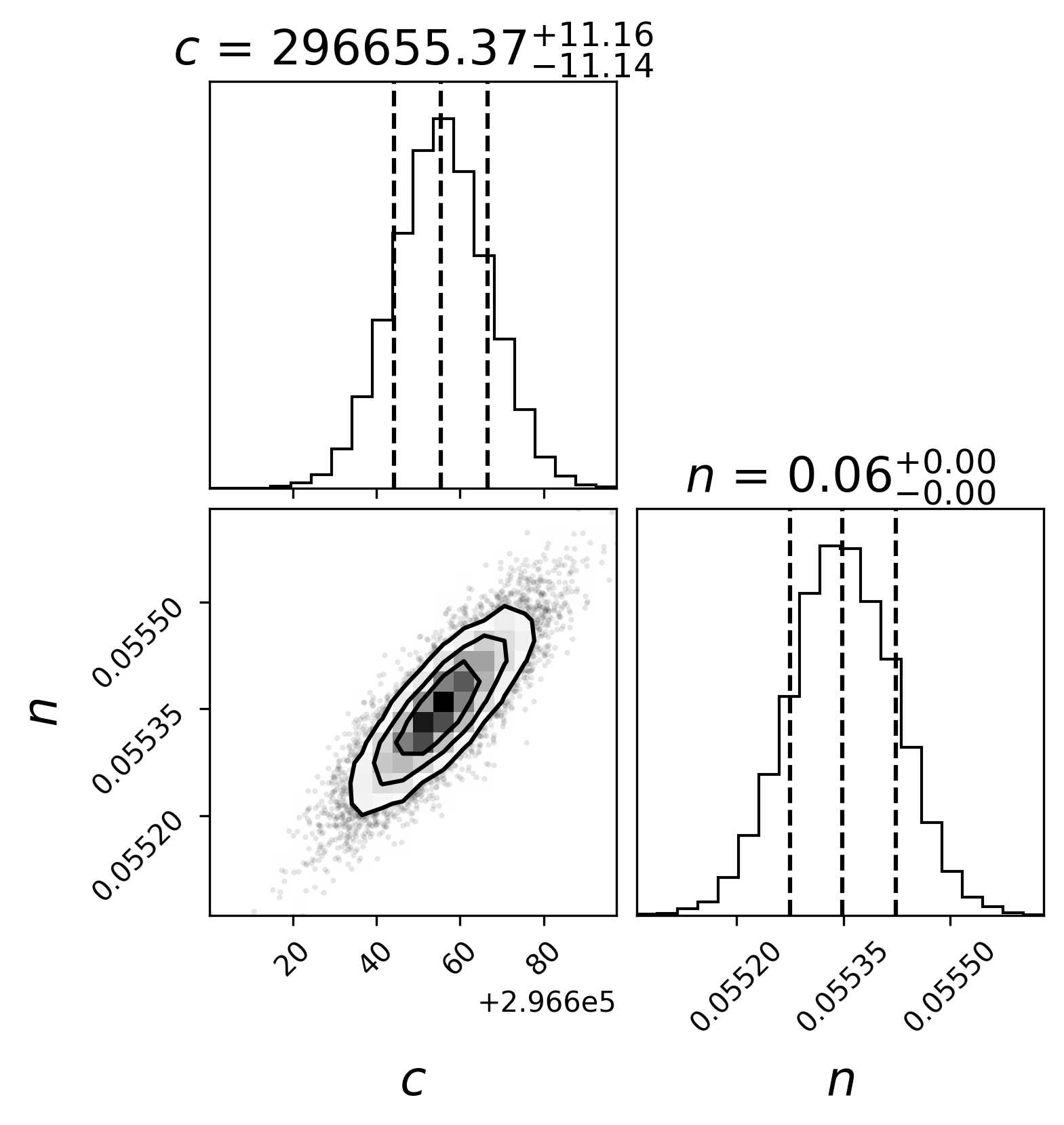}
	}
	\subfigure[]{
		\includegraphics[width=0.31\linewidth]{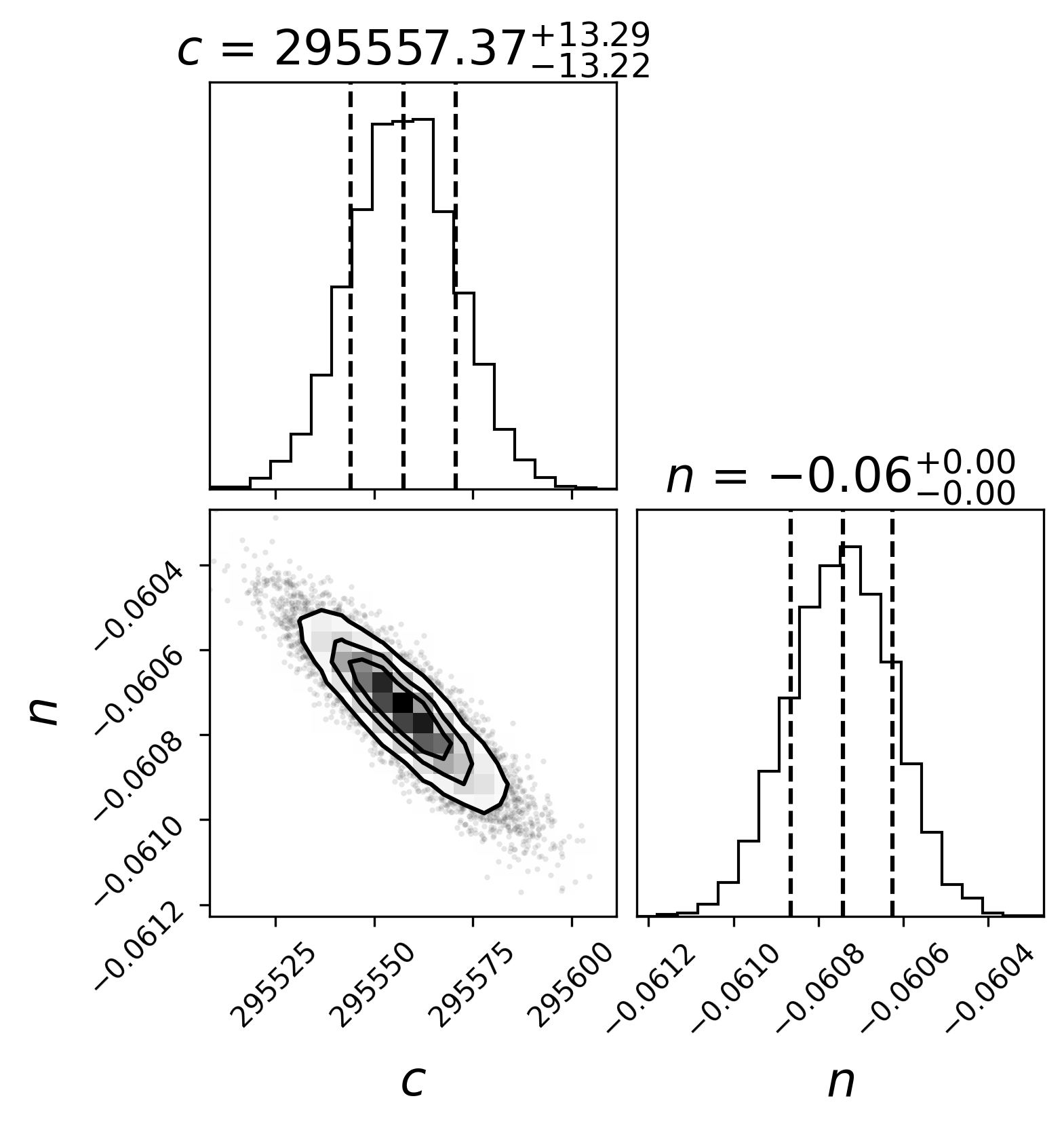}
	}
	\caption{The $68\%$, $95\%$, and $99\%$ confidence regions of the joint and marginal posterior probability distributions of $c_0$ and $n$ are estimated for (a) the ``$c$-c" model, (b) the ``$c$-cl" model, and (c) the ``$c$-CPL" model.
	}
	\label{fig:model-corner}
\end{figure*}

\begin{figure*}
	\centering
	\subfigure[]{
		\includegraphics[width=0.30\linewidth]{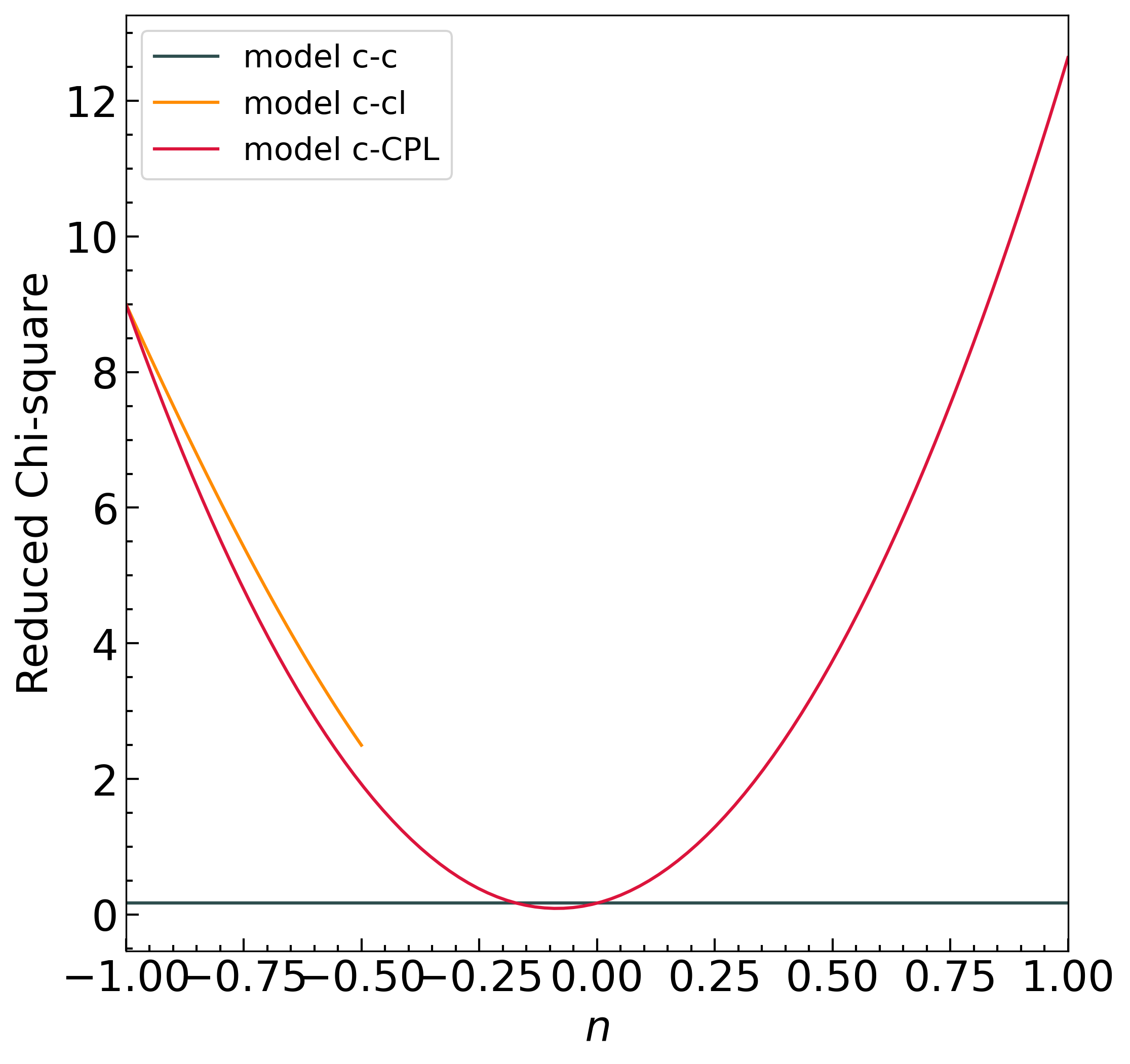}
	}
	\subfigure[]{
		\includegraphics[width=0.33\linewidth]{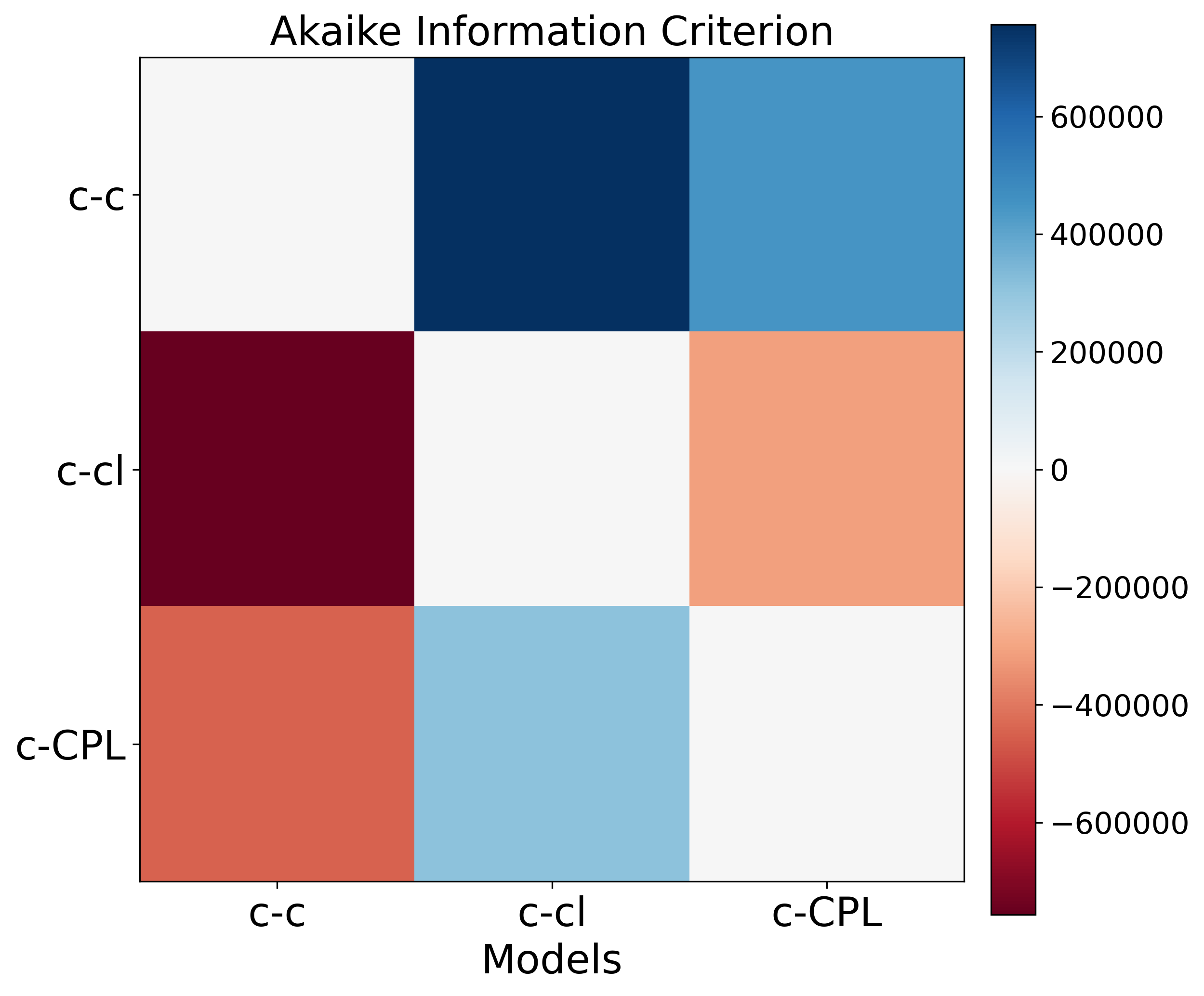}
	}
	\subfigure[]{
		\includegraphics[width=0.33\linewidth]{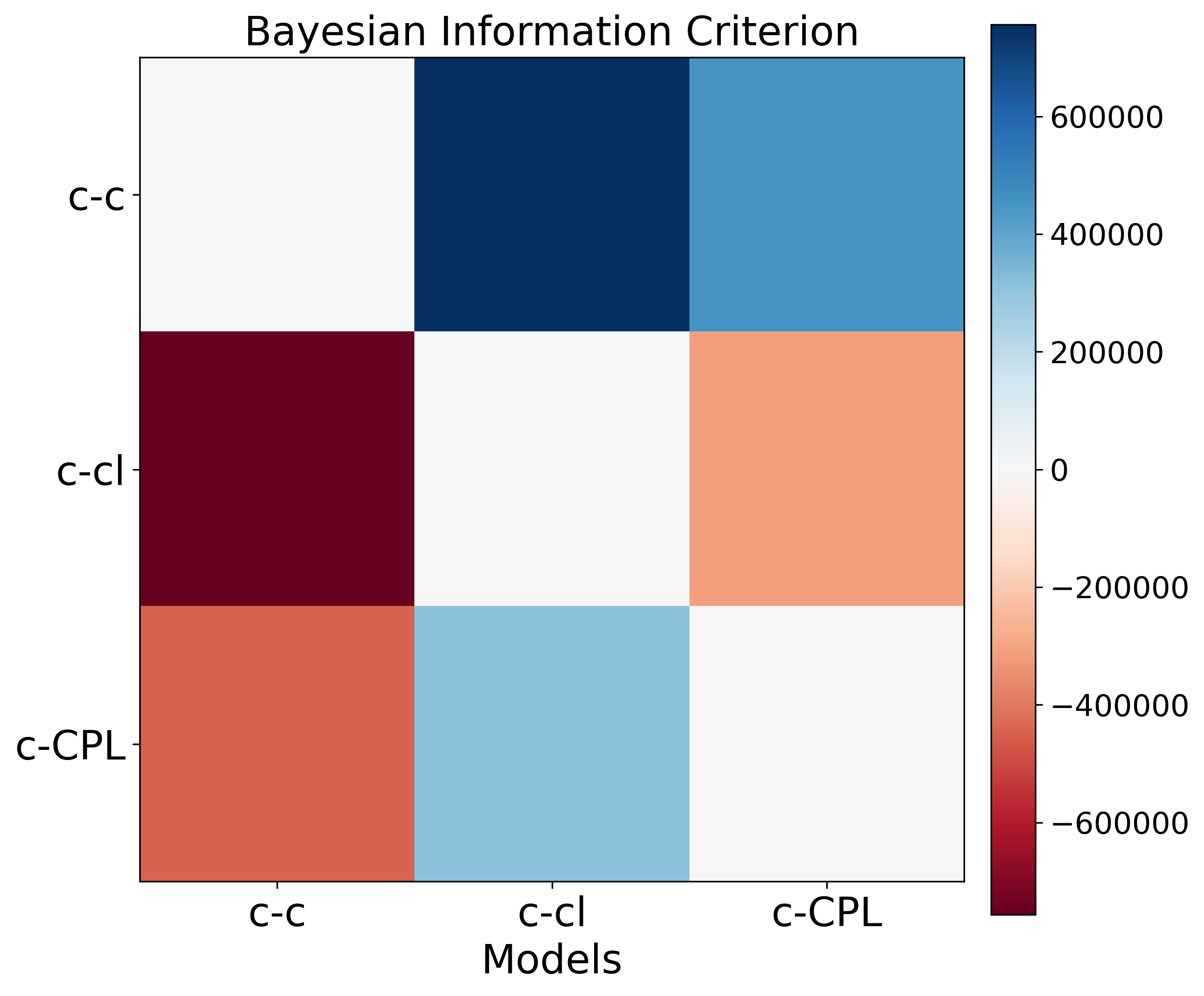}
	}
	\caption{The diagram shows (a) the values of the $\chi^2$ for three models with different $n$, (b) the model selection criteria of AIC, and (c) BIC.}
	\label{fig:model-criterion}
\end{figure*}

In addition, we can also use the reconstructed speed of light $c(z)$ to constrain the parameters in the three models, so as to apply the scope of the model. However, it should be noted that our speed of light $c(z)$ is dependent on our method and data, and other methods and data may give different results for the applicable range of the model. We assume the priors $c_0\in[0,5\times 10^{5}]$ and $n\in[-5,5]$ to constrain parameters in Markov chain Monte Carlo (MCMC) respectively. Here, we use the Python implementation of the affine-invariant ensemble sampler for Markov chain Monte Carlo (emcee) to obtain the estimated posterior \citep{2013PASP..125..306F}. The posteriors of mock data and reconstructed data are shown in Fig. \ref{fig:model-corner}. We find that (1) for the ``$c$-c" model, $c_0=29492.6 \pm^{6.2}_{5.3} \mathrm{~km} \mathrm{~s}^{-1}$. (2) For the ``$c$-cl" model, $c_0=29665.5 \pm^{11.2}_{11.4}\mathrm{~km} \mathrm{~s}^{-1}$ and $n=0.05535 \pm^{0.00008}_{0.00007}$. (3) For the ``$c$-CPL" model, $c_0=29555.7 \pm^{13.3}_{13.2} \mathrm{~km} \mathrm{~s}^{-1}$ and $n=-0.0607 \pm 0.0001$. It is worth noting that, unlike GPs, the parameter constraints obtained through likelihood functions and least squares methods describe the overall information of the data and are influenced by the global data. GPs, on the other hand, focus on reflecting the local relationships between data points. This characteristic of GPs can be observed from Equation \ref{GP-ker}, highlighting how data varies with respect to a particular variable, such as the speed of light varying with redshift in this study. In order to compare the significance of the three models, we utilize two selection model criteria: Akaike Information Criterion (AIC) \citep{1311138} and Bayesian Information Criterion (BIC) \citep{10.1214/aos/1176344136}. Both the AIC and the BIC estimate the quality of a model using a given dataset. AIC and BIC provide a measure of the relative quality between two models, estimate the missing information of a given model and consider both the goodness of fit and the simplicity of the model. A model with smaller values of AIC and BIC indicates less information loss and higher model quality. Both AIC and BIC suffer from overfitting, which they solve by adding a penalty term to the model. The difference is that the penalty term in BIC is larger than in AIC. The definitions of AIC and BIC are $\mathrm{AIC}=2 k-2 \ln (\hat{L})$, $\mathrm{BIC}=k \ln (n)-2 \ln (\hat{L})$. Where $\hat{L}$ is the maximum value of the likelihood function of the model, $k$ is the number of the estimated parameters of the model, $n$ is the sample size. Combined with the reduced chi-square given in Equation \ref{eq6}, we show the reduced chi-square, AIC and BIC of the three models with the change of parameter $n$ in Figure \ref{fig:model-criterion}. Since the parameter $n$ is not included in the ``$c$-c" model, its reduced chi-square does not change with $n$. It can be seen from Figure \ref{fig:model-criterion}(a) that the parameter $n$ of the ``$c$-CPL" model is lower than the reduced chi-square of the ``$c$-c" model in only a small range, which is more consistent with the data. The ``$c$-cl" model is slightly less consistent with the data than the other two models in its parameter range. The heatmaps in Figure \ref{fig:model-criterion}(b) and (c) can be read like this, from the Y-axis to the X-axis. For example, the first row and second column in the concrete result of each graph should be interpreted as the AIC or BIC of c-c (Y-axis) with respect to c-cl (X-axis) $\mathrm{AIC/BIC}=Y-X$. It can be concluded from both Figure \ref{fig:model-criterion}(b) and (c) that: In the three models, ``$c$-c" model is most consistent with the data, ``$c$-CPL" model is slightly less consistent with the data, and `$c$-cl" model is least consistent with the data.

\begin{figure*}
	\centering
	\subfigure[]{
		\includegraphics[width=0.20\linewidth]{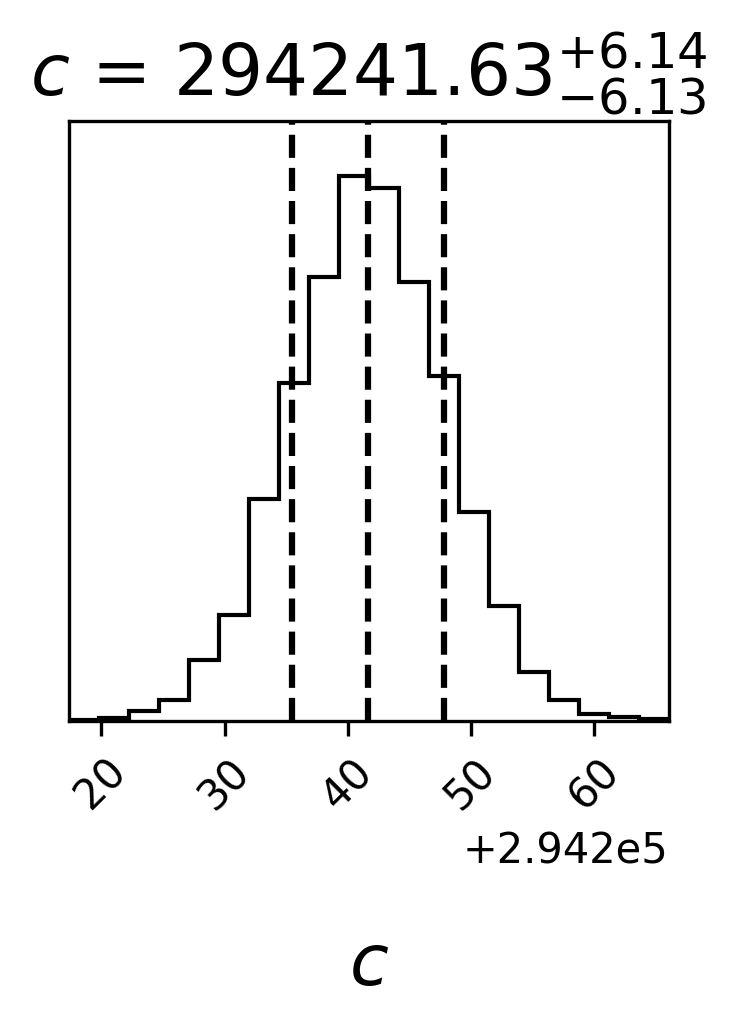}
	}
	\subfigure[]{
		\includegraphics[width=0.33\linewidth]{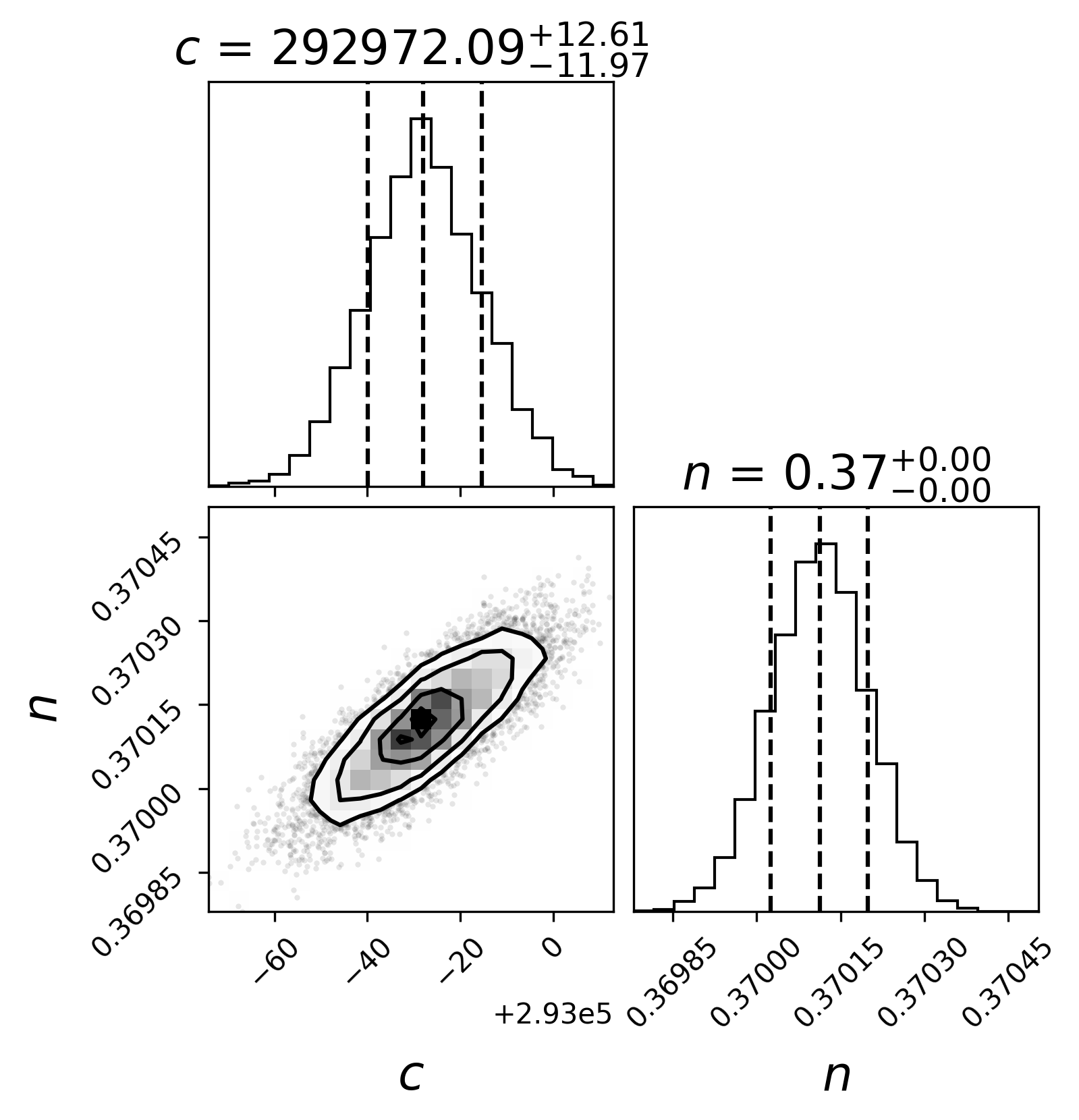}
	}
	\subfigure[]{
		\includegraphics[width=0.33\linewidth]{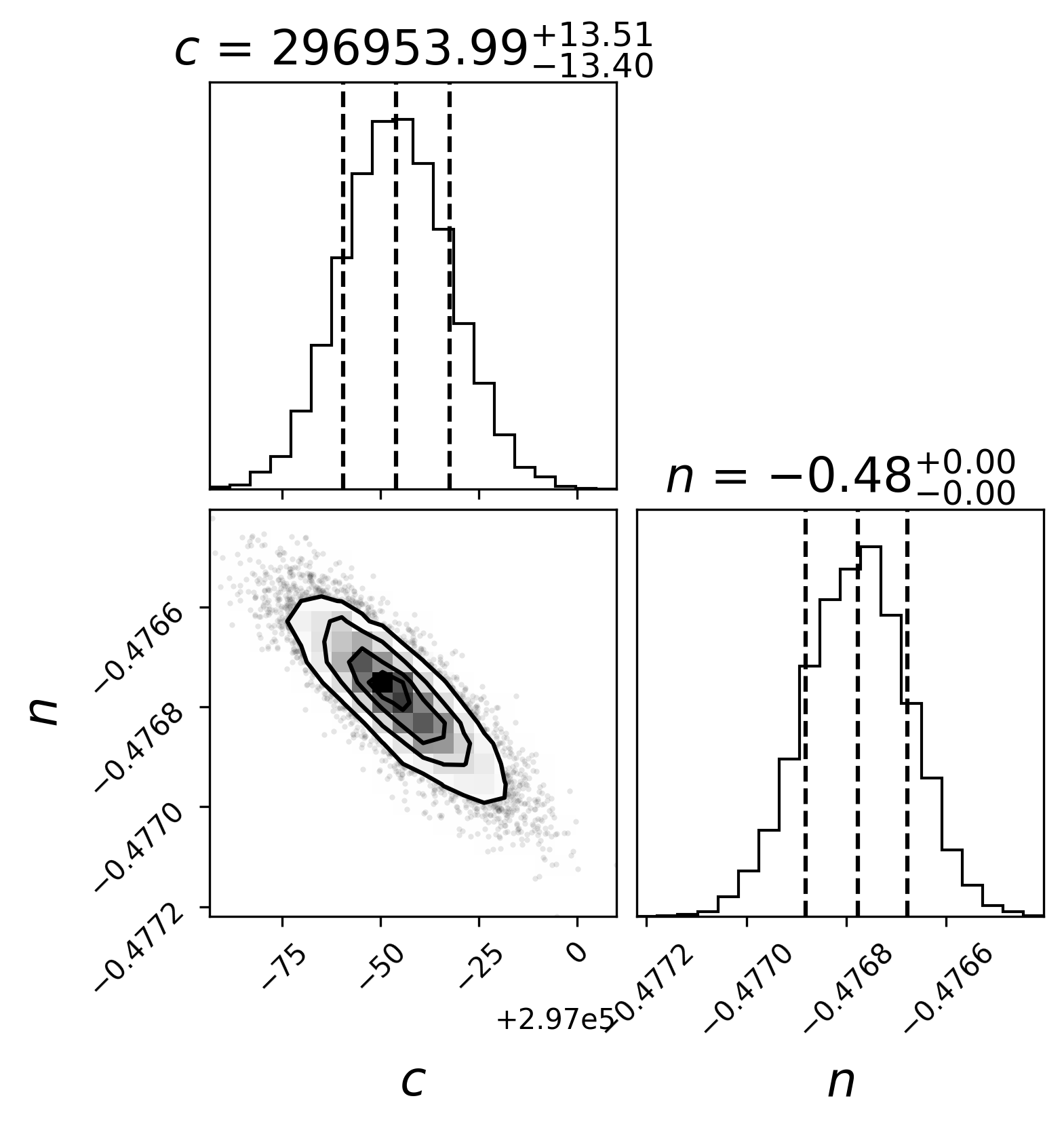}
	}
	\subfigure[]{
		\includegraphics[width=0.30\linewidth]{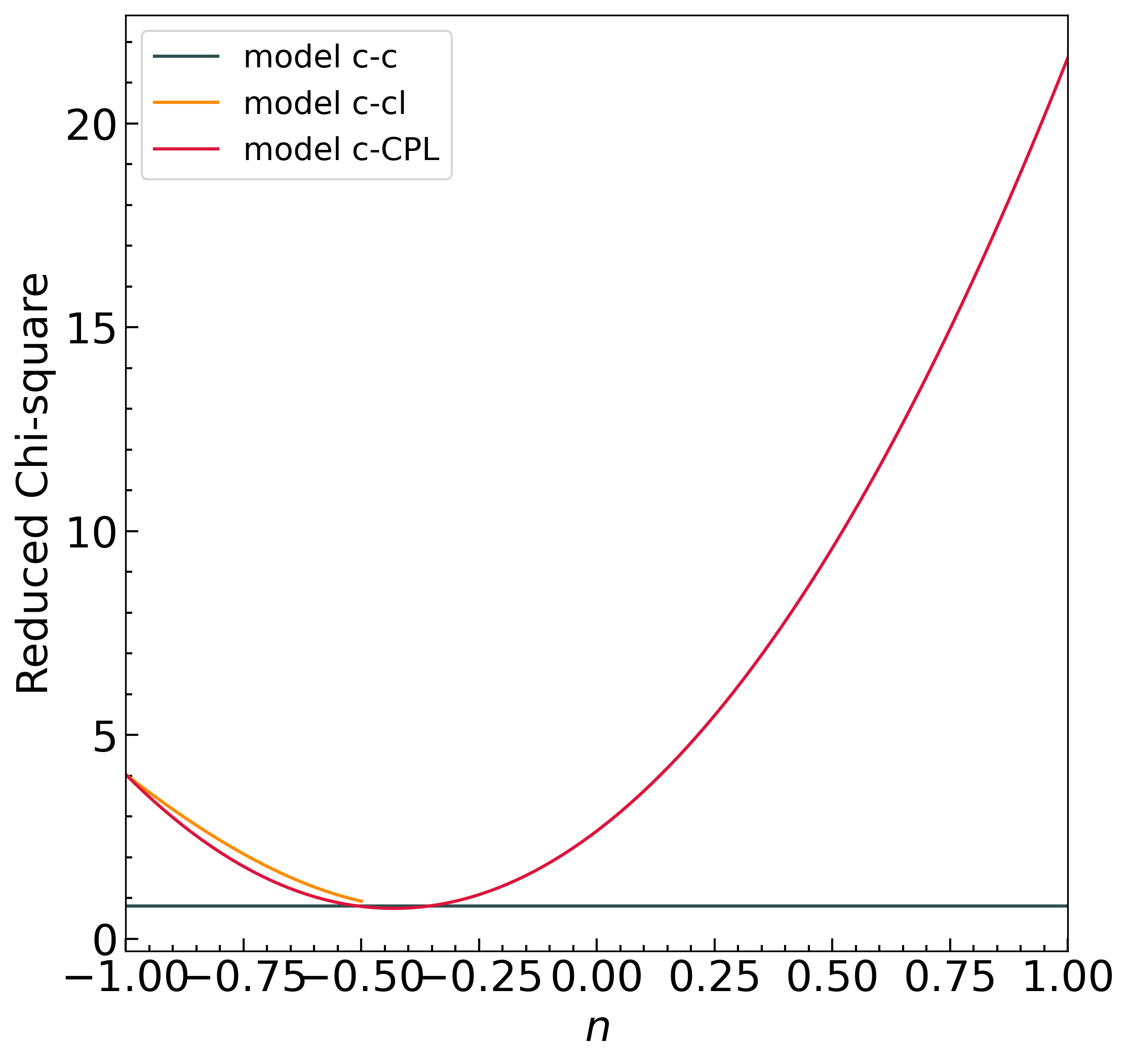}
	}
	\subfigure[]{
		\includegraphics[width=0.33\linewidth]{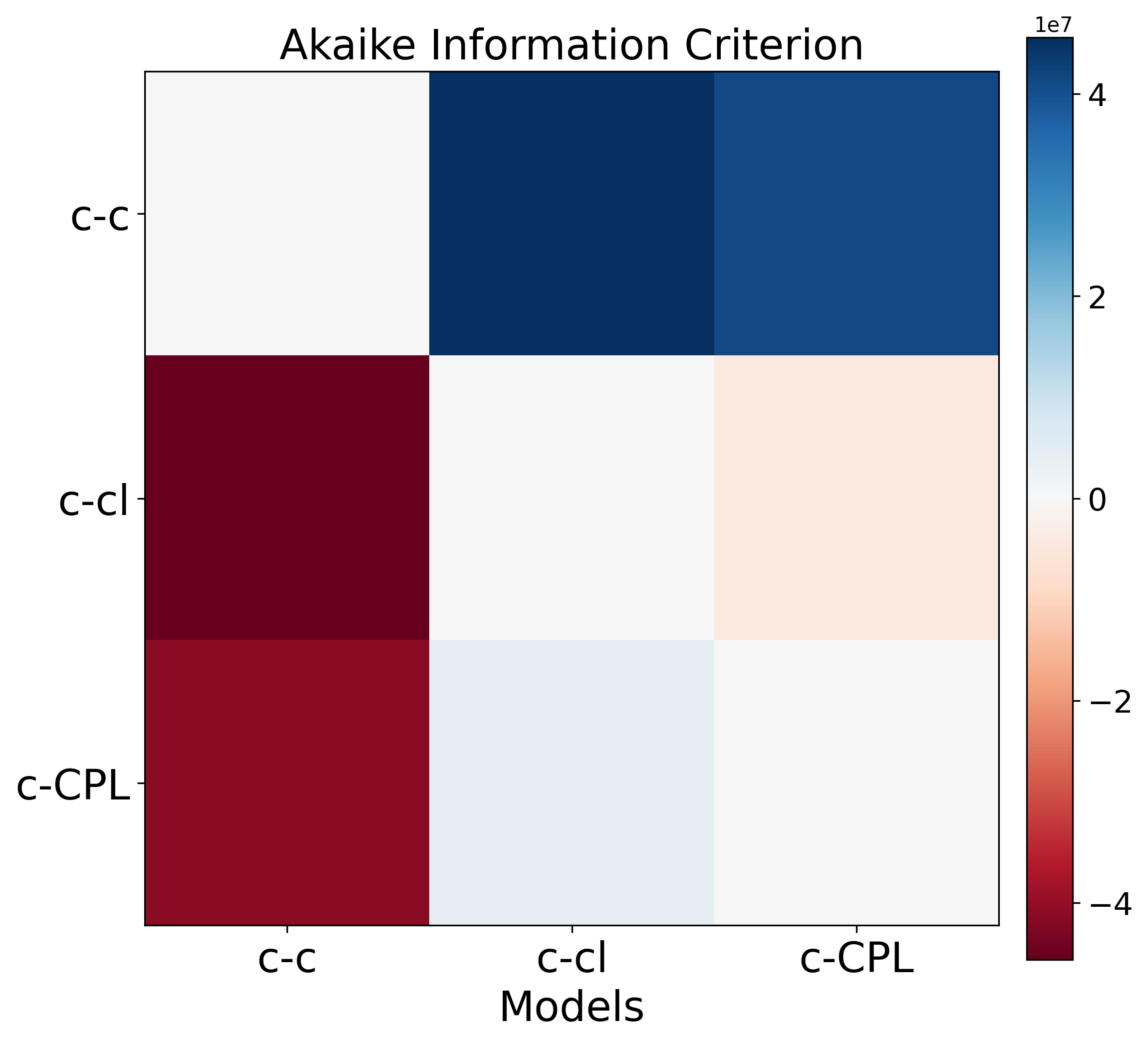}
	}
	\subfigure[]{
		\includegraphics[width=0.33\linewidth]{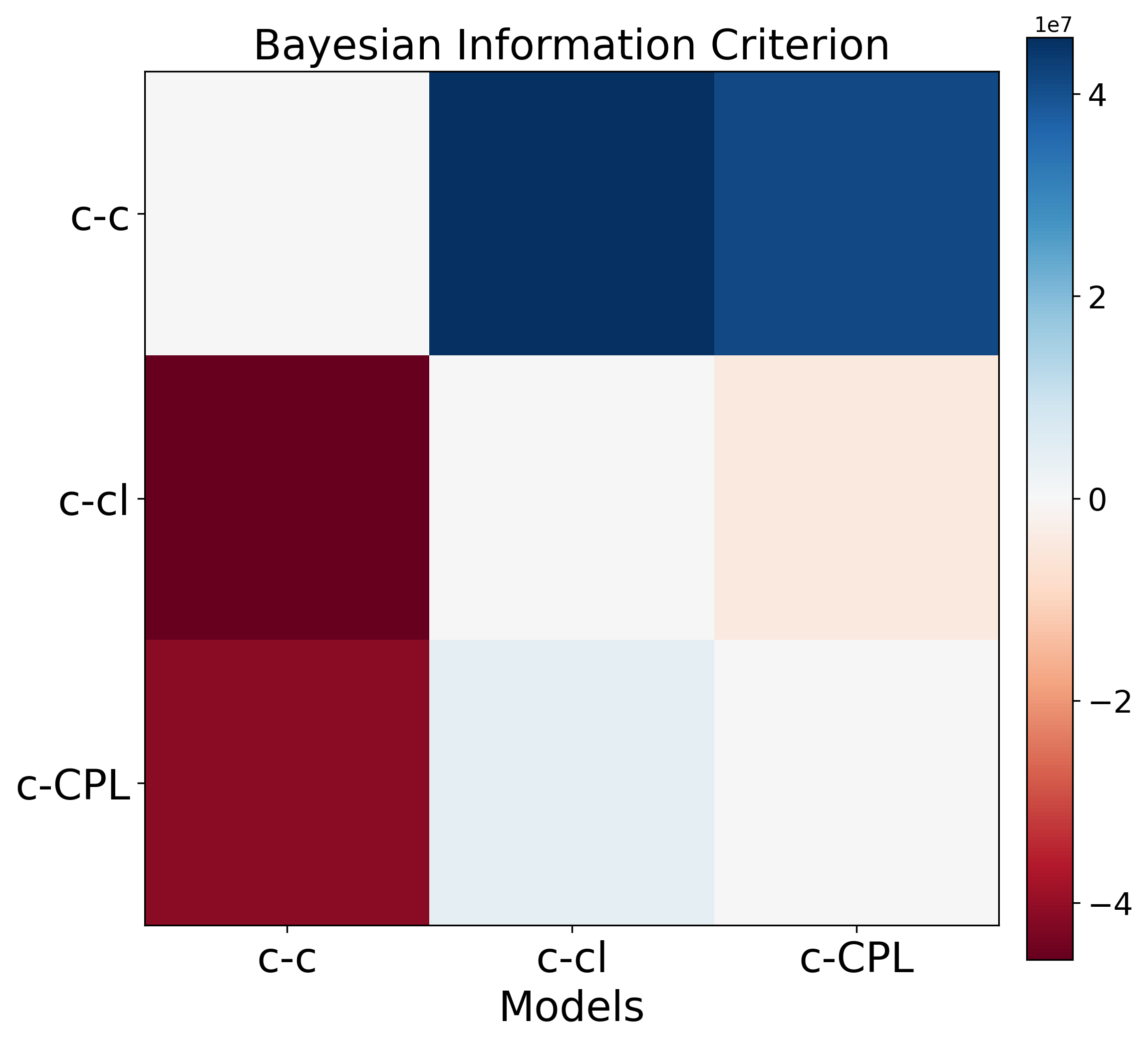}
	}
	\caption{The diagram shows that at the top panels: the $68\%$, $95\%$, and $99\%$ confidence regions of the joint and marginal posterior probability distributions of $c_0$ and $n$ are estimated for (a) the ``$c$-c" model, (b) the ``$c$-cl" model, and (c) the ``$c$-CPL" model. And at the bottom panels: (d) the values of the $\chi^2$ for three models with different $n$, (e) the model selection criteria of AIC, and (f) BIC.}
	\label{fig:redshift-cut}
\end{figure*}

It is interesting to study the applicability of the model by discussing cutting out data with high redshifts. We pointed out earlier that the reconstructed results fluctuate downward around redshift 1.5, and one possible explanation is the lack of high redshift data. Therefore, we intercept the data with high redshift, so that the redshift range of the $H(z)$ data becomes $[0.07,1.53]$, and the redshift range of the $D_A(z)$ data becomes $[0.009783,1.52]$. Repeating the MCMC, reduced chi-square, and AIC/BIC calculations above, we can obtain the parameter constraint results and model selection results for the three models. We find that from Figure \ref{fig:redshift-cut}(a), (b), and(c): (1) for the ``$c$-c" model, $c_0=29424.1 \pm^{6.14}_{6.13} \mathrm{~km} \mathrm{~s}^{-1}$. (2) For the ``$c$-cl" model, $c_0=29720.9 \pm^{12.6}_{12.0}\mathrm{~km} \mathrm{~s}^{-1}$ and $n=0.37113 \pm^{0.00009}_{0.00009}$. (3) For the ``$c$-CPL" model, $c_0=2996954.0 \pm^{13.5}_{13.4} \mathrm{~km} \mathrm{~s}^{-1}$ and $n=-0.4768 \pm 0.0001$. Since the parameter $n$ is not included in the ``$c$-c" model, its reduced chi-square does not change with $n$. It can be seen from Figure \ref{fig:redshift-cut}(d) that the parameter $n$ of the ``$c$-CPL" model is lower than the reduced chi-square of the ``$c$-c" model in only a small range, which is more consistent with the data. The ``$c$-cl" model is slightly less consistent with the data than the other two models in its parameter range. The heatmaps in Figure \ref{fig:redshift-cut}(e) and (f) can be read like Figure \ref{fig:model-criterion}. It can be concluded from both Figure \ref{fig:redshift-cut}(e) and (f) that: In the three models, ``$c$-c" model is most consistent with the data, ``$c$-CPL" model is slightly less consistent with the data, and `$c$-cl" model is least consistent with the data. Compared with the results without high redshift data interception: (1) the speed of light constraint results are reduced. (2) The reduced chi-square of ``$c$-c" model increased. The reduced chi-square of ``$c$-cl" and ``$c$-CPL" models decreased when the parameter $n$ was negative, and the gap between the two narrowed. (3) The AIC/BIC gap between the three models was increased.

The constancy of fundamental physical constants is not always guaranteed, either in the terms of spatial or temporal variations. Despite the apparent simplicity of the aforementioned proposition, it bears profound implications for numerous physical phenomena and interactions, subject to scrutiny through diverse observational methodologies. The rules governing natural phenomena are contingent upon certain fundamental constants, which include but are not limited to Newton's constant, denoted as $G$, the speed of light, denoted as $c$, and the elementary charge of an electron, denoted as $e$. The values of these constants have been obtained by empirical experimentation, but ideally they should be derived directly from the fundamental theory. Therefore, it is unwarranted to make the assumption that the locally established values of the fundamental constants may be directly applied to other regions of the universe or to other time periods in cosmic history \citep{Uzan:2010pm,Wong:2007ym,2017RPPh...80l6902M}. The exploration of fundamental constants and their potential spatiotemporal fluctuations holds profound significance within the discipline. Such studies provide valuable insights into physics beyond the standard model, perhaps revealing the existence of supplementary scalar fields and their interactions with the standard sector. The conceptualization not only aids in elucidating the speed of light but also facilitates the determination of several other fundamental physical constants. Leveraging Gaussian processes, alongside artificial neural networks, not only enables the reconstruction of observables but also promises a gradual refinement in the precision of constraints as observational datasets accumulate.

\section{Conclusion} \label{sec:conclusion}

In this paper, we employ GPR to reconstruct the functions $H(z)$, $D_A(z)$, and $D^{\prime}_A(z)$. By doing so, we are able to obtain the values of $c(z)$ at different redshifts. We then proceed to compare these results with several theoretical models and derive the constraints on the model parameters. We find that (1) for the ``$c$-c" model, $c_0=29492.6 \pm^{6.2}_{5.3} \mathrm{~km} \mathrm{~s}^{-1}$. (2) For the ``$c$-cl" model, $c_0=29665.5 \pm^{11.2}_{11.4}\mathrm{~km} \mathrm{~s}^{-1}$ and $n=0.05535 \pm^{0.00008}_{0.00007}$. (3) For the ``$c$-CPL" model, $c_0=29555.7 \pm^{13.3}_{13.2} \mathrm{~km} \mathrm{~s}^{-1}$ and $n=-0.0607 \pm 0.0001$. To acquire the outcomes of the speed of light measurement, the approximate Bayesian computation rejection technique is employed. This method facilitates the selection of the Gaussian kernel function suitable for two distinct observables, namely $H(z)$ and $D_A(z)$. Additionally, the likelihood function method is utilized to train the hyperparameters of the GP. After ensuring that each kernel function carries equally sampling weight when considering three different distance functions, we conclude that M32 is the most appropriate kernel function for the two observables. This determination is based on the approximate Bayesian computation rejection posterior distribution and the Bayes factor $\mathcal{B}_f$. Based on the assumption of a constant speed of light $c$, it may be inferred that the fitted outcome exhibits superior performance compared to the traditional VSL model, as provided by \cite{PhysRevD.59.043515}. Nevertheless, it is important to consider the theoretical limitations on the parameter $n$ inside the $c(z)$ function of the CPL model before completely dismissing its relevance.

Currently, it can be inferred that it is possible for us to constrain the speed of light roughly based on OHD data and $D_A(z)$ data, and the results are basically consistent with the speed of light being constant and also with some other VSL models (which cannot be ruled out), but it has been possible to rule out some of the VSL model parameters. It is evident that the reconstruction of $c(z)$ does not exhibit the anticipated constant, which may be due to the scarcity of data points and the fact that we do not introduce additional cosmological models and cosmological information in the reconstruction from the beginning of the data to the result. Moreover, the curve of $c(z)$ shows an aberrant decline. This is an unexpected result that is in disagreement with the constancy of $c$ and even runs counter to most of the famous VSL models that are being investigated. This phenomenon arises because of the observed sensitivity of various kernel functions to the reconstruction outcomes of $D^{\prime}_A(z)$ during reconstruction. When the redshift $z$ evolves to the redshift $z_m$'s neighborhood of the maximum of $D^{\prime}_A(z_m)$, large derivative estimates will cause an obvious shake in the measurement value of the speed of light $c(z)$. It is hypothesized that enhancing the reconstruction of variable $c$ may be achieved by the acquisition of additional data points within the interval $z\in[1.52,2.33]$. This indicates that our observations of OHD and BAO data at high redshifts are still inadequate. Therefore, we should still enhance the scale and precision of our galaxy surveys to obtain richer and more accurate $D_A(z)$ and OHD observations. In addition to the traditional $D_A(z)$ and OHD data obtained from galactic observations, gravitational waves and fast radio bursts can also provide $D_A(z)$ and OHD from the standard siren and the dispersion measure of the intergalactic medium. These data can be used as a source of new cosmological observations, providing a wider choice of constraints for the speed of light and other cosmological parameters.

In forthcoming research, our intention is to employ artificial neural networks for the purpose of reconstructing the desired observables' functions. Additionally, we aim to investigate the measurement outcomes of various physical constants under a reconstruction hypothesis that deviates from the Gauss process. This endeavor is undertaken with the objective of minimizing the occurrence of peculiar phenomena resulting from the reconstruction methodology. Furthermore, our research aims to develop a versatile observation design capable of accommodating multiple observations at a consistent redshift. This approach will effectively mitigate the intricate systematic errors that arise when comparing datasets from different observations. Additionally, this design will simplify the calculation of covariance and eliminate the need for reconstructing the function to obtain the final result.

\section*{Acknowledgements}

We sincerely appreciate Kang Jiao, Jing Niu, and Hao Zhang for their kind help. This work was supported by the National SKA Program of China (2022SKA0110202)，China Manned Space Program through its Space Application System and National Science Foundation of China (Grants No. 11929301).

%%%%%%%%%%%%%%%%%%%%%%%%%%%%%%%%%%%%%%%%%%%%%%%%%%
\section*{Data Availability}

The data underlying this article are available in the article from Table 1 and 2.



\bibliographystyle{mnras}
\bibliography{ref} % if your bibtex file is called example.bib

\begin{thebibliography}{}
\makeatletter
\relax
\def\mn@urlcharsother{\let\do\@makeother \do\$\do\&\do\#\do\^\do\_\do\%\do\~}
\def\mn@doi{\begingroup\mn@urlcharsother \@ifnextchar [ {\mn@doi@}
  {\mn@doi@[]}}
\def\mn@doi@[#1]#2{\def\@tempa{#1}\ifx\@tempa\@empty \href
  {http://dx.doi.org/#2} {doi:#2}\else \href {http://dx.doi.org/#2} {#1}\fi
  \endgroup}
\def\mn@eprint#1#2{\mn@eprint@#1:#2::\@nil}
\def\mn@eprint@arXiv#1{\href {http://arxiv.org/abs/#1} {{\tt arXiv:#1}}}
\def\mn@eprint@dblp#1{\href {http://dblp.uni-trier.de/rec/bibtex/#1.xml}
  {dblp:#1}}
\def\mn@eprint@#1:#2:#3:#4\@nil{\def\@tempa {#1}\def\@tempb {#2}\def\@tempc
  {#3}\ifx \@tempc \@empty \let \@tempc \@tempb \let \@tempb \@tempa \fi \ifx
  \@tempb \@empty \def\@tempb {arXiv}\fi \@ifundefined
  {mn@eprint@\@tempb}{\@tempb:\@tempc}{\expandafter \expandafter \csname
  mn@eprint@\@tempb\endcsname \expandafter{\@tempc}}}

\bibitem[\protect\citeauthoryear{{Abazajian} et~al.,}{{Abazajian}
  et~al.}{2003}]{2003AJ....126.2081A}
{Abazajian} K.,  et~al., 2003, \mn@doi [\aj] {10.1086/378165}, \href
  {https://ui.adsabs.harvard.edu/abs/2003AJ....126.2081A} {126, 2081}

\bibitem[\protect\citeauthoryear{{Abazajian} et~al.,}{{Abazajian}
  et~al.}{2004}]{2004AJ....128..502A}
{Abazajian} K.,  et~al., 2004, \mn@doi [\aj] {10.1086/421365}, \href
  {https://ui.adsabs.harvard.edu/abs/2004AJ....128..502A} {128, 502}

\bibitem[\protect\citeauthoryear{{Abazajian} et~al.,}{{Abazajian}
  et~al.}{2005}]{2005AJ....129.1755A}
{Abazajian} K.,  et~al., 2005, \mn@doi [\aj] {10.1086/427544}, \href
  {https://ui.adsabs.harvard.edu/abs/2005AJ....129.1755A} {129, 1755}

\bibitem[\protect\citeauthoryear{{Abazajian} et~al.,}{{Abazajian}
  et~al.}{2009}]{2009ApJS..182..543A}
{Abazajian} K.~N.,  et~al., 2009, \mn@doi [\apjs]
  {10.1088/0067-0049/182/2/543}, \href
  {https://ui.adsabs.harvard.edu/abs/2009ApJS..182..543A} {182, 543}

\bibitem[\protect\citeauthoryear{{Abbott} et~al.,}{{Abbott}
  et~al.}{2022}]{2022PhRvD.105d3512A}
{Abbott} T.~M.~C.,  et~al., 2022, \mn@doi [\prd] {10.1103/PhysRevD.105.043512},
  \href {https://ui.adsabs.harvard.edu/abs/2022PhRvD.105d3512A} {105, 043512}

\bibitem[\protect\citeauthoryear{Abdessalem, Dervilis, Wagg  \&
  Worden}{Abdessalem et~al.}{2017}]{10.3389/fbuil.2017.00052}
Abdessalem A.~B.,  Dervilis N.,  Wagg D.~J.,   Worden K.,  2017, \mn@doi
  [Frontiers in Built Environment] {10.3389/fbuil.2017.00052}, 3

\bibitem[\protect\citeauthoryear{{Abdurro'uf} et~al.,}{{Abdurro'uf}
  et~al.}{2022}]{2022ApJS..259...35A}
{Abdurro'uf} et~al., 2022, \mn@doi [\apjs] {10.3847/1538-4365/ac4414}, \href
  {https://ui.adsabs.harvard.edu/abs/2022ApJS..259...35A} {259, 35}

\bibitem[\protect\citeauthoryear{{Abolfathi} et~al.,}{{Abolfathi}
  et~al.}{2018}]{2018ApJS..235...42A}
{Abolfathi} B.,  et~al., 2018, \mn@doi [\apjs] {10.3847/1538-4365/aa9e8a},
  \href {https://ui.adsabs.harvard.edu/abs/2018ApJS..235...42A} {235, 42}

\bibitem[\protect\citeauthoryear{{Adelman-McCarthy} et~al.,}{{Adelman-McCarthy}
  et~al.}{2006}]{2006ApJS..162...38A}
{Adelman-McCarthy} J.~K.,  et~al., 2006, \mn@doi [\apjs] {10.1086/497917},
  \href {https://ui.adsabs.harvard.edu/abs/2006ApJS..162...38A} {162, 38}

\bibitem[\protect\citeauthoryear{{Adelman-McCarthy} et~al.,}{{Adelman-McCarthy}
  et~al.}{2007}]{2007ApJS..172..634A}
{Adelman-McCarthy} J.~K.,  et~al., 2007, \mn@doi [\apjs] {10.1086/518864},
  \href {https://ui.adsabs.harvard.edu/abs/2007ApJS..172..634A} {172, 634}

\bibitem[\protect\citeauthoryear{{Adelman-McCarthy} et~al.,}{{Adelman-McCarthy}
  et~al.}{2008}]{2008ApJS..175..297A}
{Adelman-McCarthy} J.~K.,  et~al., 2008, \mn@doi [\apjs] {10.1086/524984},
  \href {https://ui.adsabs.harvard.edu/abs/2008ApJS..175..297A} {175, 297}

\bibitem[\protect\citeauthoryear{{Aguado} et~al.,}{{Aguado}
  et~al.}{2019}]{2019ApJS..240...23A}
{Aguado} D.~S.,  et~al., 2019, \mn@doi [\apjs] {10.3847/1538-4365/aaf651},
  \href {https://ui.adsabs.harvard.edu/abs/2019ApJS..240...23A} {240, 23}

\bibitem[\protect\citeauthoryear{{Ahn} et~al.,}{{Ahn}
  et~al.}{2012}]{2012ApJS..203...21A}
{Ahn} C.~P.,  et~al., 2012, \mn@doi [\apjs] {10.1088/0067-0049/203/2/21}, \href
  {https://ui.adsabs.harvard.edu/abs/2012ApJS..203...21A} {203, 21}

\bibitem[\protect\citeauthoryear{{Ahn} et~al.,}{{Ahn}
  et~al.}{2014}]{2014ApJS..211...17A}
{Ahn} C.~P.,  et~al., 2014, \mn@doi [\apjs] {10.1088/0067-0049/211/2/17}, \href
  {https://ui.adsabs.harvard.edu/abs/2014ApJS..211...17A} {211, 17}

\bibitem[\protect\citeauthoryear{{Ahumada} et~al.,}{{Ahumada}
  et~al.}{2020}]{2020ApJS..249....3A}
{Ahumada} R.,  et~al., 2020, \mn@doi [\apjs] {10.3847/1538-4365/ab929e}, \href
  {https://ui.adsabs.harvard.edu/abs/2020ApJS..249....3A} {249, 3}

\bibitem[\protect\citeauthoryear{{Aihara} et~al.,}{{Aihara}
  et~al.}{2011}]{2011ApJS..193...29A}
{Aihara} H.,  et~al., 2011, \mn@doi [\apjs] {10.1088/0067-0049/193/2/29}, \href
  {https://ui.adsabs.harvard.edu/abs/2011ApJS..193...29A} {193, 29}

\bibitem[\protect\citeauthoryear{{Alam} et~al.,}{{Alam}
  et~al.}{2015}]{2015ApJS..219...12A}
{Alam} S.,  et~al., 2015, \mn@doi [\apjs] {10.1088/0067-0049/219/1/12}, \href
  {https://ui.adsabs.harvard.edu/abs/2015ApJS..219...12A} {219, 12}

\bibitem[\protect\citeauthoryear{{Alam} et~al.,}{{Alam}
  et~al.}{2021}]{2021PhRvD.103h3533A}
{Alam} S.,  et~al., 2021, \mn@doi [\prd] {10.1103/PhysRevD.103.083533}, \href
  {https://ui.adsabs.harvard.edu/abs/2021PhRvD.103h3533A} {103, 083533}

\bibitem[\protect\citeauthoryear{{Albareti} et~al.,}{{Albareti}
  et~al.}{2017}]{2017ApJS..233...25A}
{Albareti} F.~D.,  et~al., 2017, \mn@doi [\apjs] {10.3847/1538-4365/aa8992},
  \href {https://ui.adsabs.harvard.edu/abs/2017ApJS..233...25A} {233, 25}

\bibitem[\protect\citeauthoryear{{Albrecht} \& {Magueijo}}{{Albrecht} \&
  {Magueijo}}{1999}]{1999PhRvD..59d3516A}
{Albrecht} A.,  {Magueijo} J.,  1999, \mn@doi [\prd]
  {10.1103/PhysRevD.59.043516}, \href
  {https://ui.adsabs.harvard.edu/abs/1999PhRvD..59d3516A} {59, 043516}

\bibitem[\protect\citeauthoryear{{Almeida} et~al.,}{{Almeida}
  et~al.}{2023}]{2023ApJS..267...44A}
{Almeida} A.,  et~al., 2023, \mn@doi [\apjs] {10.3847/1538-4365/acda98}, \href
  {https://ui.adsabs.harvard.edu/abs/2023ApJS..267...44A} {267, 44}

\bibitem[\protect\citeauthoryear{Barrow}{Barrow}{1999}]{PhysRevD.59.043515}
Barrow J.~D.,  1999, \mn@doi [Phys. Rev. D] {10.1103/PhysRevD.59.043515}, 59,
  043515

\bibitem[\protect\citeauthoryear{Barrow \& Magueijo}{Barrow \&
  Magueijo}{1998}]{Barrow:1998df}
Barrow J.~D.,  Magueijo J.,  1998, \mn@doi [Phys. Lett. B]
  {10.1016/S0370-2693(98)01294-5}, 443, 104

\bibitem[\protect\citeauthoryear{Barrow \& Magueijo}{Barrow \&
  Magueijo}{1999a}]{Barrow:1999jq}
Barrow J.~D.,  Magueijo J.,  1999a, \mn@doi [Class. Quant. Grav.]
  {10.1088/0264-9381/16/4/030}, 16, 1435

\bibitem[\protect\citeauthoryear{Barrow \& Magueijo}{Barrow \&
  Magueijo}{1999b}]{Barrow:1998he}
Barrow J.~D.,  Magueijo J.,  1999b, \mn@doi [Phys. Lett. B]
  {10.1016/S0370-2693(99)00008-8}, 447, 246

\bibitem[\protect\citeauthoryear{Barrow \& Magueijo}{Barrow \&
  Magueijo}{2000}]{Barrow:1999st}
Barrow J.~D.,  Magueijo J.,  2000, \mn@doi [Astrophys. J. Lett.]
  {10.1086/312572}, 532, L87

\bibitem[\protect\citeauthoryear{{Bautista} et~al.,}{{Bautista}
  et~al.}{2021}]{2021MNRAS.500..736B}
{Bautista} J.~E.,  et~al., 2021, \mn@doi [\mnras] {10.1093/mnras/staa2800},
  \href {https://ui.adsabs.harvard.edu/abs/2021MNRAS.500..736B} {500, 736}

\bibitem[\protect\citeauthoryear{{Bernardo} \& {Levi Said}}{{Bernardo} \& {Levi
  Said}}{2021}]{2021JCAP...08..027B}
{Bernardo} R.~C.,  {Levi Said} J.,  2021, \mn@doi [\jcap]
  {10.1088/1475-7516/2021/08/027}, \href
  {https://ui.adsabs.harvard.edu/abs/2021JCAP...08..027B} {2021, 027}

\bibitem[\protect\citeauthoryear{{Beutler} et~al.,}{{Beutler}
  et~al.}{2011}]{2011MNRAS.416.3017B}
{Beutler} F.,  et~al., 2011, \mn@doi [\mnras]
  {10.1111/j.1365-2966.2011.19250.x}, \href
  {https://ui.adsabs.harvard.edu/abs/2011MNRAS.416.3017B} {416, 3017}

\bibitem[\protect\citeauthoryear{{Beutler} et~al.,}{{Beutler}
  et~al.}{2017a}]{2017MNRAS.464.3409B}
{Beutler} F.,  et~al., 2017a, \mn@doi [\mnras] {10.1093/mnras/stw2373}, \href
  {https://ui.adsabs.harvard.edu/abs/2017MNRAS.464.3409B} {464, 3409}

\bibitem[\protect\citeauthoryear{{Beutler} et~al.,}{{Beutler}
  et~al.}{2017b}]{2017MNRAS.466.2242B}
{Beutler} F.,  et~al., 2017b, \mn@doi [\mnras] {10.1093/mnras/stw3298}, \href
  {https://ui.adsabs.harvard.edu/abs/2017MNRAS.466.2242B} {466, 2242}

\bibitem[\protect\citeauthoryear{{Blake} et~al.,}{{Blake}
  et~al.}{2011}]{2011MNRAS.418.1707B}
{Blake} C.,  et~al., 2011, \mn@doi [\mnras] {10.1111/j.1365-2966.2011.19592.x},
  \href {https://ui.adsabs.harvard.edu/abs/2011MNRAS.418.1707B} {418, 1707}

\bibitem[\protect\citeauthoryear{Blake et~al.,}{Blake
  et~al.}{2012a}]{blake-2012-33470}
Blake C.,  et~al., 2012a, {WiggleZ Dark Energy Survey Baryon Acoustic
  Oscillation Random Catalogues}, \mn@doi{10.5281/zenodo.33470}, \url
  {https://doi.org/10.5281/zenodo.33470}

\bibitem[\protect\citeauthoryear{{Blake} et~al.,}{{Blake}
  et~al.}{2012b}]{2012MNRAS.425..405B}
{Blake} C.,  et~al., 2012b, \mn@doi [\mnras]
  {10.1111/j.1365-2966.2012.21473.x}, \href
  {https://ui.adsabs.harvard.edu/abs/2012MNRAS.425..405B} {425, 405}

\bibitem[\protect\citeauthoryear{{Blomqvist} et~al.,}{{Blomqvist}
  et~al.}{2019}]{2019AA...629A..86B}
{Blomqvist} M.,  et~al., 2019, \mn@doi [\aap] {10.1051/0004-6361/201935641},
  \href {https://ui.adsabs.harvard.edu/abs/2019A&A...629A..86B} {629, A86}

\bibitem[\protect\citeauthoryear{{Bonamente}, {Joy}, {LaRoque}, {Carlstrom},
  {Reese}  \& {Dawson}}{{Bonamente} et~al.}{2006}]{2006ApJ...647...25B}
{Bonamente} M.,  {Joy} M.~K.,  {LaRoque} S.~J.,  {Carlstrom} J.~E.,  {Reese}
  E.~D.,   {Dawson} K.~S.,  2006, \mn@doi [\apj] {10.1086/505291}, \href
  {https://ui.adsabs.harvard.edu/abs/2006ApJ...647...25B} {647, 25}

\bibitem[\protect\citeauthoryear{{Borghi}, {Moresco}  \& {Cimatti}}{{Borghi}
  et~al.}{2022}]{2022ApJ...928L...4B}
{Borghi} N.,  {Moresco} M.,   {Cimatti} A.,  2022, \mn@doi [\apjl]
  {10.3847/2041-8213/ac3fb2}, \href
  {https://ui.adsabs.harvard.edu/abs/2022ApJ...928L...4B} {928, L4}

\bibitem[\protect\citeauthoryear{Buitinck et~al.,}{Buitinck
  et~al.}{2013}]{sklearn-api}
Buitinck L.,  et~al., 2013, in ECML PKDD Workshop: Languages for Data Mining
  and Machine Learning. pp 108--122

\bibitem[\protect\citeauthoryear{{Cai}, {Guo}  \& {Yang}}{{Cai}
  et~al.}{2016}]{2016JCAP...08..016C}
{Cai} R.-G.,  {Guo} Z.-K.,   {Yang} T.,  2016, \mn@doi [\jcap]
  {10.1088/1475-7516/2016/08/016}, \href
  {https://ui.adsabs.harvard.edu/abs/2016JCAP...08..016C} {2016, 016}

\bibitem[\protect\citeauthoryear{{Chen}, {Chapman}, {Wolz}  \&
  {Mazumder}}{{Chen} et~al.}{2023}]{2023MNRAS.524.3724C}
{Chen} Z.,  {Chapman} E.,  {Wolz} L.,   {Mazumder} A.,  2023, \mn@doi [\mnras]
  {10.1093/mnras/stad2102}, \href
  {https://ui.adsabs.harvard.edu/abs/2023MNRAS.524.3724C} {524, 3724}

\bibitem[\protect\citeauthoryear{{Chevallier} \& {Polarski}}{{Chevallier} \&
  {Polarski}}{2001}]{2001IJMPD..10..213C}
{Chevallier} M.,  {Polarski} D.,  2001, \mn@doi [International Journal of
  Modern Physics D] {10.1142/S0218271801000822}, \href
  {https://ui.adsabs.harvard.edu/abs/2001IJMPD..10..213C} {10, 213}

\bibitem[\protect\citeauthoryear{{Dainotti}, {Sharma}, {Narendra}, {Levine},
  {Rinaldi}, {Pollo}  \& {Bhatta}}{{Dainotti}
  et~al.}{2023}]{2023ApJS..267...42D}
{Dainotti} M.~G.,  {Sharma} R.,  {Narendra} A.,  {Levine} D.,  {Rinaldi} E.,
  {Pollo} A.,   {Bhatta} G.,  2023, \mn@doi [\apjs] {10.3847/1538-4365/acdd07},
  \href {https://ui.adsabs.harvard.edu/abs/2023ApJS..267...42D} {267, 42}

\bibitem[\protect\citeauthoryear{{Dawson} et~al.,}{{Dawson}
  et~al.}{2013}]{2013AJ....145...10D}
{Dawson} K.~S.,  et~al., 2013, \mn@doi [\aj] {10.1088/0004-6256/145/1/10},
  \href {https://ui.adsabs.harvard.edu/abs/2013AJ....145...10D} {145, 10}

\bibitem[\protect\citeauthoryear{{Drinkwater} et~al.,}{{Drinkwater}
  et~al.}{2010}]{2010MNRAS.401.1429D}
{Drinkwater} M.~J.,  et~al., 2010, \mn@doi [\mnras]
  {10.1111/j.1365-2966.2009.15754.x}, \href
  {https://ui.adsabs.harvard.edu/abs/2010MNRAS.401.1429D} {401, 1429}

\bibitem[\protect\citeauthoryear{Drinkwater et~al.,}{Drinkwater
  et~al.}{2017}]{10.1093/mnras/stx2963}
Drinkwater M.~J.,  et~al., 2017, \mn@doi [Monthly Notices of the Royal
  Astronomical Society] {10.1093/mnras/stx2963}, 474, 4151

\bibitem[\protect\citeauthoryear{{Einstein}}{{Einstein}}{1911}]{1911AnP...340..898E}
{Einstein} A.,  1911, \mn@doi [Annalen der Physik] {10.1002/andp.19113401005},
  \href {https://ui.adsabs.harvard.edu/abs/1911AnP...340..898E} {340, 898}

\bibitem[\protect\citeauthoryear{{Font-Ribera} et~al.,}{{Font-Ribera}
  et~al.}{2014}]{2014JCAP...05..027F}
{Font-Ribera} A.,  et~al., 2014, \mn@doi [\jcap]
  {10.1088/1475-7516/2014/05/027}, \href
  {https://ui.adsabs.harvard.edu/abs/2014JCAP...05..027F} {2014, 027}

\bibitem[\protect\citeauthoryear{{Foreman-Mackey}, {Hogg}, {Lang}  \&
  {Goodman}}{{Foreman-Mackey} et~al.}{2013}]{2013PASP..125..306F}
{Foreman-Mackey} D.,  {Hogg} D.~W.,  {Lang} D.,   {Goodman} J.,  2013, \mn@doi
  [\pasp] {10.1086/670067}, \href
  {https://ui.adsabs.harvard.edu/abs/2013PASP..125..306F} {125, 306}

\bibitem[\protect\citeauthoryear{{Gil-Mar{\'\i}n} et~al.,}{{Gil-Mar{\'\i}n}
  et~al.}{2018}]{2018MNRAS.477.1604G}
{Gil-Mar{\'\i}n} H.,  et~al., 2018, \mn@doi [\mnras] {10.1093/mnras/sty453},
  \href {https://ui.adsabs.harvard.edu/abs/2018MNRAS.477.1604G} {477, 1604}

\bibitem[\protect\citeauthoryear{{Gonz{\'a}lez}, {Alcaniz}  \&
  {Carvalho}}{{Gonz{\'a}lez} et~al.}{2016}]{2016JCAP...04..016G}
{Gonz{\'a}lez} J.~E.,  {Alcaniz} J.~S.,   {Carvalho} J.~C.,  2016, \mn@doi
  [\jcap] {10.1088/1475-7516/2016/04/016}, \href
  {https://ui.adsabs.harvard.edu/abs/2016JCAP...04..016G} {2016, 016}

\bibitem[\protect\citeauthoryear{{Grieb} et~al.,}{{Grieb}
  et~al.}{2017}]{2017MNRAS.467.2085G}
{Grieb} J.~N.,  et~al., 2017, \mn@doi [\mnras] {10.1093/mnras/stw3384}, \href
  {https://ui.adsabs.harvard.edu/abs/2017MNRAS.467.2085G} {467, 2085}

\bibitem[\protect\citeauthoryear{{Hemantha}, {Wang}  \& {Chuang}}{{Hemantha}
  et~al.}{2014}]{2014MNRAS.445.3737H}
{Hemantha} M. D.~P.,  {Wang} Y.,   {Chuang} C.-H.,  2014, \mn@doi [\mnras]
  {10.1093/mnras/stu1997}, \href
  {https://ui.adsabs.harvard.edu/abs/2014MNRAS.445.3737H} {445, 3737}

\bibitem[\protect\citeauthoryear{{Hong}, {Jiao}, {Wang}, {Zhang}  \&
  {Zhang}}{{Hong} et~al.}{2023}]{2023ApJS..268...67H}
{Hong} W.,  {Jiao} K.,  {Wang} Y.-C.,  {Zhang} T.,   {Zhang} T.-J.,  2023,
  \mn@doi [\apjs] {10.3847/1538-4365/acf654}, \href
  {https://ui.adsabs.harvard.edu/abs/2023ApJS..268...67H} {268, 67}

\bibitem[\protect\citeauthoryear{{Hou} et~al.,}{{Hou}
  et~al.}{2021}]{2021MNRAS.500.1201H}
{Hou} J.,  et~al., 2021, \mn@doi [\mnras] {10.1093/mnras/staa3234}, \href
  {https://ui.adsabs.harvard.edu/abs/2021MNRAS.500.1201H} {500, 1201}

\bibitem[\protect\citeauthoryear{{Hwang}, {L'Huillier}, {Keeley}, {Jee}  \&
  {Shafieloo}}{{Hwang} et~al.}{2023}]{2023JCAP...02..014H}
{Hwang} S.-g.,  {L'Huillier} B.,  {Keeley} R.~E.,  {Jee} M.~J.,   {Shafieloo}
  A.,  2023, \mn@doi [\jcap] {10.1088/1475-7516/2023/02/014}, \href
  {https://ui.adsabs.harvard.edu/abs/2023JCAP...02..014H} {2023, 014}

\bibitem[\protect\citeauthoryear{{Icaza-Lizaola} et~al.,}{{Icaza-Lizaola}
  et~al.}{2020}]{2020MNRAS.492.4189I}
{Icaza-Lizaola} M.,  et~al., 2020, \mn@doi [\mnras] {10.1093/mnras/stz3602},
  \href {https://ui.adsabs.harvard.edu/abs/2020MNRAS.492.4189I} {492, 4189}

\bibitem[\protect\citeauthoryear{{Im} et~al.,}{{Im}
  et~al.}{2017}]{2017ApJ...849L..16I}
{Im} M.,  et~al., 2017, \mn@doi [\apjl] {10.3847/2041-8213/aa9367}, \href
  {https://ui.adsabs.harvard.edu/abs/2017ApJ...849L..16I} {849, L16}

\bibitem[\protect\citeauthoryear{{Jee}, {Komatsu}  \& {Suyu}}{{Jee}
  et~al.}{2015}]{2015JCAP...11..033J}
{Jee} I.,  {Komatsu} E.,   {Suyu} S.~H.,  2015, \mn@doi [\jcap]
  {10.1088/1475-7516/2015/11/033}, \href
  {https://ui.adsabs.harvard.edu/abs/2015JCAP...11..033J} {2015, 033}

\bibitem[\protect\citeauthoryear{Jeffreys}{Jeffreys}{1998}]{10.1093/oso/9780198503682.001.0001}
Jeffreys H.,  1998, {Theory of Probability}.
Oxford University Press, \mn@doi{10.1093/oso/9780198503682.001.0001}, \url
  {https://doi.org/10.1093/oso/9780198503682.001.0001}

\bibitem[\protect\citeauthoryear{{Jennings} \& {Madigan}}{{Jennings} \&
  {Madigan}}{2017}]{2017A&C....19...16J}
{Jennings} E.,  {Madigan} M.,  2017, \mn@doi [Astronomy and Computing]
  {10.1016/j.ascom.2017.01.001}, \href
  {https://ui.adsabs.harvard.edu/abs/2017A&C....19...16J} {19, 16}

\bibitem[\protect\citeauthoryear{{Jiao}, {Borghi}, {Moresco}  \&
  {Zhang}}{{Jiao} et~al.}{2023}]{2023ApJS..265...48J}
{Jiao} K.,  {Borghi} N.,  {Moresco} M.,   {Zhang} T.-J.,  2023, \mn@doi [\apjs]
  {10.3847/1538-4365/acbc77}, \href
  {https://ui.adsabs.harvard.edu/abs/2023ApJS..265...48J} {265, 48}

\bibitem[\protect\citeauthoryear{{Jimenez} \& {Loeb}}{{Jimenez} \&
  {Loeb}}{2002}]{2002ApJ...573...37J}
{Jimenez} R.,  {Loeb} A.,  2002, \mn@doi [\apj] {10.1086/340549}, \href
  {https://ui.adsabs.harvard.edu/abs/2002ApJ...573...37J} {573, 37}

\bibitem[\protect\citeauthoryear{{Jimenez}, {Moresco}, {Verde}  \&
  {Wandelt}}{{Jimenez} et~al.}{2023}]{2023JCAP...11..047J}
{Jimenez} R.,  {Moresco} M.,  {Verde} L.,   {Wandelt} B.~D.,  2023, \mn@doi
  [\jcap] {10.1088/1475-7516/2023/11/047}, \href
  {https://ui.adsabs.harvard.edu/abs/2023JCAP...11..047J} {2023, 047}

\bibitem[\protect\citeauthoryear{{Jones} et~al.,}{{Jones}
  et~al.}{2004}]{2004MNRAS.355..747J}
{Jones} D.~H.,  et~al., 2004, \mn@doi [\mnras]
  {10.1111/j.1365-2966.2004.08353.x}, \href
  {https://ui.adsabs.harvard.edu/abs/2004MNRAS.355..747J} {355, 747}

\bibitem[\protect\citeauthoryear{{Jones}, {Saunders}, {Read}  \&
  {Colless}}{{Jones} et~al.}{2005}]{2005PASA...22..277J}
{Jones} D.~H.,  {Saunders} W.,  {Read} M.,   {Colless} M.,  2005, \mn@doi
  [\pasa] {10.1071/AS05018}, \href
  {https://ui.adsabs.harvard.edu/abs/2005PASA...22..277J} {22, 277}

\bibitem[\protect\citeauthoryear{{Jones} et~al.,}{{Jones}
  et~al.}{2009}]{2009MNRAS.399..683J}
{Jones} D.~H.,  et~al., 2009, \mn@doi [\mnras]
  {10.1111/j.1365-2966.2009.15338.x}, \href
  {https://ui.adsabs.harvard.edu/abs/2009MNRAS.399..683J} {399, 683}

\bibitem[\protect\citeauthoryear{{Kazin} et~al.,}{{Kazin}
  et~al.}{2014}]{2014MNRAS.441.3524K}
{Kazin} E.~A.,  et~al., 2014, \mn@doi [\mnras] {10.1093/mnras/stu778}, \href
  {https://ui.adsabs.harvard.edu/abs/2014MNRAS.441.3524K} {441, 3524}

\bibitem[\protect\citeauthoryear{{Kugel} et~al.,}{{Kugel}
  et~al.}{2023}]{2023arXiv230605492K}
{Kugel} R.,  et~al., 2023, \mn@doi [arXiv e-prints]
  {10.48550/arXiv.2306.05492}, \href
  {https://ui.adsabs.harvard.edu/abs/2023arXiv230605492K} {p. arXiv:2306.05492}

\bibitem[\protect\citeauthoryear{{Liao}}{{Liao}}{2019}]{2019ApJ...883....3L}
{Liao} K.,  2019, \mn@doi [\apj] {10.3847/1538-4357/ab39e6}, \href
  {https://ui.adsabs.harvard.edu/abs/2019ApJ...883....3L} {883, 3}

\bibitem[\protect\citeauthoryear{{Liberati} \& {Maccione}}{{Liberati} \&
  {Maccione}}{2009}]{2009ARNPS..59..245L}
{Liberati} S.,  {Maccione} L.,  2009, \mn@doi [Annual Review of Nuclear and
  Particle Science] {10.1146/annurev.nucl.010909.083640}, \href
  {https://ui.adsabs.harvard.edu/abs/2009ARNPS..59..245L} {59, 245}

\bibitem[\protect\citeauthoryear{Linder}{Linder}{2003}]{PhysRevLett.90.091301}
Linder E.~V.,  2003, \mn@doi [Phys. Rev. Lett.]
  {10.1103/PhysRevLett.90.091301}, 90, 091301

\bibitem[\protect\citeauthoryear{Magueijo}{Magueijo}{2000}]{Magueijo:2000zt}
Magueijo J.,  2000, \mn@doi [Phys. Rev. D] {10.1103/PhysRevD.62.103521}, 62,
  103521

\bibitem[\protect\citeauthoryear{{Magueijo}}{{Magueijo}}{2003}]{2003RPPh...66.2025M}
{Magueijo} J.,  2003, \mn@doi [Reports on Progress in Physics]
  {10.1088/0034-4885/66/11/R04}, \href
  {https://ui.adsabs.harvard.edu/abs/2003RPPh...66.2025M} {66, 2025}

\bibitem[\protect\citeauthoryear{{Martins}}{{Martins}}{2017}]{2017RPPh...80l6902M}
{Martins} C.~J.~A.~P.,  2017, \mn@doi [Reports on Progress in Physics]
  {10.1088/1361-6633/aa860e}, \href
  {https://ui.adsabs.harvard.edu/abs/2017RPPh...80l6902M} {80, 126902}

\bibitem[\protect\citeauthoryear{{Moffat}}{{Moffat}}{1993}]{1993IJMPD...2..351M}
{Moffat} J.~W.,  1993, \mn@doi [International Journal of Modern Physics D]
  {10.1142/S0218271893000246}, \href
  {https://ui.adsabs.harvard.edu/abs/1993IJMPD...2..351M} {2, 351}

\bibitem[\protect\citeauthoryear{{Moresco}}{{Moresco}}{2015}]{2015MNRAS.450L..16M}
{Moresco} M.,  2015, \mn@doi [\mnras] {10.1093/mnrasl/slv037}, \href
  {https://ui.adsabs.harvard.edu/abs/2015MNRAS.450L..16M} {450, L16}

\bibitem[\protect\citeauthoryear{{Moresco} et~al.,}{{Moresco}
  et~al.}{2012}]{2012JCAP...08..006M}
{Moresco} M.,  et~al., 2012, \mn@doi [\jcap] {10.1088/1475-7516/2012/08/006},
  \href {https://ui.adsabs.harvard.edu/abs/2012JCAP...08..006M} {2012, 006}

\bibitem[\protect\citeauthoryear{{Moresco} et~al.,}{{Moresco}
  et~al.}{2016}]{2016JCAP...05..014M}
{Moresco} M.,  et~al., 2016, \mn@doi [\jcap] {10.1088/1475-7516/2016/05/014},
  \href {https://ui.adsabs.harvard.edu/abs/2016JCAP...05..014M} {2016, 014}

\bibitem[\protect\citeauthoryear{Morey, Romeijn  \& Rouder}{Morey
  et~al.}{2016}]{MOREY20166}
Morey R.~D.,  Romeijn J.-W.,   Rouder J.~N.,  2016, \mn@doi [Journal of
  Mathematical Psychology] {https://doi.org/10.1016/j.jmp.2015.11.001}, 72, 6

\bibitem[\protect\citeauthoryear{{Mukherjee} \& {Banerjee}}{{Mukherjee} \&
  {Banerjee}}{2022}]{2022PDU....3600998M}
{Mukherjee} P.,  {Banerjee} N.,  2022, \mn@doi [Physics of the Dark Universe]
  {10.1016/j.dark.2022.100998}, \href
  {https://ui.adsabs.harvard.edu/abs/2022PDU....3600998M} {36, 100998}

\bibitem[\protect\citeauthoryear{{Parkinson} et~al.,}{{Parkinson}
  et~al.}{2012}]{2012PhRvD..86j3518P}
{Parkinson} D.,  et~al., 2012, \mn@doi [\prd] {10.1103/PhysRevD.86.103518},
  \href {https://ui.adsabs.harvard.edu/abs/2012PhRvD..86j3518P} {86, 103518}

\bibitem[\protect\citeauthoryear{Pedregosa et~al.,}{Pedregosa
  et~al.}{2011}]{scikit-learn}
Pedregosa F.,  et~al., 2011, Journal of Machine Learning Research, 12, 2825

\bibitem[\protect\citeauthoryear{{Perivolaropoulos} \&
  {Skara}}{{Perivolaropoulos} \& {Skara}}{2022}]{2022NewAR..9501659P}
{Perivolaropoulos} L.,  {Skara} F.,  2022, \mn@doi [\nar]
  {10.1016/j.newar.2022.101659}, \href
  {https://ui.adsabs.harvard.edu/abs/2022NewAR..9501659P} {95, 101659}

\bibitem[\protect\citeauthoryear{{Rasmussen} \& {Williams}}{{Rasmussen} \&
  {Williams}}{2006}]{2006gpml.book.....R}
{Rasmussen} C.~E.,  {Williams} C. K.~I.,  2006, {Gaussian Processes for Machine
  Learning}

\bibitem[\protect\citeauthoryear{{Ratsimbazafy}, {Loubser}, {Crawford},
  {Cress}, {Bassett}, {Nichol}  \& {V{\"a}is{\"a}nen}}{{Ratsimbazafy}
  et~al.}{2017}]{2017MNRAS.467.3239R}
{Ratsimbazafy} A.~L.,  {Loubser} S.~I.,  {Crawford} S.~M.,  {Cress} C.~M.,
  {Bassett} B.~A.,  {Nichol} R.~C.,   {V{\"a}is{\"a}nen} P.,  2017, \mn@doi
  [\mnras] {10.1093/mnras/stx301}, \href
  {https://ui.adsabs.harvard.edu/abs/2017MNRAS.467.3239R} {467, 3239}

\bibitem[\protect\citeauthoryear{{Reid} et~al.,}{{Reid}
  et~al.}{2012}]{2012MNRAS.426.2719R}
{Reid} B.~A.,  et~al., 2012, \mn@doi [\mnras]
  {10.1111/j.1365-2966.2012.21779.x}, \href
  {https://ui.adsabs.harvard.edu/abs/2012MNRAS.426.2719R} {426, 2719}

\bibitem[\protect\citeauthoryear{{Rodrigues} \& {Bengaly}}{{Rodrigues} \&
  {Bengaly}}{2022}]{2022JCAP...07..029R}
{Rodrigues} G.,  {Bengaly} C.,  2022, \mn@doi [\jcap]
  {10.1088/1475-7516/2022/07/029}, \href
  {https://ui.adsabs.harvard.edu/abs/2022JCAP...07..029R} {2022, 029}

\bibitem[\protect\citeauthoryear{Salzano, Dabrowski  \& Lazkoz}{Salzano
  et~al.}{2015}]{PhysRevLett.114.101304}
Salzano V.,  Dabrowski M.~P.,   Lazkoz R.,  2015, \mn@doi [Phys. Rev. Lett.]
  {10.1103/PhysRevLett.114.101304}, 114, 101304

\bibitem[\protect\citeauthoryear{{Samushia} et~al.,}{{Samushia}
  et~al.}{2014}]{2014MNRAS.439.3504S}
{Samushia} L.,  et~al., 2014, \mn@doi [\mnras] {10.1093/mnras/stu197}, \href
  {https://ui.adsabs.harvard.edu/abs/2014MNRAS.439.3504S} {439, 3504}

\bibitem[\protect\citeauthoryear{{S{\'a}nchez} et~al.,}{{S{\'a}nchez}
  et~al.}{2017}]{2017MNRAS.464.1640S}
{S{\'a}nchez} A.~G.,  et~al., 2017, \mn@doi [\mnras] {10.1093/mnras/stw2443},
  \href {https://ui.adsabs.harvard.edu/abs/2017MNRAS.464.1640S} {464, 1640}

\bibitem[\protect\citeauthoryear{{Satpathy} et~al.,}{{Satpathy}
  et~al.}{2017}]{2017MNRAS.469.1369S}
{Satpathy} S.,  et~al., 2017, \mn@doi [\mnras] {10.1093/mnras/stx883}, \href
  {https://ui.adsabs.harvard.edu/abs/2017MNRAS.469.1369S} {469, 1369}

\bibitem[\protect\citeauthoryear{Schwarz}{Schwarz}{1978}]{10.1214/aos/1176344136}
Schwarz G.,  1978, \mn@doi [The Annals of Statistics] {10.1214/aos/1176344136},
  6, 461

\bibitem[\protect\citeauthoryear{{Seikel}, {Clarkson}  \& {Smith}}{{Seikel}
  et~al.}{2012}]{2012JCAP...06..036S}
{Seikel} M.,  {Clarkson} C.,   {Smith} M.,  2012, \mn@doi [\jcap]
  {10.1088/1475-7516/2012/06/036}, \href
  {https://ui.adsabs.harvard.edu/abs/2012JCAP...06..036S} {2012, 036}

\bibitem[\protect\citeauthoryear{Shafieloo, Kim  \& Linder}{Shafieloo
  et~al.}{2012a}]{PhysRevD.85.123530}
Shafieloo A.,  Kim A.~G.,   Linder E.~V.,  2012a, \mn@doi [Phys. Rev. D]
  {10.1103/PhysRevD.85.123530}, 85, 123530

\bibitem[\protect\citeauthoryear{{Shafieloo}, {Kim}  \& {Linder}}{{Shafieloo}
  et~al.}{2012b}]{2012PhRvD..85l3530S}
{Shafieloo} A.,  {Kim} A.~G.,   {Linder} E.~V.,  2012b, \mn@doi [\prd]
  {10.1103/PhysRevD.85.123530}, \href
  {https://ui.adsabs.harvard.edu/abs/2012PhRvD..85l3530S} {85, 123530}

\bibitem[\protect\citeauthoryear{{Simon}, {Verde}  \& {Jimenez}}{{Simon}
  et~al.}{2005}]{2005PhRvD..71l3001S}
{Simon} J.,  {Verde} L.,   {Jimenez} R.,  2005, \mn@doi [\prd]
  {10.1103/PhysRevD.71.123001}, \href
  {https://ui.adsabs.harvard.edu/abs/2005PhRvD..71l3001S} {71, 123001}

\bibitem[\protect\citeauthoryear{{Slosar} et~al.,}{{Slosar}
  et~al.}{2013}]{2013JCAP...04..026S}
{Slosar} A.,  et~al., 2013, \mn@doi [\jcap] {10.1088/1475-7516/2013/04/026},
  \href {https://ui.adsabs.harvard.edu/abs/2013JCAP...04..026S} {2013, 026}

\bibitem[\protect\citeauthoryear{{Stern}, {Jimenez}, {Verde}, {Stanford}  \&
  {Kamionkowski}}{{Stern} et~al.}{2010}]{2010ApJS..188..280S}
{Stern} D.,  {Jimenez} R.,  {Verde} L.,  {Stanford} S.~A.,   {Kamionkowski} M.,
   2010, \mn@doi [\apjs] {10.1088/0067-0049/188/1/280}, \href
  {https://ui.adsabs.harvard.edu/abs/2010ApJS..188..280S} {188, 280}

\bibitem[\protect\citeauthoryear{Stoica \& Selen}{Stoica \&
  Selen}{2004}]{1311138}
Stoica P.,  Selen Y.,  2004, \mn@doi [IEEE Signal Processing Magazine]
  {10.1109/MSP.2004.1311138}, 21, 36

\bibitem[\protect\citeauthoryear{{Stoughton} et~al.,}{{Stoughton}
  et~al.}{2002}]{2002AJ....123..485S}
{Stoughton} C.,  et~al., 2002, \mn@doi [\aj] {10.1086/324741}, \href
  {https://ui.adsabs.harvard.edu/abs/2002AJ....123..485S} {123, 485}

\bibitem[\protect\citeauthoryear{{Sun}, {Jiao}  \& {Zhang}}{{Sun}
  et~al.}{2021}]{2021ApJ...915..123S}
{Sun} W.,  {Jiao} K.,   {Zhang} T.-J.,  2021, \mn@doi [\apj]
  {10.3847/1538-4357/ac05b8}, \href
  {https://ui.adsabs.harvard.edu/abs/2021ApJ...915..123S} {915, 123}

\bibitem[\protect\citeauthoryear{{Tomasetti} et~al.,}{{Tomasetti}
  et~al.}{2023}]{2023AA...679A..96T}
{Tomasetti} E.,  et~al., 2023, \mn@doi [\aap] {10.1051/0004-6361/202346992},
  \href {https://ui.adsabs.harvard.edu/abs/2023A&A...679A..96T} {679, A96}

\bibitem[\protect\citeauthoryear{{Toni} \& {Stumpf}}{{Toni} \&
  {Stumpf}}{2009}]{2009arXiv0910.4472T}
{Toni} T.,  {Stumpf} M. P.~H.,  2009, \mn@doi [arXiv e-prints]
  {10.48550/arXiv.0910.4472}, \href
  {https://ui.adsabs.harvard.edu/abs/2009arXiv0910.4472T} {p. arXiv:0910.4472}

\bibitem[\protect\citeauthoryear{Turner \& {Van Zandt}}{Turner \& {Van
  Zandt}}{2012}]{TURNER201269}
Turner B.~M.,  {Van Zandt} T.,  2012, \mn@doi [Journal of Mathematical
  Psychology] {https://doi.org/10.1016/j.jmp.2012.02.005}, 56, 69

\bibitem[\protect\citeauthoryear{Uzan}{Uzan}{2011}]{Uzan:2010pm}
Uzan J.-P.,  2011, \mn@doi [Living Rev. Rel.] {10.12942/lrr-2011-2}, 14, 2

\bibitem[\protect\citeauthoryear{{Wang} et~al.,}{{Wang}
  et~al.}{2020}]{2020MNRAS.498.3470W}
{Wang} Y.,  et~al., 2020, \mn@doi [\mnras] {10.1093/mnras/staa2593}, \href
  {https://ui.adsabs.harvard.edu/abs/2020MNRAS.498.3470W} {498, 3470}

\bibitem[\protect\citeauthoryear{{Wang}, {Xie}, {Zhang}, {Huang}, {Zhang}  \&
  {Liu}}{{Wang} et~al.}{2021}]{2021ApJS..254...43W}
{Wang} Y.-C.,  {Xie} Y.-B.,  {Zhang} T.-J.,  {Huang} H.-C.,  {Zhang} T.,
  {Liu} K.,  2021, \mn@doi [\apjs] {10.3847/1538-4365/abf8aa}, \href
  {https://ui.adsabs.harvard.edu/abs/2021ApJS..254...43W} {254, 43}

\bibitem[\protect\citeauthoryear{Wong, Moss  \& Scott}{Wong
  et~al.}{2008}]{Wong:2007ym}
Wong W.~Y.,  Moss A.,   Scott D.,  2008, \mn@doi [Mon. Not. Roy. Astron. Soc.]
  {10.1111/j.1365-2966.2008.13092.x}, 386, 1023

\bibitem[\protect\citeauthoryear{{Xu}, {Cuesta}, {Padmanabhan}, {Eisenstein}
  \& {McBride}}{{Xu} et~al.}{2013}]{2013MNRAS.431.2834X}
{Xu} X.,  {Cuesta} A.~J.,  {Padmanabhan} N.,  {Eisenstein} D.~J.,   {McBride}
  C.~K.,  2013, \mn@doi [\mnras] {10.1093/mnras/stt379}, \href
  {https://ui.adsabs.harvard.edu/abs/2013MNRAS.431.2834X} {431, 2834}

\bibitem[\protect\citeauthoryear{{Yahya}, {Seikel}, {Clarkson}, {Maartens}  \&
  {Smith}}{{Yahya} et~al.}{2014}]{2014PhRvD..89b3503Y}
{Yahya} S.,  {Seikel} M.,  {Clarkson} C.,  {Maartens} R.,   {Smith} M.,  2014,
  \mn@doi [\prd] {10.1103/PhysRevD.89.023503}, \href
  {https://ui.adsabs.harvard.edu/abs/2014PhRvD..89b3503Y} {89, 023503}

\bibitem[\protect\citeauthoryear{{Ye}, {Jiang}  \& {Piao}}{{Ye}
  et~al.}{2023}]{2023arXiv230518873Y}
{Ye} G.,  {Jiang} J.-Q.,   {Piao} Y.-S.,  2023, \mn@doi [arXiv e-prints]
  {10.48550/arXiv.2305.18873}, \href
  {https://ui.adsabs.harvard.edu/abs/2023arXiv230518873Y} {p. arXiv:2305.18873}

\bibitem[\protect\citeauthoryear{{Yu}, {Yuan}  \& {Zhang}}{{Yu}
  et~al.}{2013}]{2013PhRvD..88j3528Y}
{Yu} H.-R.,  {Yuan} S.,   {Zhang} T.-J.,  2013, \mn@doi [\prd]
  {10.1103/PhysRevD.88.103528}, \href
  {https://ui.adsabs.harvard.edu/abs/2013PhRvD..88j3528Y} {88, 103528}

\bibitem[\protect\citeauthoryear{{Zarrouk} et~al.,}{{Zarrouk}
  et~al.}{2018}]{2018MNRAS.477.1639Z}
{Zarrouk} P.,  et~al., 2018, \mn@doi [\mnras] {10.1093/mnras/sty506}, \href
  {https://ui.adsabs.harvard.edu/abs/2018MNRAS.477.1639Z} {477, 1639}

\bibitem[\protect\citeauthoryear{Zhang, Zhang, Yuan, Zhang  \& Sun}{Zhang
  et~al.}{2014}]{Zhang:2012mp}
Zhang C.,  Zhang H.,  Yuan S.,  Zhang T.-J.,   Sun Y.-C.,  2014, \mn@doi [Res.
  Astron. Astrophys.] {10.1088/1674-4527/14/10/002}, 14, 1221

\bibitem[\protect\citeauthoryear{{Zhang}, {Wang}, {Zhang}  \& {Zhang}}{{Zhang}
  et~al.}{2023}]{2023ApJS..266...27Z}
{Zhang} H.,  {Wang} Y.-C.,  {Zhang} T.-J.,   {Zhang} T.,  2023, \mn@doi [\apjs]
  {10.3847/1538-4365/accb92}, \href
  {https://ui.adsabs.harvard.edu/abs/2023ApJS..266...27Z} {266, 27}

\bibitem[\protect\citeauthoryear{de Sainte~Agathe et~al.}{de~Sainte~Agathe
  et~al.}{2019}]{deSainteAgathe:2019voe}
de Sainte~Agathe V.,  et~al., 2019, \mn@doi [Astron. Astrophys.]
  {10.1051/0004-6361/201935638}, 629, A85

\makeatother
\end{thebibliography}

%\begin{thebibliography}{99}
%\end{thebibliography}

%%%%%%%%%%%%%%%%%%%%%%%%%%%%%%%%%%%%%%%%%%%%%%%%%%

%%%%%%%%%%%%%%%%% APPENDICES %%%%%%%%%%%%%%%%%%%%%

%%%%%%%%%%%%%%%%%%%%%%%%%%%%%%%%%%%%%%%%%%%%%%%%%%

% Don't change these lines
\bsp	% typesetting comment
\label{lastpage}
\end{CJK*}
\end{document}